\documentclass{aa}  
\usepackage{graphicx}
\usepackage{txfonts}
\usepackage{cancel}
\usepackage{amsmath}
\usepackage{lscape} 
\usepackage{longtable}

\newcommand{\bandg}{\textit{g}}
\newcommand{\bandr}{\textit{r}}
\newcommand{\bandi}{\textit{i}}
\newcommand{\bandz}{\textit{z}}
\newcommand{\HA}{H$\alpha$}
\newcommand{\Nii}{[\ion{N}{ii}]}
\newcommand{\HAN}{H$\alpha$ + [\ion{N}{ii}]}
\newcommand{\Sii}{[\ion{S}{ii}]$\lambda\lambda$6716,6731}
\newcommand{\z}{$z$}
\newcommand{\zphot}{$z_{\rm phot}$}

\newcommand{\otelodeep}{OTELO$_{\rm deep}$}
\newcommand{\oteloint}{OTELO$_{\rm Int}$}
\newcommand{\zotelo}{$z_{\rm OTELO}$}
\newcommand{\gmenosi}{$(g\,-\,i)$}
\newcommand{\hst}{\textit{HST}}
\newcommand{\hstacs}{\textit{HST}-ACS}
\newcommand{\msun}{$M_\odot$}
\newcommand{\mstar}{$M_*$}
\newcommand{\logmass}{$\log M_*$}
\newcommand{\sedbest}{SED$_{\rm best}$}
\newcommand{\nmean}{$\overline{n}$}
\newcommand{\absMi}{$M_i$}
\newcommand{\zguess}{$z_{\rm GUESS}$}

\defcitealias{OteloI}{OTELO-I}
\defcitealias{OteloII}{OTELO-II}
\defcitealias{OteloIII}{OTELO-III}
\defcitealias{LopezSanjuan2018}{LS18}

\begin{document}

        \title{The OTELO survey: \\ Nature and mass-metallicity relation for \HA\ emitters at $z\sim\,0.4$.}
        
        \titlerunning{MZR of the OTELO H$\alpha$ sources}
        
        \author{Jakub Nadolny\inst{1,2}
                \and Maritza A. Lara-L\'opez\inst{3} 
                \and Miguel Cervi\~no\inst{4}   
                \and \'Angel Bongiovanni\inst{5,1,2,6}
                \and Jordi Cepa\inst{1,2,6}
                \and Jos\'e A. de Diego \inst{7}
                \and Ana Mar\'ia P\'erez Garc\'ia\inst{4}
                \and Ricardo P\'erez Mart\'inez\inst{6,8}
                \and Miguel S\'anchez-Portal\inst{5,6}
                \and Emilio Alfaro\inst{9}
                \and H\'ector O. Casta\~neda \inst{10}
                \and Jes\'us Gallego \inst{11}
                \and J. Jes\'us Gonz\'alez \inst{7}
                \and J. Ignacio Gonz\'alez-Serrano \inst{12,6}
                \and Carmen P. Padilla Torres \inst{1,2,13}
                \and Irene Pintos-Castro \inst{14}
                \and Mirjana Povi\'c \inst{15}
        }
        
        \institute{Instituto de Astrof\'isica de Canarias, E-38205 La Laguna, Tenerife, Spain 
                \and Universidad de La Laguna, Dept. Astrof\'isica, E-38206 La Laguna, Tenerife, Spain 
                \and  DARK, Niels Bohr Institute, University of Copenhagen, Lyngbyvej 2, Copenhagen DK-2100, Denmark 
                \and Centro de Astrobiolog\'ia (CSIC/INTA), 28850 Torrej\'on de Ardoz, Madrid, Spain 
                \and Instituto de Radioastronom\'ia Milim\'etrica (IRAM), Av. Divina Pastora 7, N\'ucleo Central, E-18012 Granada, Spain 
                \and Asociaci\'on Astrof\'isica para la Promoci\'on de la Investigaci\'on, Instrumentaci\'on y su Desarrollo, ASPID, E-38205 La Laguna, Tenerife, Spain 
                \and Instituto de Astronom\'ia, Universidad Nacional Aut\'onoma de M\'exico, Apdo. Postal 70-264, 04510 Ciudad de M\'exico, Mexico 
                \and ISDEFE for European Space Astronomy Centre (ESAC)/ESA, P.O. Box 78, E-28690, Villanueva de la Ca\~nada, Madrid, Spain 
                \and Instituto de Astrof\'isica de Andaluc\'ia, CSIC, E-18080, Granada, Spain  
                \and Departamento de F\'isica, Escuela Superior de F\'isica y Matem\'aticas, Instituto Polit\'ecnico Nacional, M\'exico D.F., Mexico 
        \and Departamento de F\'isica de la Tierra y Astrof\'isica, Facultad CC F\'isicas, Instituto de F\'isica de Part\'iculas y del Cosmos, IPARCOS, Universidad Complutense de Madrid 
                \and Instituto de F\'isica de Cantabria (CSIC-Universidad de Cantabria), E-39005 Santander, Spain 
                \and INAF, Telescopio Nazionale Galileo, Apartado de Correos 565, E-38700 Santa Cruz de la Palma, Spain 
                \and Department of Astronomy \& Astrophysics, University of Toronto, Canada 
                \and Ethiopian Space Science and Technology Institute (ESSTI), Entoto Observatory and Research Center (EORC), Astronomy and Astrophysics Research Division, PO Box 33679, Addis Abbaba, Ethiopia 
                \\
                \email{jnadolny@iac.es; quba.nadolny@gmail.com}
        }
        
        \date{Received 28 June 2019; accepted 01 March 2020}

        \abstract
        {A sample of low-mass H$\alpha$ emission line sources (ELS) at $z\,\sim\,0.4$ was studied in the context of the mass-metallicty relation (MZR) and its possible evolution. We drew our sample from the OSIRIS Tunable Emission Line Object (OTELO) survey, which exploits the red tunable filter of OSIRIS at the Gran Telescopio Canarias to perform a blind narrow-band spectral scan in a selected field of the Extended Groth Strip. We were able to directly measure emission line fluxes and equivalent widths from the analysis of OTELO pseudo-spectra.}
        {This study aims to explore the MZR in the very low-mass regime. Our sample reaches stellar masses (\mstar) as low as $10^{6.8}$\msun, where 63\% of the sample have \mstar$\,<10^9\,$\msun. We also explore the relation of the star formation rate (SFR) and specific SFR (sSFR) with \mstar\ and gas-phase oxygen abundances, as well as the \mstar-size relation and the morphological classification.}
        {The \mstar\ were estimated using synthetic rest-frame colours. Using an $\chi^2$ minimization method, we separated the contribution of \Nii$\lambda$6583 to the \HA\ emission lines. Using the N2 index, we separated active galactic nuclei from star-forming galaxies (SFGs) and estimated the gas metallicity. We studied the morphology of the sampled galaxies qualitatively (visually) and quantitatively (automatically) using high-resolution data from the \textit{Hubble Space Telescope}-ACS. The physical size of the galaxies was derived from the morphological analysis using \texttt{GALAPAGOS2/GALFIT}, where we fit a single-S\'ersic 2D model to each source.}
        {We find no evidence for an MZR evolution from comparing our very low-mass sample with local SFGs from the Sloan Digital Sky Survey. Furthermore, the same is true for \mstar-size and \mstar-SFR relations, as we deduce from comparison with recent literature. Morphologically, our sample is mostly (63\%) populated by late-type galaxies, with 13\% of early-type sources. For the first time, we identify one possible candidate outlier in the MZR at $z\,=\,0.4$. The stellar-mass, metallicity, colour, morphology, and SFR of this source suggest that it is compatible with a transitional dwarf galaxy.}
        {}
        
        
        \maketitle
        
        \section{Introduction}
        \label{Introduction}
        
        The stellar mass (\mstar) and gas-phase metallicity (\textit{Z}) are among the most fundamental properties of galaxies and tracers of galaxy formation and evolution.  While the stellar mass provides information about the amount of gas that is locked up into stars, the metallicity gives indications on the history of star formation and the exchange of gas between an object and its environment.
        
        The mass-metallicity relation (MZR) has been studied for several decades, beginning with the pioneering work of \citet{Lequeux1979}. 
        Since then, several authors have investigated the origin and behaviour of this relationship in low and high redshifts,  and in different environments and mass regimes. \citet{Tremonti2004ApJ...613..898T} studied this relation for $>53\,000$ star-forming galaxies (SFGs) in a demonstration of the statistical power of the Sloan Digital Sky Survey (SDSS). They found a tight ($\pm\,0.1$ dex) correlation between stellar mass and metallicity spanning over three orders of magnitude in stellar mass (down to $10^{8.5}$\msun) and a factor of 10 in metallicity, with a median redshift of 0.1. 
        Possible explanations of the origin of the MZR include
        (i) gas outflows. In this scenario,  low-mass galaxies suffer a stronger effect because they have lower escape velocities and can more easily loose enriched gas through stellar winds than massive galaxies. This scenario was first proposed by \citet[][see also \citealt{Spitoni2010A&A...514A..73S} and \citealt{Tremonti2004ApJ...613..898T}]{Larson_1974MNRAS.169..229L}. Furthermore, \citet{Belfiore2019} recently found that the inferred outflow loading factor decreases with stellar mass.
        (ii) The second possible explanation might be the downsizing effect: massive galaxies process their gas faster, on shorter timescales and at earlier epochs than low-mass galaxies, where the star formation is slower and extends over longer periods \citep[e.g.][]{Cowie1996AJ....112..839C,Thomas2010MNRAS.404.1775T}. 
        (iii) Another possible mechanism for the MZR is the dependence of stellar mass or star formation rate (SFR) on the initial mass function  \citep[IMF; e.g.][]{Koppen2007MNRAS.375..673K,Gunawardhana2011MNRAS.415.1647G}. However, it is still a matter of debate which of these mechanisms or which combination of them plays a more important role in the formation and evolution of galaxies.
        
        With the advent of large surveys, there has been an effort to observe fainter and higher redshift objects because they would provide strong constraints on our understanding of how galaxies evolve. For instance, the Galaxy and Mass Assembly \citep[GAMA;][]{GAMA_Driver2011MNRAS.413..971D, Liske2015} survey observed $\sim\, 300\,000$ galaxies two orders of magnitude deeper than SDSS, covering an area of $\sim\,286$ deg$^2$. In the survey, an evolution of metallicity of $\sim\,0.1$ dex was found for massive galaxies at redshift  $\sim\,0.35$ \citep{Maritza2013MNRAS.433L..35L}. At higher redshifts, the VIMOS Very Large Telescope (VLT) Deep Survey \citep[VVDS;][]{VVDS2013A&A...559A..14L} provides spectroscopic data of > 35,000 galaxies up to redshift \z$\sim$ 6.7 over a total of > 8 deg$^2$ down to $i_{\rm AB}=24.75$. Using VVDS data, \citet{Lamareille2009A&A...495...53L} studied the MZR up to $z\,\sim\,0.9$. They defined two samples: a wide ($i_{\rm AB}$ $<$ 22.5) and a deep sample ($i_{\rm AB}\,<24$). They used three different metallicity calibrators depending on the redshift range. All methods were then normalized to the \citet{Charlot_Longhetti2001MNRAS.323..887C} calibrator, which is in agreement with the \citet{Tremonti2004ApJ...613..898T} metallicities. Assuming that the MZR shape is constant with redshift, they found a stronger metallicity evolution in the wide sample and concluded that the MZR is flatter at higher redshift. They arrived at the same conclusion for the luminosity-metallicity relation (LZR) in the wide sample.
        
        There is an incentive to also analyse the low-mass end of the MZR because a flattening in this relation has been reported. For instance, \citet{Jimmy2015ApJ...812...98J} found a flattening of the MZR relation using a sample of local (D < 20 Mpc) dwarf galaxies with the VIMOS integral field unit (IFU) spectrograph on the VLT. They also found a clear dependence of the MZR and LZR on SFR and \ion{H}{i}-gas mass. They used this finding to explain the observed scatter at the low-mass ($<\,10^8$\msun) end of these relations.
Using a sample of 25 dwarf galaxies, \citet{Lee2006ApJ...647..970L} also extended the MZR for galaxies down to a \mstar $\sim\,10^6$ \msun, roughly 2.5 dex lower in stellar masses than \citet{Tremonti2004ApJ...613..898T}. Using \textit{Spitzer} mid-IR (MIR) (4.5 $\mu$m) photometry, they found a comparable scatter in the MZR over 5.5 dex in stellar mass. They concluded that if galactic winds are responsible for the MZR, increasing dispersion is expected, as they observed in their analysis.
        
        A clear extreme in the MZR is given by extremely metal-deficient (XMD) galaxies ($\leq$ 0.1 $Z_{\rm \odot}$). As found by \citet{Thuan2016}, XMD galaxies show a clear trend of increasing gas mass fraction with decreasing metallicity, mass, and luminosity. On the same line, \citet{SanchezAlmeida2014, SanchezAlmeida2015} have argued that the low metallicities of the XMD galaxies are an indicator of infall of pristine gas. However, \citet{Thuan2005ApJS..161..240T} found that low metallicity alone cannot explain the hard ionising radiation observed in blue compact dwarf (BCD) galaxies, and proposed fast radiative shocks (velocity > 450 km s$^{-1}$) associated with supernovae explosions of massive Population III stars.
        \citet{Ekta2010MNRAS.406.1238E} examined the MZR of XMD galaxies and compared it with the MZR of BCD and dwarf irregular galaxies. They proposed that XMD galaxies are extremely metal depleted due to better mixing of the inter-stellar medium (ISM).

        The fact that there are outliers of the MZR makes the situation even more interesting.
        \citet{Peeples2008ApJ...685..904P} found a sample of 41 local ($z\,<\,0.048$), low-mass, high-oxygen abundance outliers from the \citet{Tremonti2004ApJ...613..898T} SDSS-based relation. They argued that these compact, isolated, and morphological undisturbed sources may be transitional dwarf galaxies. Their red colours (approaching or on the red sequence) and low gas fraction suggest that they are in the final stage of their star formation period. Moreover, using a larger sample of galaxies from the SDSS and DEEP2 surveys with stellar masses down to $10^{7}$\msun, \citet{Zahid2012ApJ...750..120Z}  found that the properties of low-mass metal-rich galaxies (the outliers from MZR) are, in agreement with previous works, possible transitional objects between gas-rich dwarf irregular and gas-poor dwarf spheroidal and elliptical galaxies. 
        
        In consideration of previous studies, we wish to shed light on the low-mass end of the MZR. To this aim, we used data from the OSIRIS Tunable filter Emission-Line Object survey reported by \citet[OTELO;][hereafter OTELO-I]{OteloI} to study a sample of low-mass \HA\ emission line sources (ELS) at redshift $\sim\,0.4$, firstly described in \citet[][hereafter OTELO-II]{OteloII}. As demonstrated in \citet{Otelo-OIII}, OTELO has allowed validation of the narrow-band (NB) scan technique using tunable filters (TFs) for finding faint populations of SFGs, which makes this survey the most sensitive to date in terms of minimum emission-line flux ($\sim\,5\,\times\,10^{-19}$ erg s$^{-1}$ cm$^{-2}$) and observed equivalent width (EW$\approx\,6$ \AA) achieved.
        
        This article is organized as follows. In Section \ref{sec:data} we present the data and sample selection process together with its characteristics and the external catalogues we used. In Section \ref{sec:methods} we describe the methods for estimating stellar mass, gas metallicity, SFR, and morphology determination from OTELO data for the science case of this study. In Section \ref{sec:analysis} we present our analysis and discussion, followed with our conclusions in Section \ref{sec:conclusion}. When necessary, we adopt the cosmology from \citet{Planck2016A&A...594A..13P} with $\Omega_{\Lambda}\,=0.69$, $\Omega_{m}\,=0.31$, and H$_{0}\,=67.8$ km s$^{-1}$ Mpc$^{-1}$.

        \section{Data and sample selection}
        \label{sec:data}
        We study a sample of \HA\ ELSs with stellar masses in the range of $10^{6.8} <\, M_*/M_{\rm \odot}\,< 10^{10}$, which places our sample in the low-mass regime. To this aim, we used data products of OTELO. A brief description follows.
        
        The core of the OTELO survey consists of red tunable filter (RTF) observations, which were carried out in a total of 108 dark hours with the OSIRIS \citep{OSIRIS_2000SPIE.4008..623C} instrument at the 10.4 m Gran Telescopio Canarias\footnote{\url{http://www.gtc.iac.es/gtc/gtc.php}} (GTC). These observations cover a $\sim$56 arcmin$^2$ section of the Extended Groth Strip (EGS) region in six dithered pointings. The nominal central wavelengths at the centre of the field of view (FOV) range from 9070 to 9280\AA\ with $\Delta \lambda$ of 6\AA\ steps. A full width at half-maximum (FWHM) of 12\AA\ was adopted for each one of the resulting 36 RTF tunings. This sampling of the RTF is optimal for deblending the \HA\ and \Nii$\lambda6583$ lines \citep{LL+2010Ha} for systems at \z$\sim0.4$. In addition, all RTF images were combined in a single \otelodeep\ image.
        A source detection on this image was performed using the {\tt SExtractor} software \citep{SEx1996}, providing 11\,237 sources. The integrated flux (\oteloint) of the sources detected in this image has a limiting magnitude of 26.4 [AB] (50\% completeness).
        
        The \otelodeep\ image was also used to measure the point spread function (psf)-model fluxes of these 11\,237 sources on registered and resampled public images in the optical and near-infrared (i.e. $u$, $g$, $r$, $i$, $z$, $J$, $H$ and $Ks$ bands) obtained from the AEGIS collaboration\footnote{\url{http://aegis.ucolick.org}}. These data were cross-matched with public catalogues in X-rays, far-UV (FUV) and near-UV (NUV)\footnote{\tt http://www.galex.caltech.edu}, as well as those from {\it Herschel} and {\it Spitzer} IR bands to obtain additional photometric data for the sources so correlated. In  addition, high-resolution\ F6006W and F814W image stamps from the ACS of the \textit{Hubble Space Telescope} (\hst) of each source, when found, were obtained for further studies, in particular for morphological classification.

        Related to the scope of this work, the resulting spectral energy distributions (SED) for each OTELO source were processed with the \textit{LePhare}\footnote{\url{http://www.cfht.hawaii.edu/~arnouts/LEPHARE/lephare.html}} code \citep{Arnouts1999,Ilbert2006} using a galaxy template library with the four standard Hubble types E/S0, Sbc, Scd, and Irr from \citet{Coleman1980} and six SFGs templates from \citet{Kinney1996}, and an active galactic nucleus/quasi-stellar object ({\tt AGN/QSOs)} library with two Seyfert, three type-1 and two type-2 QSOs and three composite (starburst+QSO) templates created by \citet{Polletta2007}. This procedure allowed us to obtain photometric redshift solutions \zphot, which were independently calculated including and excluding the \otelodeep\ photometry in the input SED. This provided a set of possible \zphot\ solutions, along with the two best-fit SED templates and the corresponding extinction $E(B-V)$ estimates. The final best solution was obtained by the analysis of the best $\chi^2$ solution obtained together with the estimated redshift error, $\sigma_{\rm Zphot}$, defined by
                \begin{equation}
                \sigma_{\rm Zphot} = \lvert \rm {\tt Z\_BEST\_LOW\_deepX} - {\tt Z\_BEST\_HIGH\_deepX} \rvert,
                \end{equation}
                where {\tt Z\_BEST\_LOW\_deepX} and {\tt Z\_BEST\_HIGH\_deepX} are the 1$\sigma$ deviations of the estimated redshift for the given object, while {\tt X} stands for {\tt Y} or {\tt N,} including or excluding the \otelodeep\ photometry in the SED fitting, respectively. Finally, the astrometric and photometric data for each source, along with the photometric redshift estimates and the redshift data from other surveys DEEP2 \citep{Newman_Deep2_2013ApJS..208....5N} and CFHTLS T0004 Deep3 \citep{2009A&A...500..981C}, were included in the dubbed OTELO multi-wavelength catalogue.
        
        The RTF scan data were also used to provide a pseudo-spectrum (PS) for each entry in the OTELO multi-wavelength catalogue. The PS are properly calibrated in flux and wavelength and differ from conventional spectra in that the former are affected by the distinctive RTF transmission profile. Such PS typically contain 36 wavelength points (R$\sim$700), but the actual number of points and the wavelength range covered by each PS depend on the position of the object in the FOV. Some PS examples are given in Figure \ref{fig:ps}. For each PS a pseudo-continuum flux ($f_\mathrm{c}$), which itself is defined as the median flux of the PS, and a deviation around $f_\mathrm{c}$ ($\sigma_{\rm med,\,c\,}$) were estimated. An object was qualified as an ELS candidate when either two adjacent points of the PS were above $f_\mathrm{c} + 2\sigma_{\rm med,\,c\,}$, or only one was above it but it has an adjacent point above $f_\mathrm{c} + \sigma_{\rm med,\,c\,}$. With these criteria, a list 5\,322 preliminary ELS candidates for the whole survey was obtained. Further details about the survey design, observations, data reduction, photometric redshift estimations, the construction of the multi-wavelength catalogue, the PS preparation, and the preliminary ELS selection are given in \citetalias{OteloI}.

        \subsection{\HAN\ sample selection}
        \label{sec:sample_selection}
        We used the OTELO data described above to select the \HA\ sample as follows:
        
	(a) According to \citetalias{OteloII}, we selected the sources where the best \zphot\ (including and excluding the \otelodeep\ photometry) was in the range  $0.2 < z_{\rm phot}< 0.5$ with the further constraint that $\sigma_{z_{\rm phot}}< 0.2\times(1+z_{\rm phot})$. The total number of sources in this redshift window is 2\,289. In this selection we lost some \HAN\ candidates whose best \zphot\ would be obtained with {\tt AGN/QSOs} libraries instead of galaxy templates. This does not affect the goals of this work. In addition, we are also aware that our selection includes possible \Sii\ at $z\sim0.3$. The best-fit templates correspond to Scd, Im, or SFGs types, with the exception of two sources where the best \zphot\ was provided by an E/S0 galaxy type. Within this set of 2\,289 sources, we found 73 preliminary ELS candidates to \HAN.
        \newline
        
        \begin{figure*}[t!]
                \begin{center}
                        \includegraphics[width=0.33\textwidth]{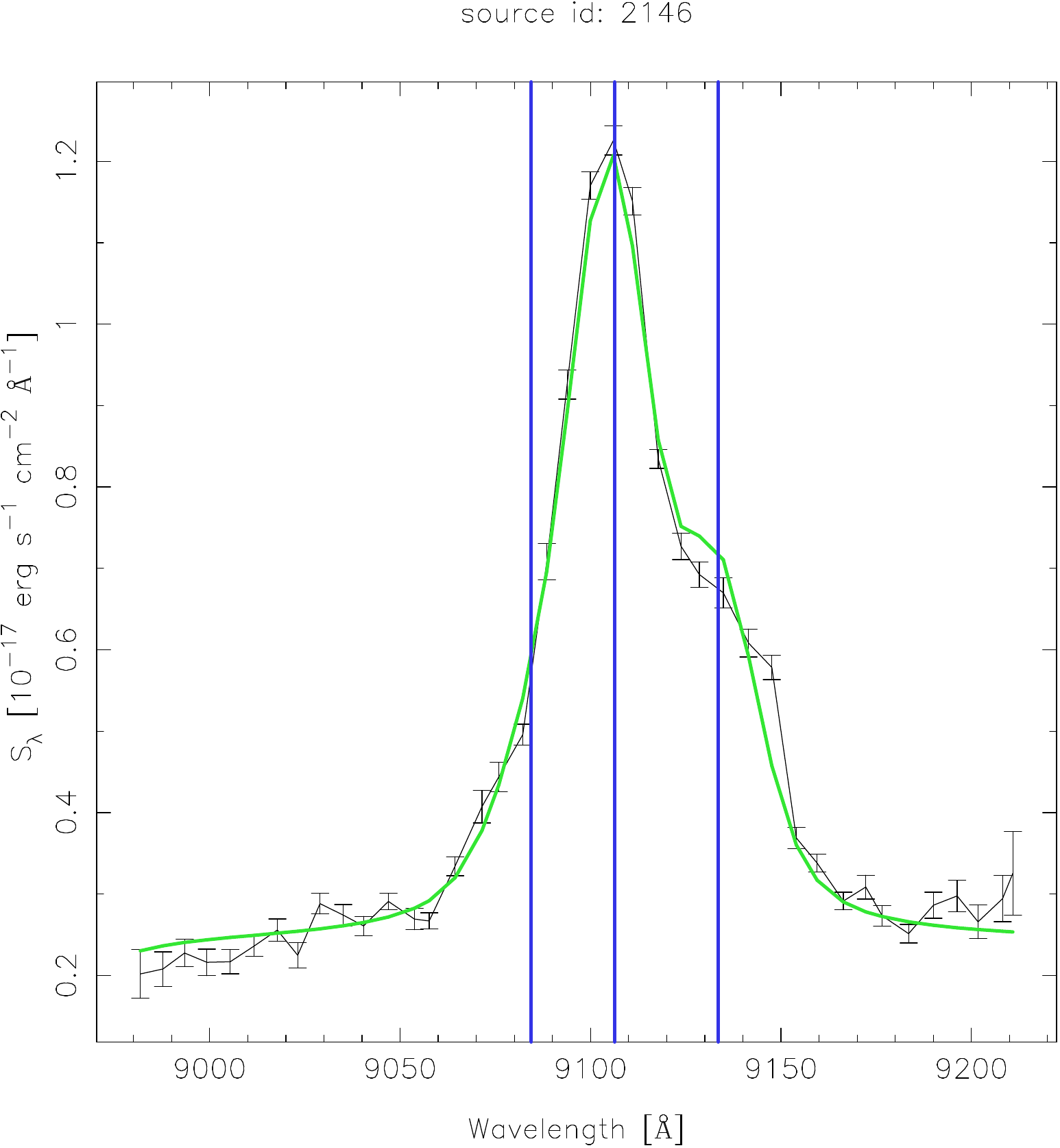}
                        \includegraphics[width=0.33\textwidth]{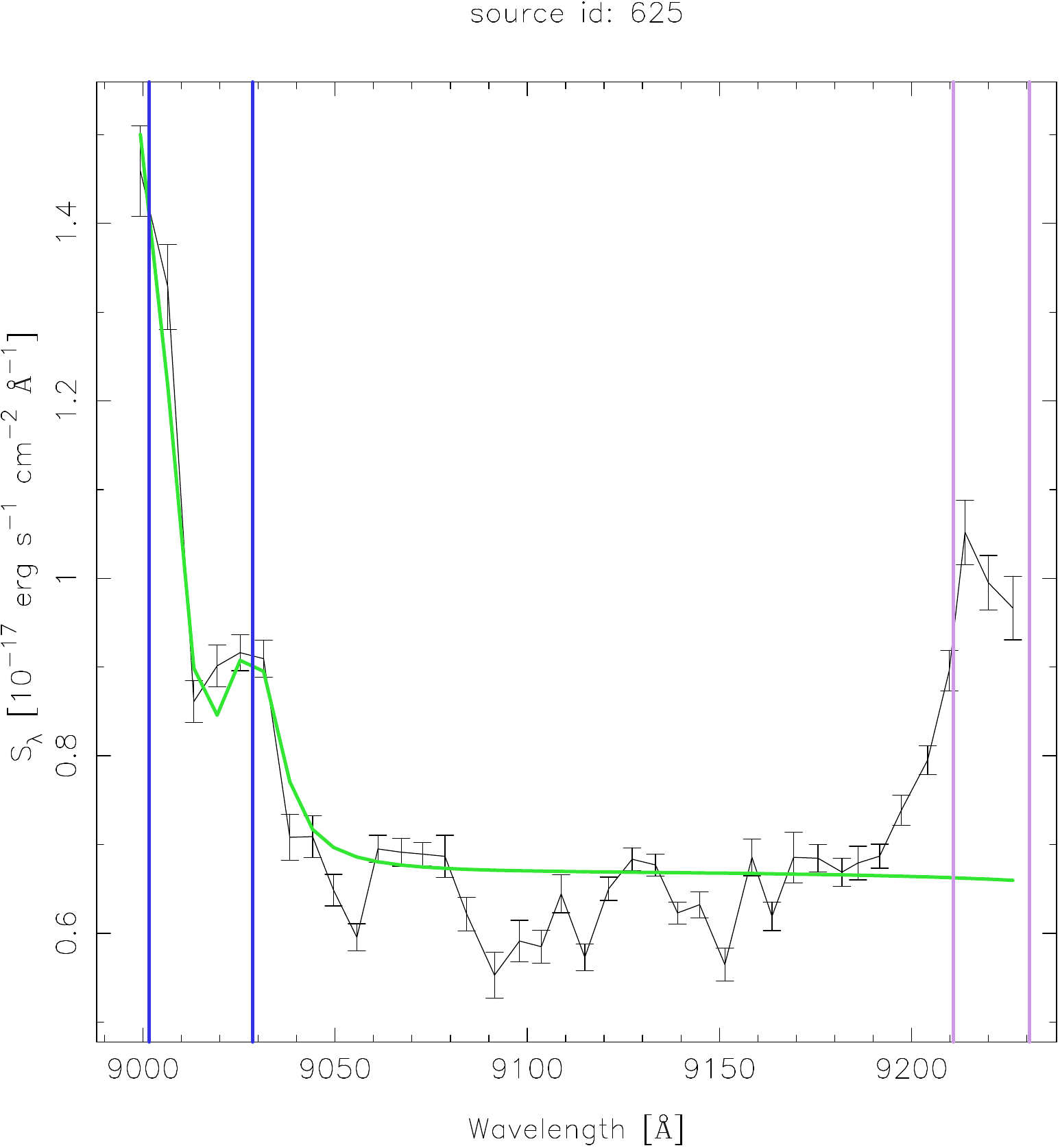}
                        \includegraphics[width=0.33\textwidth]{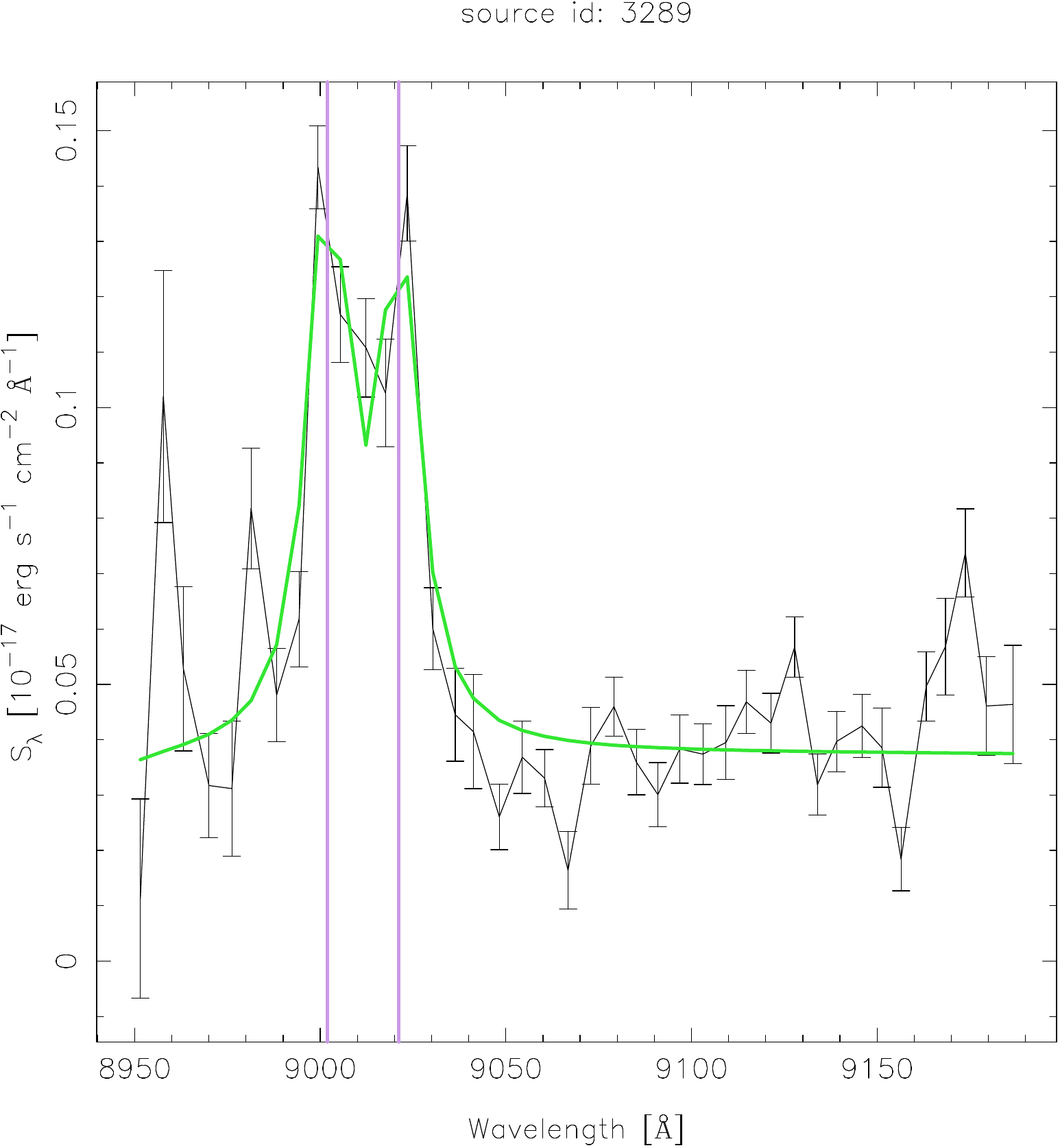}
                        \caption{\label{fig:ps} Pseudo-spectra of selected sources. From left: Source {\tt id: 2146} with a clear \HAN\ emission line complex at \z=0.387; source {\tt id: 625} with \HA\ (truncated in the blue side) and \Nii\ emission lines and [\ion{S}{ii}]$\lambda$6716 (on the red side of the spectrum) at \z=0.371; and source {\tt id: 3289,} which exhibits the \Sii\ doublet at \z=0.340. The green line shows the best-fit synthetic spectrum after the convolution process. See Sect. \ref{sec:metallicity} for details.}
                \end{center}
        \end{figure*}
                
        (b) We visually inspected the 73 candidates using our web-based graphic user interface (GUI)\footnote{{\tt http://research.iac.es/projecto/otelo}} where all data about a user-selected source can be displayed; in particular, it includes the PS, the thumbnails of the images in available bands, and the {\tt SExtractor} segmentation maps together with high-resolution \hst\ images, \zphot\ solution including and excluding \otelodeep\ photometry obtained from our photo-{\z} analysis and their comparison with the input photometry, the \zphot\ solution from the CFHTLS T0004 Deep3 photo-{\z} catalogue, and the $z_\mathrm{spec}$ from DEEP2.
        
        The high-resolution \hstacs\ images allow us to identify possible contamination of the PS from neighbouring sources that are not resolved in the ground-based images. The comparison of the best-fit SED and the source data, together with the additional redshift estimates (obtained in our analysis, from the CFHTLS \zphot\ catalogue, and from the objects with DEEP2 $\mathrm{z}_\mathrm{spec}$) allows us to evaluate the confidence in our \zphot\ solution. Finally, the PS allows us to distinguish between \HAN\ SFG sources, \Sii,\ or AGN galaxies with \HAN\ emission. We took advantage of the fact that the \Nii$\lambda 6583$ over \HA\ ratio must be lower than 0.4 \cite[using the extreme case used by][see below]{Pettini2004} in SFGs. In addition, the [\ion{S}{ii}]$\lambda$6716 over [\ion{S}{ii}]$\lambda$6731 line ratio must range between 1.49 and 0.44 (see \citealt{McCall84} and \citealt{Bernabe2013} in the context of tunable filters). This means that any doublet where the redward  line (\Nii$\lambda 6583$ or [\ion{S}{ii}]$\lambda$6731) is stronger than about half of the blueward line (\HA\ or [\ion{S}{ii}]$\lambda$6716) should be a \HAN\ AGN or a \Sii\ source. This process reduced the sample of 73 candidates to a sample with 18 sources where \HA\ and \Nii$\lambda 6583$ can be measured. In addition, we also obtained a \zguess\ value (from the web-based GUI), which was obtained by the association of the brightest RTF slice and the \HA\ emission line at a given observed wavelength.
        \newline
        
        (c) As a final cross-check, we searched for objects with DEEP2 spectroscopic redshift in the range of $0.35 \leq \mathrm{z}_\mathrm{spec}  \leq 0.42$, which covers the range where \HA\ line would be detected in the OTELO spectral window. This redshift range would overlap with \Sii\ sources ($0.32 \leq$ \z\ $\leq 0.38$). DEEP2 data provides 18 sources in common with our selection in point (b), 9 of them do not show any emission feature in the OTELO PS, and 7 of them are in common with our selection from point (b). The remaining 2 objects had a best \zphot\ solution for the galaxy library outside of our redshift selection range (although one of them provides a \zphot\ solution in the \HAN\ range using the {\tt AGN/QSOs} library with a Sy-2 galaxy template). After visual inspection of the PS, we verified that one source of this set is indeed a \HAN\ emitter ({\tt id: 625}) and the second one displays only the \Nii$\lambda 6583$ line (and \Sii\ lines on the red side, {\tt id: 6059}) in its PS. We therefore included only {\tt id: 625} source in the final sample, which increases to 19 objects. We note that this additional source was not included in the preliminary ELS list because the emission line is truncated in one extremity of the PS (see the middle panel of Figure \ref{fig:ps}), but it was selected as a \HAN\ SFG candidate with the colour-magnitude ELS selection technique that is traditionally used in deep NB surveys \citep{2007PASP..119...30P}. In our case, a ($z-$\oteloint) colour was obtained for all the OTELO sources detected in $z$ band. See \citetalias{OteloI} for details about this ELS selection procedure. 
        
        The recovery of a preliminary ELS through a colour-magnitude diagram instead of using the flux-excess in PS, as indicated above, is an exception. As can be inferred from the description of this survey, OTELO combines the advantages of the deep NB surveys that imperatively use the colour-magnitude selection technique, but whose sensitivity to low values of observed EW is constrained by the passband of the NB filter used, with the strengths of the conventional spectroscopic surveys but without their selection biases and long observing times required. This has allowed us to reach the line flux and EW sensitivities reported in Section \ref{Introduction}.
        
        In summary, the final sample of \HA\ ELSs SFG candidates consists of 19 sources; we refer to this hereafter as the \HA\ sample. Further refinements (i.e. the flux and metallicity estimations) are described in the following sections. The examples given in Figure \ref{fig:ps} show a PS for a \HAN\ source ({\tt id: 2146}), a source with a truncated \HA\ ({\tt id: 625}), and [\ion{S}{ii}]$\lambda$6716 emission and a \Sii\ source ({\tt id: 3289}, not in the \HA\ sample). The top part of Table \ref{tab:source_number} shows a summary of the number of sources at different steps of the selection process. 
        In the Appendix we show figures with the PS of the sources that compose the \HA\ sample.

        \begin{table} 
                \centering
                \caption{Source number in selected \HA\ samples.}
                {
                        \addtolength{\tabcolsep}{-2.5pt}
                        \begin{tabular}{rc} 
                                sample & source number\\
                                \hline
                                Total & 11237 \\
                                Preliminary ELS & 5322 \\
                                Preliminary ELS in \zphot range & 2289 \\
                                \HA\ ELS candidates & 73 \\
                                used \HA\ ELS  & 19 \\\hline
                                &\\
                                &\\
                                \HA\ sample &\\
                                \hline
                                AGN & 3 \\
                                \HA\ with metallicity & 11 \\
                                \HA\ with $f_{\mathrm{[N~II]}~\lambda 6583}$ limit & 5 \\
                                (low-flux sources) &  \\
                                \hline\hline
                        \end{tabular}} \\[0.2cm]
                        \raggedright\small{}
                        \label{tab:source_number}
                \end{table}
                
                \begin{figure}[t!]
                        \centering
                        \includegraphics[width=\columnwidth]{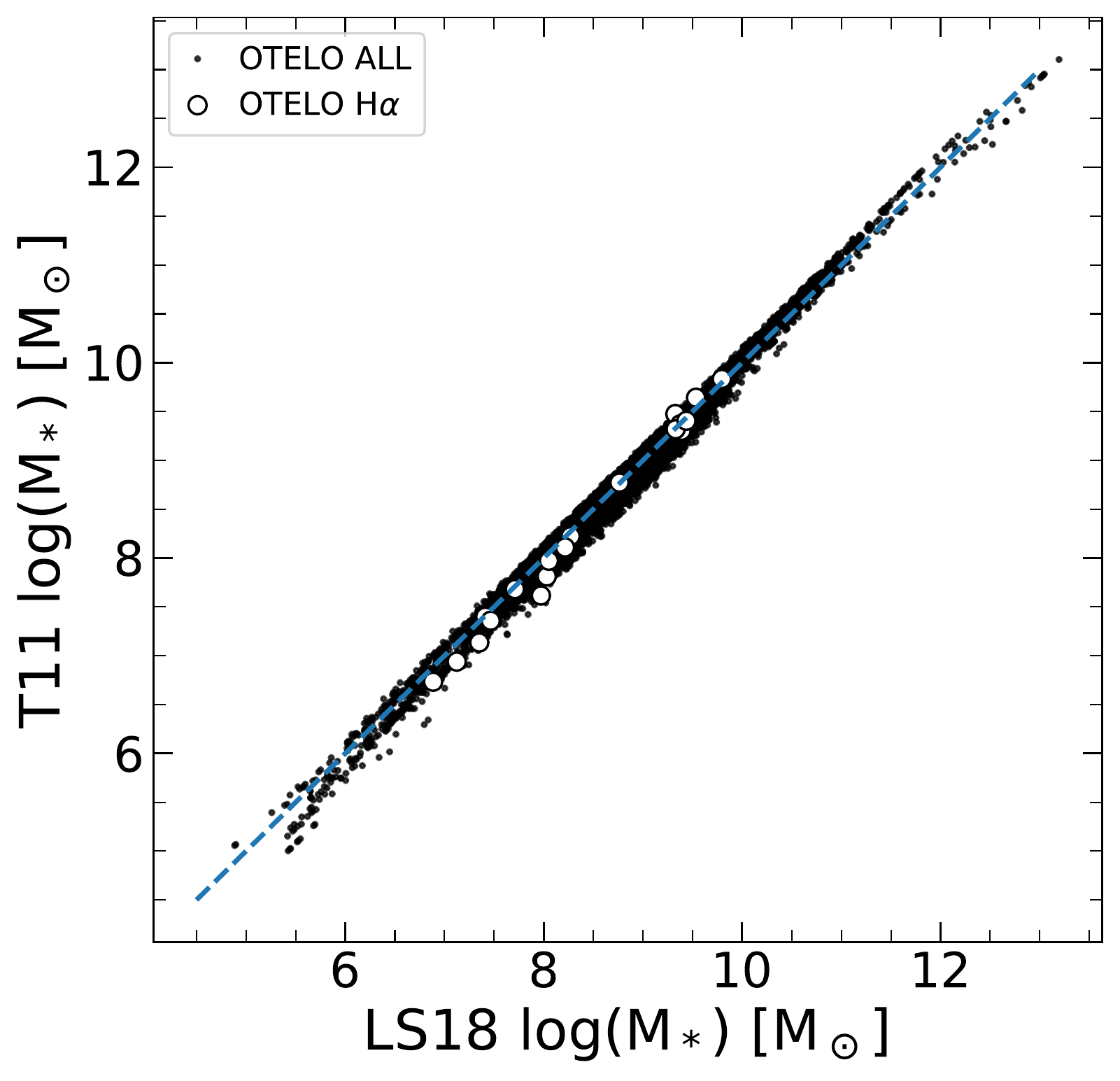}
                        \caption{\label{fig:hist_m_and_z} Comparison of computed stellar masses \mstar\ using prescriptions from  \citet{LopezSanjuan2018} and \citet{Taylor2011}. The overall OTELO and \HA\ sample are shown as black dots and black empty circles, respectively. Dashed-line in blue represent 1:1 line.} 
                \end{figure}
                
                \subsection{Data from previous surveys}
                \label{sec:external_surveys}
                In order to compare our results with those from previous studies, we used the SDSS-Data Release 7 (SDSS-DR7, \citealt{SDSS_Abazajian2009ApJS..182..543A}) and the GAMA survey \citep{GAMA_Driver2011MNRAS.413..971D}. From both datasets we considered only the SFGs selected through the BPT diagram \citep{BPT1981PASP...93....5B} using the criteria of \citet{Kauffmann2003MNRAS.346.1055K}. In both surveys, the stellar masses were estimated using SED templates from \citet{Bruzual2003MNRAS.344.1000B} and a \citet{Chabrier2003ApJ...586L.133C}  IMF. SDSS-DR7 fit templates to the observed SEDs, while stellar masses in GAMA were estimated using \gmenosi\ colours and a mass--luminosity relation \citep{Taylor2011}.
                
                We selected the following sub-samples from the SDSS-DR7 (local and high-\z\ ) and the GAMA surveys: (i) A local SDSS sample with $0.005\,<\,z\,<\,0.1$ and a signal-to-noise ratio (S/N) > 3 for \HA\ and S/N > 4 for the \Nii\ emission line. (ii) A higher redshift sample (high-\z\  SDSS and GAMA) with $0.3\,<\,z\,<\,0.33$ (upper-\z\ limit due to the H$\alpha$ line visibility in both surveys). The emission line S/N imposed for both high-\z\ samples was the same as in (i).

                Furthermore, to ensure completeness for the higher redshift samples, we imposed a cut in stellar mass of 10$^{10}$\msun. This cut is given by the limit in completeness of SDSS (see \citealt{Weigel2016}), and it approximately corresponds to an I-band absolute magnitude limit of \absMi$\sim$-21 [AB].
                These selection processes guarantee that we used robust and complete samples from both surveys.\newline
                
                We also used the data from publicly available catalogues of the VVDS\footnote{\url{http://cesam.lam.fr/vvds/index.php}}, in particular, from VVDS-Deep and VVDS-Ultra-Deep (VVDS-F0226-04). Data used from both surveys were selected as follows: redshift quality flag 1 $<$ \texttt{ZFLAG} $<$ 11 (this removes possible AGN from the sample, as described in \citealt{VVDS_Garilli_2008A&A...486..683G}); the redshift range of 0.3(Deep)/0.2 (Ultra-Deep) $<$\z$<$0.42, we decided to use \z$=$ 0.2 as the lower redshift limit for Ultra-Deep in order to increase the final number of sources.

                \begin{figure}[t!]
                        \begin{center}
                                \includegraphics[width=\columnwidth]{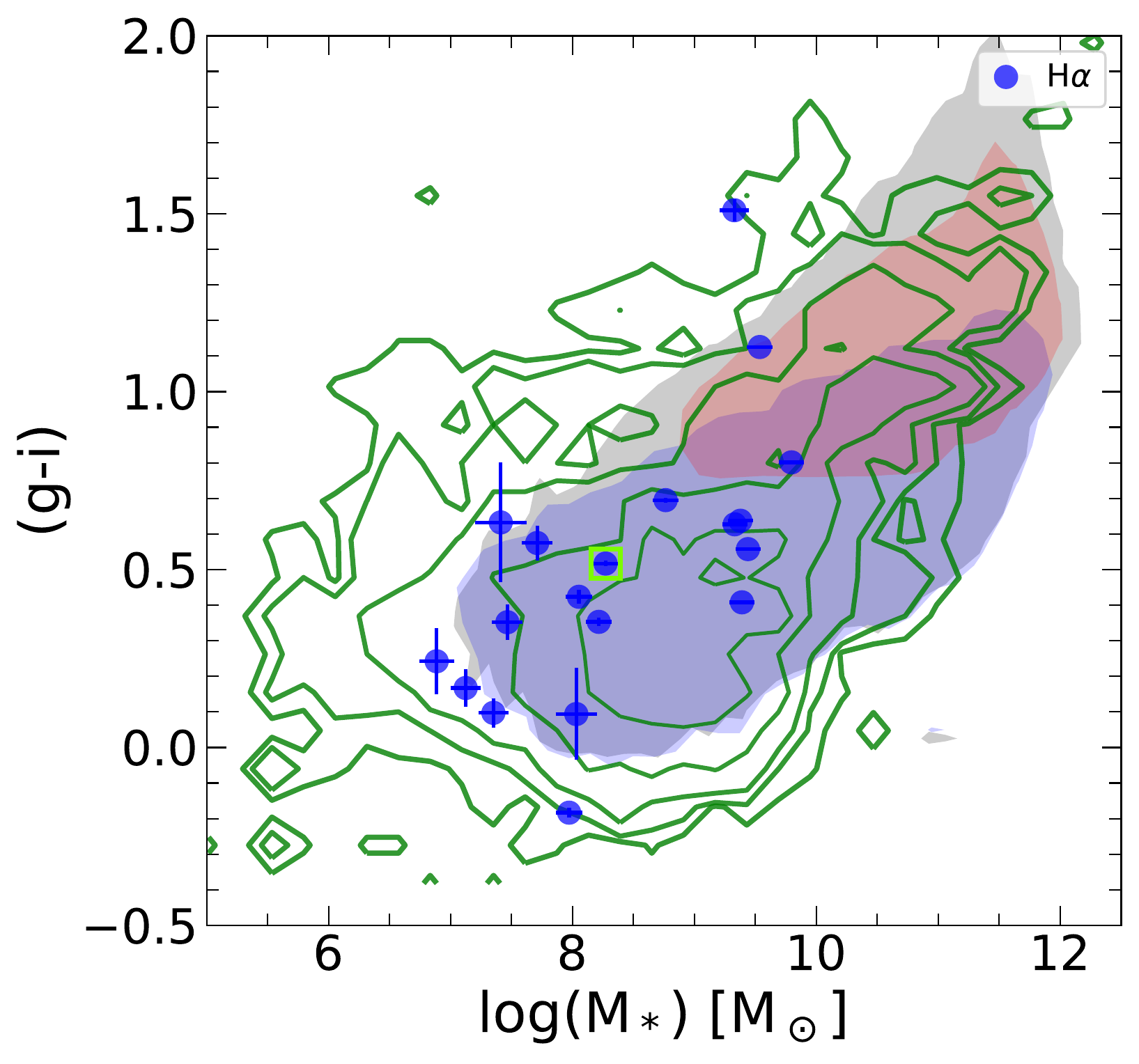}
                                \caption{\label{fig:LogMass} Colour--\mstar\ diagram. Blue points show \HA\ ELS, and green contours represent the overall OTELO sample. Filled contours in grey show the envelope of SDSS-DR7 data, and filled red and blue contours show the red and blue cloud separated with the empirically estimated limit of \citet{Bluck2014MNRAS.441..599B}. Green squares mark MZR outlier galaxies (see Section \ref{sec:MZR_analysis}).
                                }
                        \end{center}
                \end{figure}

                \section{Methods}
                \label{sec:methods}
                
                \subsection{Stellar masses}
                Stellar masses \mstar\ for the overall OTELO sample were computed using the \citeauthor{LopezSanjuan2018} (\citeyear{LopezSanjuan2018}; hereafter LS18) mass-to-light ratio for quiescent and SFGs (their equations 10 and 11, respectively) employing rest-frame \bandg-\ and \bandi-band magnitudes, together with absolute \bandi-band magnitude \absMi. The rest-frame photometry was computed by shifting the \sedbest\ (see Section \ref{sec:data}) to $z\,=\,0$, and calculating the flux below the transmission curves for \bandg, \bandr, \bandi, \bandz\ (CFHTLS) and for the \hstacs\ F606W and F814W filters. To estimate \absMi\ we used our \zotelo\  (see Sect. \ref{sec:metallicity}). A second-order correction accounted for the Galactic dust extinction towards the EGS using the Galactic dust map from \citet{Schlafly2011}. The obtained stellar masses \mstar\ for the \HA\ sample are in the range of $\sim10^{6.8}\,<\,$\mstar$\,<\,10^{10}\,$ \msun. This places our sample in a very low-mass regime where 63\% of our sample has \mstar$\,<\,10^9\,$ \msun. To test its consistency, we also calculated \mstar\ using the \citeauthor{Taylor2011} (\citeyear{Taylor2011}; hereafter T11) mass-to-light ratio. Both methods agree, with the most noticeable difference for blue SFGs, for which \citetalias{LopezSanjuan2018} provides an additional quadratic factor that depends on the \gmenosi\ colour. In Figure \ref{fig:hist_m_and_z} we compare the computed stellar masses using prescriptions from \citetalias{LopezSanjuan2018} and T11 for all the OTELO sources (median value of \mstar$=\,10^{8.9}\,$\msun) and for the \HA\ sample (median value of \mstar $=10^{8.4}$ \msun).   In Figure \ref{fig:LogMass} we show the obtained stellar masses as a function of rest-frame \gmenosi\ colour for the overall OTELO and the \HA\ ELS samples. For comparison, in the same figure, we show the SDSS overall sample together with the blue and red cloud empirical colour division of \citet{Bluck2014MNRAS.441..599B}:
                \begin{equation}
                (g -r) = 0.06 \times \log M [M_\odot] - 0.01,
                \end{equation}
                obtained using data from SDSS-DR7. After we identified blue and red clouds in SDSS using the $(g\,-\,r)$ colour, we represented them in terms of \gmenosi. Because it is only for illustrative purposes, we did not correct the blue and red cloud selection in terms of colour used.
                
                Furthermore, we compared the obtained masses with the COSMOS2015 catalogue \citep{COSMOS2015_2016ApJS..224...24L} finding good agreement: our mass and colour estimates fall within the COSMOS2015 imprint. For the sake of clarity, we decided to omit this in Figure \ref{fig:LogMass}, where  a similar comparison is shown using data from SDSS-DR7.
The resulting stellar masses of the sample we used and their associated uncertainties are discussed in Appendix \ref{App_uncert_mass} and shown in the fourth column of Table \ref{tab:Results}.
                
                \subsection{Deconvolution of PS, metallicity, SFR, and sSFR}
                \label{sec:metallicity}
                
                To obtain gas-phase metallicities and AGNs contamination, we used the N2 index:
                
                \begin{equation}
                \mathrm{N2} = \log \frac{f(\mathrm{[N~II] \;\lambda 6583)}}{f(\mathrm{H}\alpha),}
                \end{equation}
                
                \noindent which requires a flux estimate. To this end, we used  the so-called inverse deconvolution of the PS. In short, we assumed rest-frame model spectra defined by a constant continuum level, and Gaussian profiles of the \Nii$\lambda$6548, \HA,\ and \Nii$\lambda$6583 lines defined by their amplitude and a common line width for the three lines. The model to be compared with observational data is thus defined by the parameters $f_\mathrm{mod}(z,f_c,\sigma_\mathrm{line},f_{\mathrm{[N~II]}~\lambda 6548}, f_\mathrm{H\alpha}, f_{\mathrm{[N~II]}~\lambda 6583})$. We performed $10^6$ Monte Carlo simulations for each object and varied the different parameters. The different parameters varied in following ranges. The redshift was varied in the range {\zguess}$\pm 0.001$. The line width $\sigma_\mathrm{line}$ varied from 20 to 500 km/s rest-frame values, which translates into a range of 0.43 to 11 \AA\ in spectral units at the H$\alpha$ wavelength. The continuum $f_c$ level varied in the range $f_{med,c} \pm \sigma_{\mathrm{med,c}}$ with $f_\mathrm{med,c}$ the median of the PS and $\sigma_{\mathrm{med,c}}$ the root mean square around $f_\mathrm{med,c}$ , as defined in Sect.~\ref{sec:sample_selection}. The rest-frame $f_\mathrm{H\alpha}$ amplitude varied between $f_\mathrm{med,c}$ and the maximum value observed in the PS, $f_{\mathrm{PS}}^{\mathrm{max}}$ taking into account the observational error $\sigma(f_{\mathrm{PS}}^{\mathrm{max}})$, and taking into account the assumed Gaussian profile. This is is given by Eq. \ref{eq.RFHA}.  The rest-frame $f_{\mathrm{[N~II]}~\lambda 6583}$ varied between $0.006 \cdot f_\mathrm{H\alpha}$ and the maximum possible value given by Eq.~\ref{eq.RFHA}. Finally, the rest-frame $f_{\mathrm{[N~II]}~\lambda 6548}$ was fixed to $1/3 \cdot f_{\mathrm{[N~II]}~\lambda 6583}$ \citep{Osterbrock2006agna.book.....O}. See Appendix \ref{App_decInv} for details.
                
                Each model was convolved with the OTELO filter response to produce a synthetic PS that was compared with the observed PS, and the corresponding $\chi^2$ was obtained. The Monte Carlo simulations are designed to sample the multiparametric probability density function (PDF) from which modal values (i.e. best-fit solutions) and different confidence intervals around the modal value were obtained for each parameter (see details in Appendix \ref{App_decInv}). In the following we consider error bars as the confidence intervals, which includes 25\% and 68\% (which would be the equivalent of $\pm 1 \sigma$ for the Gaussian case) of the total area of the PDF around the modal value as an estimate of the errors.
                The redshift value obtained for each source after deconvolution (hereafter \zotelo) are given in Table \ref{tab:Results}. Formally, the common maximum uncertainty of these values is about 0.001(1+\zotelo), as established in \citet{Otelo-OIII}.
                
                We note that the possible flux-related parameters we obtained are strongly correlated, that is, each model has its own N2 index (i.e. metallicity), line equivalent widths, and so on, because the variation in one of the parameters of the model non-trivially affect the possible best solutions of the other parameters. In this way, we also obtained the corresponding PDFs of all the associated parameters (equivalent widths, N2 indices, and oxygen abundances). For all these inferred quantities we performed a posteriori operations over the entire Monte Carlo set associated with each particular object. The corresponding PDFs (and associated errors) of the particular quantities can be obtained as shown in  Appendix \ref{App_decInv}. In the case of $f_\mathrm{H\alpha}$, we corrected the model values for possible stellar absorption of underlying older components following \citet{Hopkins2003,Hopkins2013} using a $EW_{\rm c}$ correction value of 2.5\AA:
                
                \begin{equation}
                F^{\rm cor}_{\rm H\alpha} = \frac{EW_{\rm H\alpha} + EW_{\rm c}}{EW_{\rm H\alpha}} \times F^{\rm mod}_{\rm H\alpha},
                \label{eq:stellar_abs}
                \end{equation}
                
                \noindent where $F^{\rm mod}_{\rm H\alpha}$ and $F^{\rm cor}_{\rm H\alpha}$ are the model before and after stellar absorption correction, respectively, and $EW_{\rm H\alpha}$ refers to the equivalent width of the model before correction.
                
                Finally, we computed the gas-phase oxygen abundance from the N2, with the absorption corrected $F^{\rm cor}_{\rm H\alpha}$ value by means of equation 1 of \citet{Pettini2004}  for all simulations:
                \begin{equation}
                12 + \log {\rm(O/H)} = 8.90 + 0.57 \times {\rm N2}.
                \label{eq:metallicyt}
                \end{equation}
                
                Nominal values and errors for metallicities were calculated by analysing the metallicity PDF obtained from the Monte Carlo simulations, but we did not take the intrinsic dispersion in the calibration into account, which would be $\pm 0.41$ ($\pm 0.18$) dex at 95\% (68\%) confidence interval \citep{Pettini2004}. 
                Table \ref{tab:Results} shows the resulting  $f_\mathrm{H\alpha}$, $EW_{\rm H\alpha}$ (both including the correction for stellar absorption), $f_{\mathrm{[N~II]}~\lambda 6583}$, and $12 + {\log \rm(O/H)}$ (without calibration uncertainty) values with their 68\% confidence intervals in columns 6, 8, 7, and 5, respectively.

                We note that by construction, the inverse deconvolution process always provides a solution for the intensity of the $f_{\mathrm{[N~II]}~\lambda 6583}$ value, which would correspond to a $12 + {\rm log(O/H)} = 7.63$ (before the stellar absorption correction). These values would be below the OTELO line flux limit of 5$\times10^{-19}$erg/s/cm$^2$ as occurs for five objects \citep[see][for the flux limit estimation]{Otelo-OIII}. These cases translate into a large uncertainty in the resulting N2 and metallicity values. We refer to these objects as low-flux sources and can only provide a $f_{\mathrm{[N~II]}~\lambda 6583}$ (hence metallicity) limit.
                
\begin{figure}[t]
                        \begin{center}
                                \includegraphics[width=\columnwidth]{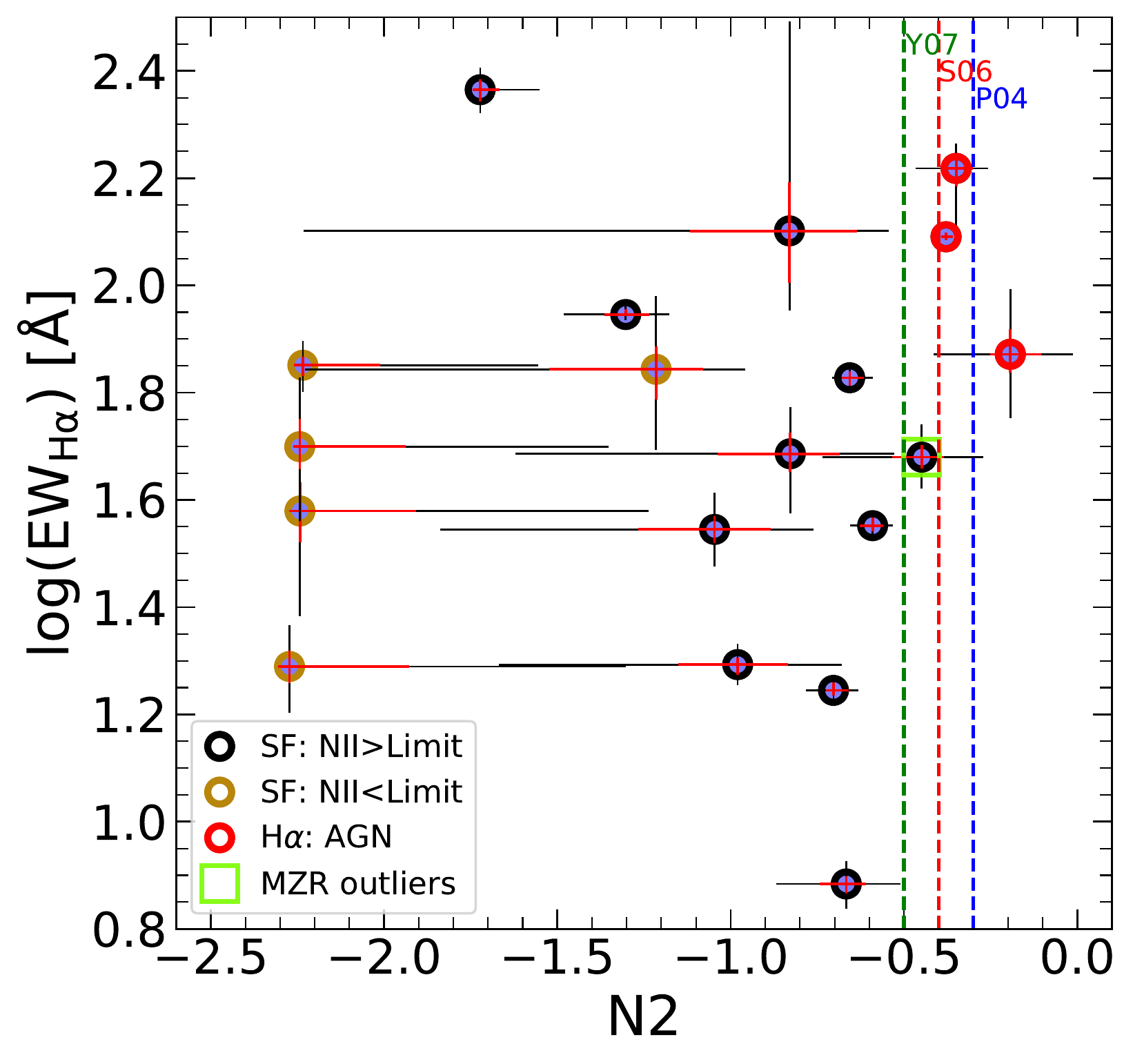}
                                \caption{\label{fig:N2}  N2 index vs. $EW_{\rm H\alpha}$. Blue points show all the \HA\ candidates. Black and golden circles represent sources with an \Nii\ line flux above and below OTELO flux limit, respectively \citep[see Section \ref{sec:metallicity} and][]{Otelo-OIII}. Red circles show AGN candidates selected using the \citet{Stasinska2006} empirical limit of N2$=-$0.4. Vertical green and blue dashed lines show the \citet{Yin2007} and \citet{Pettini2004} limits of their calibration using the N2 index, respectively. The red line shows the AGN selection limit from \citet{Stasinska2006}. The error bars in each object correspond to the 25\% and 68\% confidence interval around the most probable value. 
                                }
                        \end{center}
                        
                \end{figure}

                In addition, the \citet{Pettini2004} method is valid in the range of $-$2.5 $<$ N2 $<$ $-$0.3, which corresponds to the 7.47 to  8.73 $12 + {\rm log(O/H)}$ range.  Values of $12 + {\rm log(O/H)}$  equal to 8.67 would correspond to N2$=-0.4,$ which defines an  empirical division between SFGs and AGNs \citep{Stasinska2006}.
                In order to remove possible AGNs from our sample, we used the same method as in \citet[][as for narrow-line AGN]{OteloII}, that is, the \citet{Cid2010} diagnostic diagram ($EW_\mathrm{H\alpha}$ vs. N2) shown in Figure \ref{fig:N2}.
                In Figure \ref{fig:N2}, the green and blue dashed lines are the limits of the empirical metallicity estimations from \citet{Yin2007} and \citet{Pettini2004} of $-$0.5 and $-$0.3, respectively. In this way, we found three sources as possible AGN and excluded them from the further analysis.
                The lower part of Table \ref{tab:source_number} shows the summary of sources of different classes after the inverse deconvolution and AGN discrimination process. 
        
                Finally, we corrected the \HA\ fluxes for dust attenuation using the reddening values obtained from \texttt{LePhare} SED fitting (quoted in the second column of Table \ref{tab:Results}), and the empirical relation by \citet{Calzetti2000} estimated for galaxies at {\z}=0.5 by \citet{Ly2012ApJ...747L..16L}. We note that this is a lower value of $L(H\alpha)$ because the extinction is obtained from the fit of galaxy templates whose intrinsic extinction is unknown. 
                The derived values of $E(B-V)$ are in the range of 0.0 and 0.3 mag, which is consistent with the extinction expected for low-mass galaxies, as shown in previous studies  \citep[e.g.][]{DominguezSanchez2014}. We also note that \texttt{LePhare} does not provide an evaluation of the uncertainty in $E(B-V)$, therefore the uncertainty in this term was not taken into account to evaluate the uncertainty in $L(H\alpha)$ (hence in the SFR, see Sect. \ref{App_uncert_mass})
                The \HA\ line fluxes were used to derive SFRs following the standard calibration of \citet{Kennicutt1998ARA&A..36..189K}, but including the correction for a \citet{Chabrier2003ApJ...586L.133C} IMF  \citep[see][their Table 1]{Kennicutt2012ARA&A..50..531K} , which is the one used in this work:
                \begin{equation}
                {\rm SFR}({\rm M_\odot} yr^{-1}) = 5.37\times 10^{-42}  \rm L(H\alpha)[erg s^{-1}].
                \label{eq:sfr}
                \end{equation}
                The specific SFR (sSFR) was subsequently estimated as SFR/\mstar. The resulting SFR and its associated uncertainty are listed in column 9 in  Table \ref{tab:Results}.

                \begin{figure}[t]
                        \begin{center}
                                \includegraphics[width=\columnwidth]{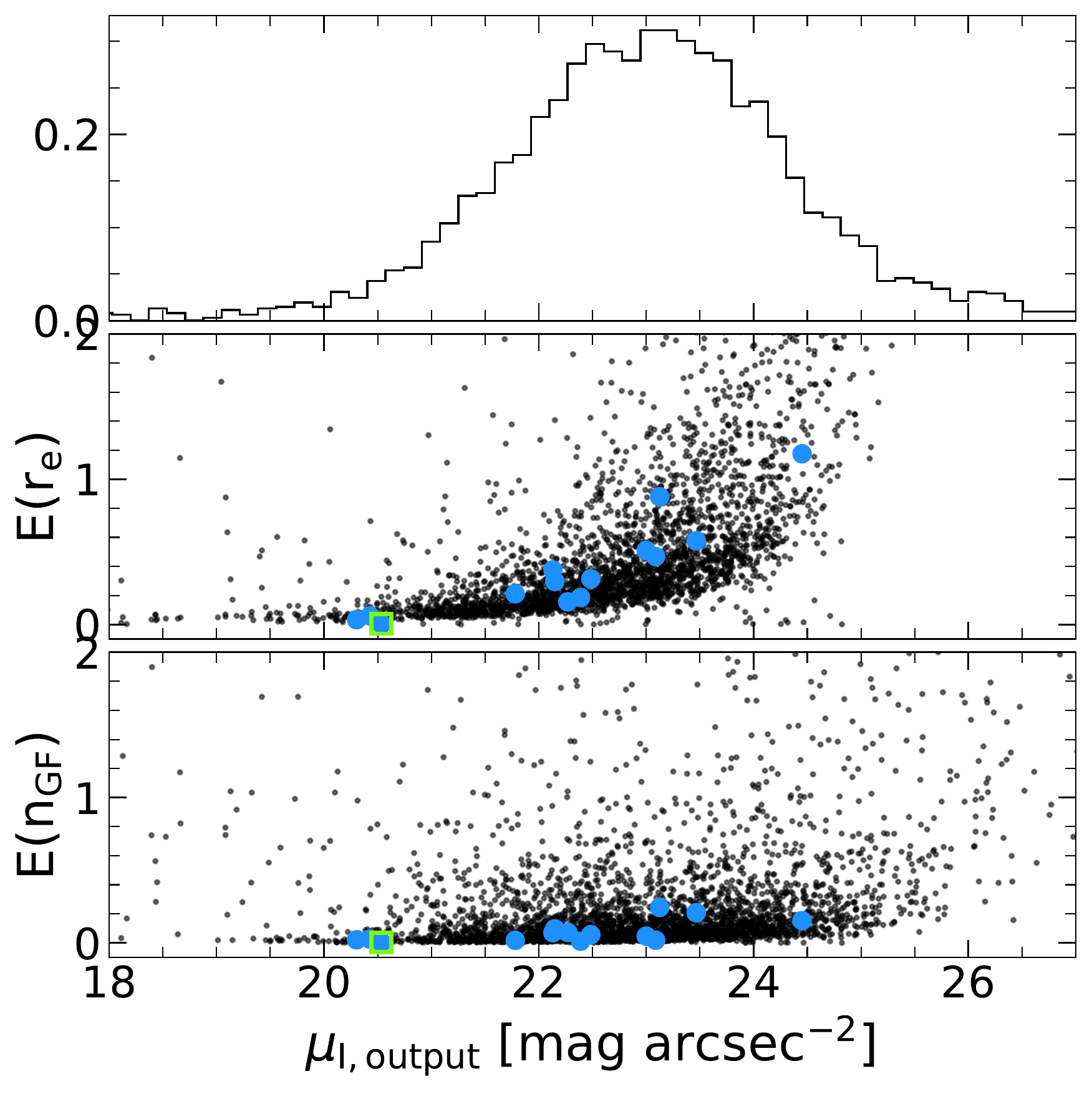}
                                \caption{\label{fig:morp_errors} Formal uncertainties of S\'ersic index and effective radius from \texttt{GALFIT}. As described in the text, these are the lower limit uncertainties. Black dots shows a sample studied in \citet{NadolnyMorpho}, and blue dots and the green square show the \HA\ sample and the MZR outlier studied in this work, respectively. 
                                }
                        \end{center}
                \end{figure}

                \begin{figure*}[t]
                        \begin{center}
                                \includegraphics[width=0.9\textwidth]{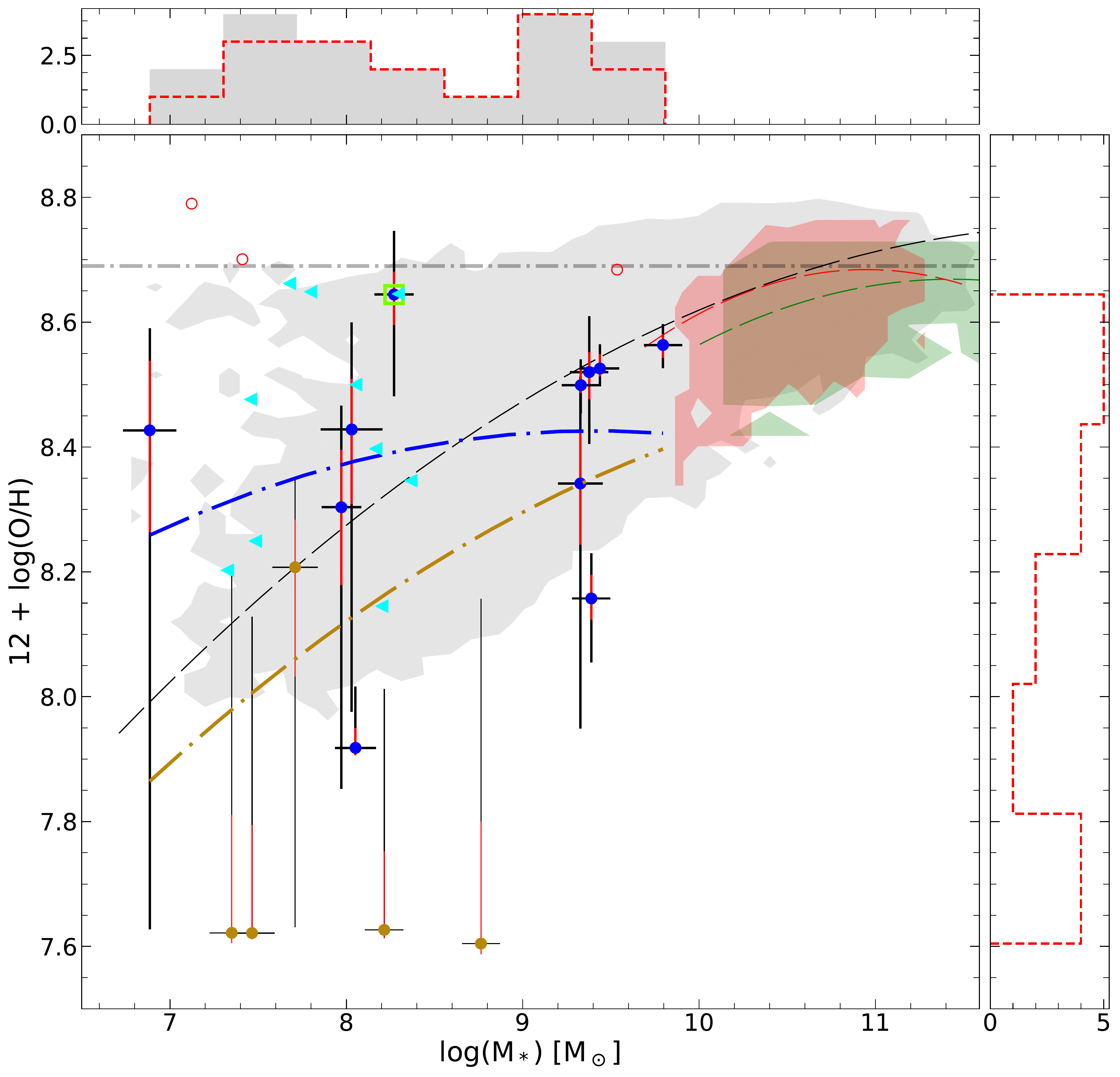}
                                \caption{\label{fig:MZ} Mass-metallicity relation for OTELO H$\alpha$ ELS. Blue and golden markers represent sources with an \Nii$\lambda$6583 line flux above and below OTELO flux limit, respectively. The light green empty square shows the MZR outlier as defined by \citet{Peeples2008ApJ...685..904P}. Cyan triangles show the VVDS-Ultra-Deep selected sample with $z_{\rm mean}\,=\,0.26$. Grey filled contours show local SDSS SFGs. Green filled contours show the high-\z\  SDSS sample. Red filled contours show the SFG sample from the GAMA survey. All fits shown in this figure represent second-order polynomial fits. The blue dot-dashed line is the OTELO MZR fit to the sources with an \Nii\ line flux above the OTELO line flux limit, and the golden dot-dashed line shows the OTELO MZR including low-flux sources; the black thin dashed line shows the fit to the local SDSS sample; and green and red dashed lines show the SDSS and GAMA samples at z$\sim$0.3, respectively.  The error bars correspond to the 25\% (red) and 68\% (black) confidence interval around the most probable value. The calibration uncertainty in metallicity is not included.}
                        \end{center}
                \end{figure*}

                \subsection{Morphology}
                
                Because the galaxies we analysed are some orders of magnitudes fainter than those in other surveys, it is mandatory to characterize this population as well as possible. Therefore, we performed a (i) quantitative and (ii) qualitative morphological classification.
                In both cases we decided to  base this on the publicly available\footnote{\url{http://aegis.ucolick.org/mosaic_page.htm}} high-resolution \hstacs\ (F606W and F814W) data. 
                
                (i) We used the \texttt{GALAPAGOS-2} software (\citealt{Haussler2007}) that was developed within the MegaMorph project\footnote{\url{https://www.nottingham.ac.uk/astronomy/megamorph/}}. \texttt{GALAPAGOS-2} is an automatic way to provide the morphological parameters with \texttt{SExtractor} \citep{SEx1996} and \texttt{GALFIT} (\citealt{Galfit2002}). In this work we provide a single-\citet{Sersic1968} fit to sources detected in \hstacs-F814W using \texttt{SExtractor} in high-dynamical range (HDR) dual mode. The HDR mode maximizes the detection of faint sources, and the dual mode ensures that we measured the same sources in all the filters. As indicated in \citet{GALFIT_2010AJ....139.2097P}, the formal uncertainties from \texttt{GALFIT} are only lower estimates. After simulations and tests of this software, \citet{Haussler2007} provided uncertainty estimates based on the comparison of input and output values as a function of surface brightness. Because we do not provide simulations like this, we cite the formal uncertainties from \texttt{GALFIT}, but we caution that these are the lower estimates. In Figure \ref{fig:morp_errors} we show the uncertainties on effective radius and S\'ersic index as a function of surface brightness defined following \citet{Haussler2007}. Even if these are the lower limit errors, we can see the expected tendency of errors to increase with surface brightness. The resulting  S\'ersic index and its uncertainty is shown in column 10 of Table \ref{tab:Results}.
                
                (ii) Visual classification was made using \texttt{MorphGUI}, which is a graphic user interface for morphological classification developed by CANDELS (see \citealt{Kartaltepe2015}). We modified the interface in order to provide additional morphological classes to extend the classical Hubble scheme, to be able to take into account peculiarities found at higher redshifts. Our classification scheme is based on the work of \citet{Elmegreen2007ApJ...658..763E}: point-like (PL), spheroidal (Sph), disc (D), spheroidal+disc (Sph+D), tadpole (T), chain (C), clumpy cluster (CC), doubles (DD), and unclassified (Unclass). 
                
                The OTELO GTC/OSIRIS data have considerably lower resolution than \hstacs\ (0.256 and 0.03 arcsec/px, respectively), therefore we performed the match of the two catalogues by allowing multiple matches when there was more than one object in the \hstacs\ F814W image inside the Kron ellipse for a source detected in an \otelodeep\ image. The classification presented in this work refers to the object closest in position (which usually also is the brightest object) detected in the F814W image of \hstacs. The detailed morphological analysis of OTELO sources up to \z=2 is the scope of a forthcoming paper \citep{NadolnyMorpho}.

                \section{Analysis and discussion}
                \label{sec:analysis}
                
                \begin{figure*}[t]
                        \begin{center}
                                \includegraphics[width=\textwidth]{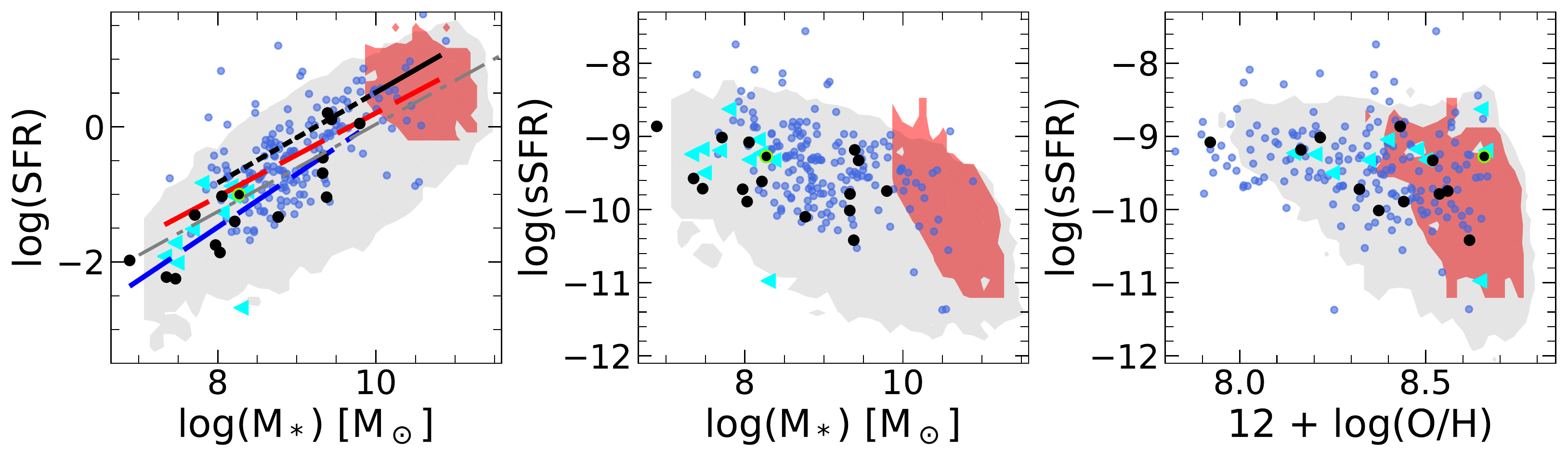}
                                \caption{\label{fig:SFR_sSFR_M_Z} Star formation rate and sSFR as a funcion of stellar mass and metallicity. In all panels, black markers show the OTELO SFG sample, green empty circles show transitional dwarf galaxies, cyan triangles show VVDS-Ultra-Deep data, blue dots show VVDS-Deep data, and contours in red and grey show the GAMA and local SDSS samples, respectively. In the left panel we show the linear fits to our SFG sample (blue dashed line), VVDS Deep+Ultra-Deep (both selected sub-samples from VVDS, red dashed-line), and the fit from \citeauthor{Noeske2007ApJ...660L..43N} (\citeyear{Noeske2007ApJ...660L..43N}; black solid line), which is given in the mass range of 10$^{10}$ to 10$^{11}$\msun, and the black dashed line is an extrapolation of their fit down to 10$^{8}$\msun. The grey dashed line shows the linear fit of the local SDSS sample. For clarity we do not plot the estimated dispersions for any fit.}
                        \end{center}
                \end{figure*}
                
                \subsection{Mass-metallicity relation}
                \label{sec:MZR_analysis}
                In this section we compare our results with the high-\z\  SDSS, GAMA, and VVDS-Deep and Ultra-Deep samples (see Section \ref{sec:external_surveys} for details on the sample selection). Finally, a comparison is made with the local SDSS sample in order to shed light on the possible evolution of the MZR at the low-mass end. 
                
                In Figure \ref{fig:MZ} we show the local SDSS sample of galaxies up to \z$\sim0.1$ together with higher redshift samples up to \z$\sim0.33$ from GAMA and SDSS with their respective second-order polynomial fits [$12 + \log (\mathrm{O/H}) = a + b  x + c x^2$ where $x =$\logmass]. Both high redshift samples were selected to be complete in mass and luminosity at a given redshift. It is worth noting that the OTELO sample is three orders of magnitude deeper in stellar mass than the two surveys at similar redshift. This is due to the depth of the OTELO survey (line flux down to 5.6$\times10^{-19}$erg/s/cm$^2$, see \citealt{Otelo-OIII}) and its potential as a blind spectroscopic survey to pre-select ELSs according to their PS and not only as a result of the colour excess in the colour-magnitude diagram. 
                In Figure \ref{fig:MZ} we also show the selected sample from the VVDS-Ultra-Deep survey: after our selection process (see Sections \ref{sec:external_surveys} and \ref{sec:metallicity}), we are left with 10 sources from a total of 1125 in the redshift range $0.2\,<\,z\,<\,0.42$ with $z_{\rm mean}\,=\,0.26$, which is much wider than our SFG sample redshift range ($0.36\,<\,z\,<\,0.42$). The highest redshift in the selected VVDS-Ultra-Deep sample is $z\,\sim\,0.33$. 
                The VVDS-Ultra-Deep survey data were processed in a consistent way with our data: (i) \HA\ fluxes were corrected for stellar absorption (eq. \ref{eq:stellar_abs}) and for intrinsic extinction (in this case, using the Balmer decrement); (ii) we estimated the N2 index, SFR (assuming a \citeauthor{Chabrier2003ApJ...586L.133C} IMF) and sSFR (using VVDS stellar masses), and (iii) we removed sources with N2$\,>\, -0.4$ as possible AGNs.
                We note that this sample spans the same range of metallicities, but covers a smaller range in stellar masses. We argue that this shows the capacity of the OTELO survey to recover the low-mass end of the observed galaxy population, even with its limited area of $\sim\,56$ arcmin$^2$ and despite the limited spectral range (230\AA\ wide, centred at 9175\AA).

                In order to shed light on the possible evolution of the MZR, we compared our result with the local SDSS sample. The fit from \citet{Tremonti2004ApJ...613..898T} is valid in the range of $8.5\,<$\logmass$\,<11.5,$ while our sample reaches masses as low as \logmass$\sim\,6.8$ (where 63\% of the sample with \logmass$\,<9$), therefore we did not use it. In Figure \ref{fig:MZ} we show our second-order polynomial fit to the local SDSS sample. From our SFG sample we identify sources with [\ion{N}{ii}]$\lambda 6583$ flux below the OTELO line flux limit (low-flux sources). 
                Because for these low-flux sources our metallicity estimate has a large uncertainty, we provide two MZR fits: the first is obtained using sources with an [\ion{N}{ii}]$\lambda 6583$ flux that is higher than the OTELO line flux limit (points and dot-dashed line in blue), and the second includes low-flux sources (points and dot-dashed line in gold). For the two fits we limited the $c$ parameter to be not greater than the $c$ parameter form fit of the local SDSS sample. Additionally, we show in this figure the MZR second-order polynomial fits to the local (\z$<0.1$) and high-\z\ ($\sim\,0.31$) SDSS and GAMA  (\z$\sim\,0.32$) samples.
                
                Our sample lies within the distribution of local SDSS MZR. Although our fit to sources whose \Nii$\lambda 6583$ line flux exceeds the OTELO line flux limit is flatter at the low-mass end, we cannot conclude that there is an evolution of the MZR because of the uncertainty in the metallicity of the low-flux sources. In the cases of the the fit that include low-flux sources (golden symbols), our fit approximates to the local SDSS MZR, especially at the low-mass end. All the fitted polynomial coefficients are listed in Table \ref{tab:poly_fit_MZR}.
                
                \begin{table} 
                        \centering
                        \caption{Polynomial coefficients of the MZR fits shown in Figure \ref{fig:MZ} in the form $12 + \log (\mathrm{O/H}) = a + b x + c x^2$, where $x = $\logmass.}
                        {   
                                \addtolength{\tabcolsep}{-1.5pt}
                                \begin{tabular}{rccc} 
                                        sample & $a$ & $b$ & $c$\\
                                        \hline
                                        OTELO SF {\z}=0.38     &  6.103 & 0.494 & -0.026 \\
                                        OTELO SF low-lim    &  4.836 & 0.620 &  -0.026 \\
                                        SDSS {\z} $<$ 0.1      &  4.798 & 0.644 & -0.026 \\
                                        SDSS {\z}=0.31         &  1.987 & 1.169 & -0.051 \\
                                        GAMA {\z}=0.32$^{(1)}$         & -0.752 & 1.723 & -0.079 \\\hline
                                \end{tabular}} \\[0.2cm]
                                \raggedright\small{$^{(1)}$ Coefficients from second-order polynomial fit of the GAMA data below \z=0.2.}
                                \label{tab:poly_fit_MZR}
                        \end{table}
                        In Figure \ref{fig:MZ} the upper histogram represents the mass distribution where the red dashed line shows the pure SFG sample and the grey filled histogram also includes possible AGNs (which are not included in the MZR). The "gap" in the \mstar\ distribution between $\sim\,10^{8.5}$ and $10^{9}\,$\msun\ is a statistical effect due to the low number of sources in the sample. The right-hand histogram shows the metallicity distribution. The peak around $\sim\,7.6$ is associated with low-flux sources with an \Nii$\lambda6583$\ line flux that is lower than the OTELO flux limit. This is the hard limit of our method when it recovers line fluxes from PS, and also for the use of N2 as a metallicity estimate.

                        \subsection{Mass, metallicity, and SFR}
                        As proposed by \citet{Maritza_FP_2010A&A...521L..53L}, the stellar mass, metallicity, and SFR are correlated and may be reduced to a plane in 3D space. In this section we study the scaling relationships between SFR, sSFR, \mstar\ , and metallicity, shown in Figure \ref{fig:SFR_sSFR_M_Z}, where we represent the selected sub-samples from local SDSS, GAMA, VVDS (Deep and Ultra-Deep), and our \HA\ SFG sample. The left panel of this figure shows the dependence of the SFR on the stellar mass. \citet{Noeske2007ApJ...660L..43N} explored this relation over a wide range of redshifts between $0.2< z < 1.1,$ with the mass complete down to $\sim10^{10}$. They found a main sequence of this relation for SFGs over cosmic time with a slope of 0.67. Our sample reaches three orders of magnitude lower in stellar mass than the work of \citet{Noeske2007ApJ...660L..43N}. We find a linear relation with a slightly higher slope of $0.78\,\pm\,0.11$, but shifted down by $\sim\,0.25$ dex. This shift is associated with the different mass ranges in the two datasets. Using data from merged VVDS Deep and Ultra-Deep survey samples, we obtain a linear fit with a slope of $0.61\,\pm\,0.03$, shown in the same figure. We note that this fit approximates ours at the high-mass end. Our fit is in agreement with previous studies within the estimated dispersion. We find no evolution of this relation from the local Universe to \z$=0.4$. Our \HA\ sample is consistent with the local SDSS sample, as can be seen not only in the spatial distribution, but also in the linear fits of the two samples. Our sample has moderate SFRs, ranging from ${-2.2} \lesssim \log{\rm (SFR)} \lesssim 0.2$ \msun yr$^{-1}$. VVDS covers this relation with a similar dispersion from the high-mass (Deep) down to the low-mass (Ultra-Deep) regime; the VVDS-Ultra-Deep sample reaches the mean mass of the OTELO \HA\ sample (\mstar$\,\sim10^{8.3}\,$\msun). 
                        
                        The middle and right panels of this figure show the relationship of sSFR with \mstar\ and metallicity, respectively. There is no clear evidence of evolution as compared with the local SDSS sample. In both cases the sSFR is anti-correlated with \mstar\ and metallicity (middle and right panel, respectively). Furthermore, these relations hold up to $z\,=\,0.4$ down to the low-mass regime for the OTELO and VVDS samples.
                        
                        \begin{figure}[t]
                                \begin{center}
                                        \includegraphics[width=\columnwidth]{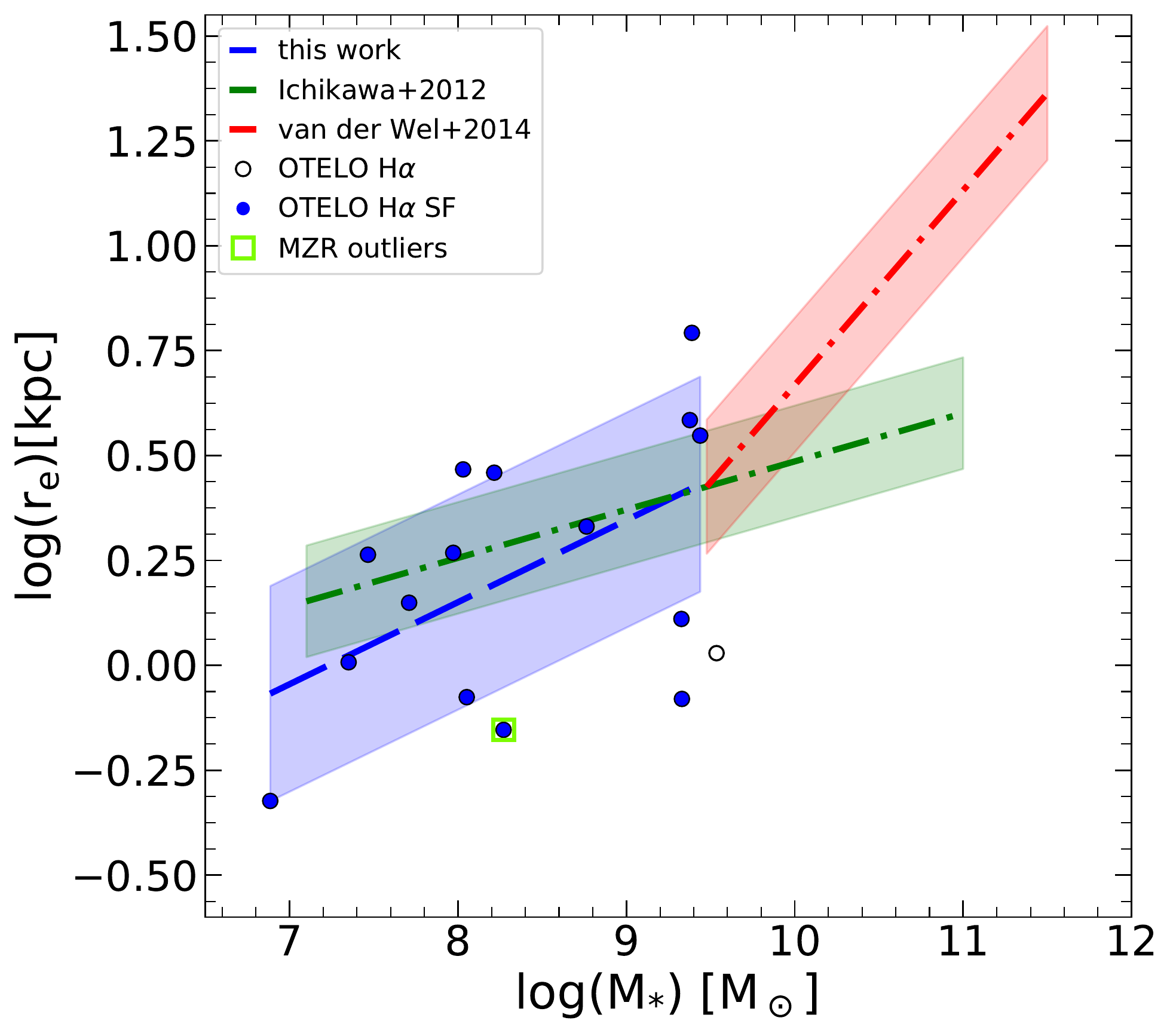}
                                        \caption{\label{fig:mass_size} Physical half-light radius r$_{\rm e}$ as a function of the stellar mass \mstar. Black empty circles show the \HA\ sample, and blue markers show SF galaxies (in both cases we show sources for which it was possible to measure both variables). The green square shows the outlier of the MZR. The blue dashed line represents the linear fit in a form y= a + bx to the \HA\ sample with a$=$-1.41$\pm$0.73 and b$=$0.19$\pm$0.09. The green dot-dashed line shows the linear fit from \citet{Ichikawa2012MNRAS.422.1014I} with the logarithmic slope of 0.115. The red dot-dashed line shows the linear fit found by \citet{vanderWel2014ApJ...788...28V} with a slope of 0.25. The green and red shaded regions represent the dispersion of the linear fit as estimated by \citet{Ichikawa2012MNRAS.422.1014I} and \citet{vanderWel2014ApJ...788...28V}, respectively, while the blue shaded region shows the dispersion of our measurement estimated as the median absolute deviation of 0.2.}
                                \end{center}
                        \end{figure}

                        \subsection{Mass-size relation}
                        \label{sec:MSR_disussion}
                        In Figure \ref{fig:mass_size} we show the \mstar--size relation (MSR) for the \HA\ sample we studied. We estimated the physical half-flux radius $r_{\rm e}$ using the effective radius from the S\'ersic model and  \zotelo\ ($r_{\rm e}$ values are given in the last column of Table \ref{tab:Results}). From a linear fit we obtain an intercept of $-1.41\,\pm\,0.73$ and a slope of $0.19\,\pm\,0.09$. In this fit we used 13 sources from the SFG sample that had available \mstar\ and $r_{\rm e}$ measurements. Here we compare our MSR with  \citet{Ichikawa2012MNRAS.422.1014I}, who studied galaxies from the Moircs Deep Survey (\citealt{Kajisawa2009ApJ...702.1393K}) in the GOODS-North region at 0.3 $<$\z$<$ 3 and in a mass range of \mstar$\sim10^{8}$-$10^{11}$\msun. They found a universal slope of the MSR (independent of redshift and galaxy activity, SFGs or quiescent, for a given mass) of R$\propto$\mstar$^{\alpha}$ with $\alpha\sim\,0.1\,-\,0.2$. Figure \ref{fig:mass_size} shows a fit from \citet{Ichikawa2012MNRAS.422.1014I} for colour-selected SFGs at $0.25\,<\,z\,<\,0.5$ with a slope of 0.115, using their Eq. 3 with the parameters given in their Table 1. In Figure \ref{fig:mass_size} we also show a linear fit from \citet{vanderWel2014ApJ...788...28V} for colour-selected SFGs from the 3D-\hst/CANDELS at $z\,\sim\,0.25$ (equation and values as given in their Table 1). We found that our results agree with those from \citet{Ichikawa2012MNRAS.422.1014I}. We argue that this is due to the mass range studied in both works.
                        
                        \subsection{Morphology}
                        \label{sec:morphology}
\begin{figure}[t]
                                \begin{center}
                                        \includegraphics[width=\columnwidth]{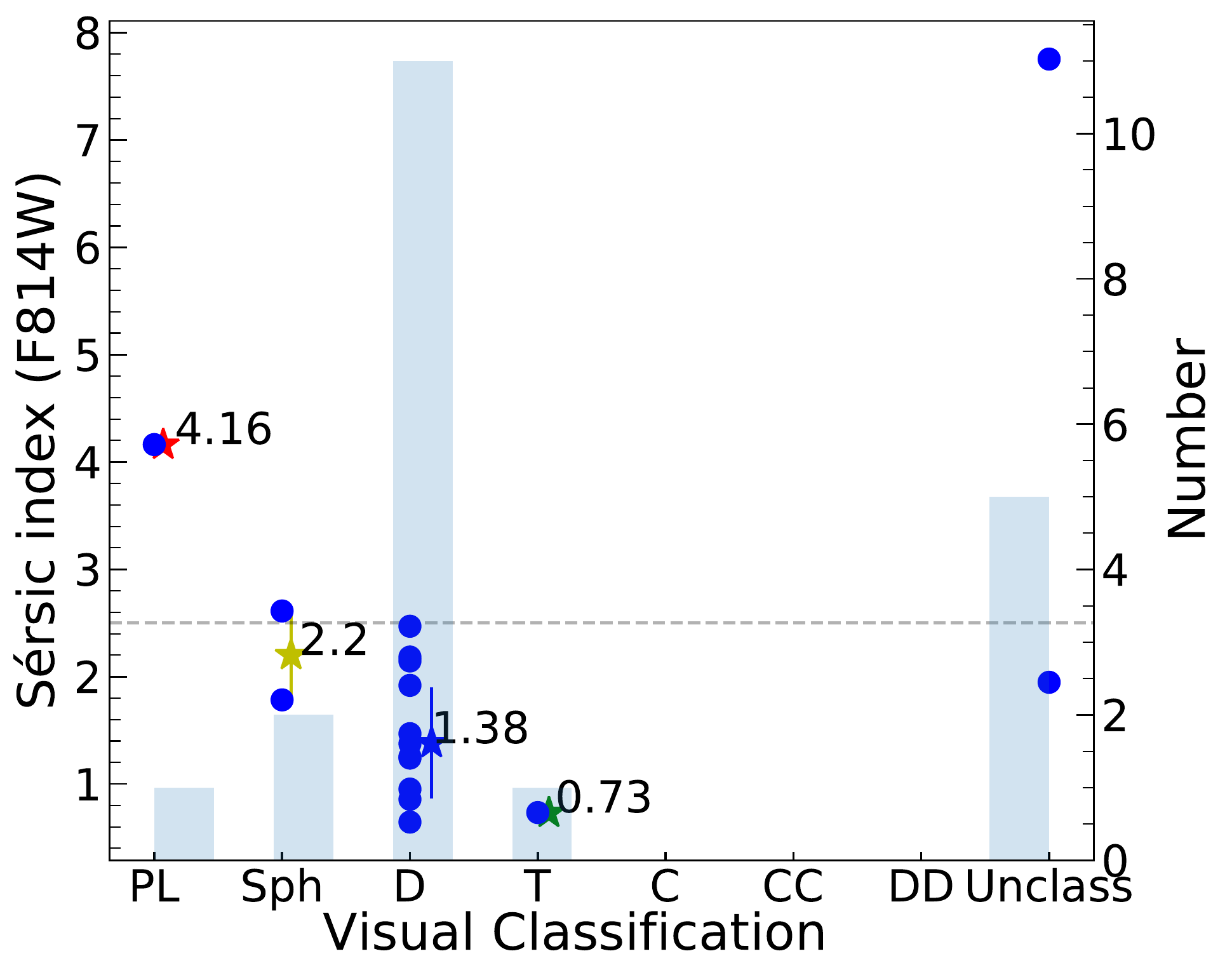}
                                        \caption{\label{fig:morphology} Comparison of the visual (x-axis) and automatic (y-axis) morphological classification or our \HA\ sample. Stars represent the mean value of the Sérsic index \nmean\ per visual type with error bars calculated as absolute standard deviation. The background histogram shows the total number of sources in each visual type. A total of five sources are visually unclassified, even though two of them have a S\'ersic model from \texttt{GALFIT}.}
                                \end{center}
                        \end{figure}

                        In Figure \ref{fig:morphology} we show the results of our morphological classification using an automatic fitting of a single-Sérsic model and a visual classification. More than a half (12 sources, 57\%; see the background histogram) of the full \HA\ sample are classified as discs with a median Sérsic index \nmean\ of 1.31. Two ($\sim\,9.5$\%) objects are classified as spheroidal with \nmean\ of 2.2. The increase of the S\'ersic index from discs to spheroids is expected because early-type galaxies are fitted with higher S\'ersic profile than late-type ones \citep[e.g.][]{Vika2013MNRAS.435..623V}. The number of sources classified as point-like and tadpole is negligible (one in each class, 9.5\%), with \nmean$\,\sim\,4.16$ and 0.7, respectively. For five (24\%) sources we are not able to provide the classification with either method. No sources are classified in the following morphological types: chain, clumpy-cluster, and doubles.

                        \subsection{Possible transitional dwarf galaxy}
                        We identified one source as a possible transitional dwarf galaxy, which is the outlier of the MZR (Figure \ref{fig:MZ}; green square, source {\tt id: 3137} shown in Fig. \ref{fig:AppepsTrans}). The selection is based on the cuts in stellar mass \logmass $<\,9$ and metallicity $12 + \log (\mathrm{O/H})\,>8.6$, following \citet{Peeples2008ApJ...685..904P}. According to these authors, it is most likely that sources with these characteristics are metal-rich dwarf galaxies. The percentage of metal-rich dwarf galaxies in our SFG sample is $\sim\,$6\% (one source in our SFG sample) compared with less than 1\% in the local SDSS sample selected with the same criteria. Morphologically, this source is visually classified as a disc  with a S\'ersic index $n\,=1.25$, it is compact (0.7 kpc), and has a close companion (at least in projection).  \citet{Peeples2008ApJ...685..904P} studied a sample of 41 local ($z\,<\,0.05$) metal-rich dwarf galaxies selected from the SDSS survey from \citet{Tremonti2004ApJ...613..898T}. They predicted that in order to have such high metallicity, these sources also need to have a lower gas fraction than objects with similar SFR and luminosities. They discussed other possible explanations for this behaviour: environment (interactions), effective yields, and finally the possibility that these sources are transitional dwarf galaxies at the end of their period of star formation. The outlier in our sample has a low stellar-mass \logmass$\,=\,8.2$, high metallicity $12 + \log (\mathrm{O/H})\,=\,8.6$, low SFR of $\log \mathrm{SFR} \sim -1$ \msun yr$^{-1}$ , and a high sSFR of $\log \mathrm{sSFR} \sim -9.2$. More information, in particular spectroscopic observations, is needed to disentangle its morphology and to shed light on its real nature.

                        \section{Conclusions}
                        \label{sec:conclusion}
                        
                        We studied the MZR and scaling relations between stellar mass, SFR, and sSFR, and the morphology of the low-mass \HA\ sample drawn from the OTELO survey at $z\,\sim0.4$. We reach three orders of magnitude deeper in stellar mass ($\sim\,10^{6.8}\,<$ \mstar/\msun $<\,10^{10}$ with 57\% of the sample with \mstar $< 10^{9}$\msun) than the SDSS and GAMA samples at similar redshift. The PS (data-product of OTELO survey) gives us the possibility to efficiently select and study the metallicity of these low-mass ELSs. About 95\% of our ELS sample was selected from spectral features present in PS through a semi-automatic procedure detailed in Section \ref{sec:data}. If the colour-excess technique used in classical NB surveys had been used instead, only half of this sample (i.e. necessarily the sources with larger observed EWs) would be recovered. For more details, see \citetalias{OteloI} and \citet{Otelo-OIII}.
                        
                        Because of the OTELO instrumental limits [as seen in the N2 index and/or $12 + \log (\mathrm{O/H})$] and the associated errors in the recovery of the line fluxes and the low number of objects in the sample, we cannot conclude about a possible MZR evolution that would be due to cosmic variance. Furthermore, our results are consistent with those obtained using the local SDSS sample (see Figure \ref{fig:MZ}). 
                        
                        We also explored the SFR as a function of \mstar\ , which has previously been studied by \citet{Noeske2007ApJ...660L..43N}, who found the so-called main sequence. We showed that this relation holds for the sample we studied, but with a systematic shift by $\sim\,0.25$ dex in SFR, which is associated with the different stellar mass ranges used in \citet{Noeske2007ApJ...660L..43N} and in this work. Our data agree with the VVDS-Deep and Ultra-Deep data used in this work. Furthermore, the global distribution of the OTELO SFG sample lies within the local SDSS distribution.
                        
                        From the automatic and visual morphological classification (see Section \ref{sec:morphology}), we find that the majority of sources in the sample are discs (57\%), followed by spheroids (9.5\%) with median S\'ersic indices of  1.31 and 2.2, respectively. Two sources are classified as point-like and tadpole (one in each class, 9.5\% in total). No classification is given for 24\% of our sample because they are not detected in the high-resolution \hstacs\ F814W image, have no data, or because the sources are just too faint to provide any reliable classification.
                        
                        We obtained the \mstar--size  relation for our \HA\ sample. Our results agrees with those of \citet{Ichikawa2012MNRAS.422.1014I}, but have a slightly higher slope (0.14 and 0.115, respectively). The difference in the results in this work and in \citet{vanderWel2014ApJ...788...28V} is probably due to the different range of stellar-masses that is covered in both studies (we reach almost three orders of magnitude deeper in \mstar\ than \citeauthor{vanderWel2014ApJ...788...28V}).

                        We identified one candidate ($\sim\,6$\% of our SFG sample) as a transitional dwarf galaxy with low mass \logmass$\,=\,8.6\,$\msun\ and high metallicity $12 + \log (\mathrm{O/H})\,= 8.64$. Considering the same criteria, we find less than 1\% of these sources in the local-SDSS sample. Morphologically, this source is classified as a compact ($\sim\,0.7$ kpc) disc galaxy ($n\,=1.25$) with blue \gmenosi\ colour. The study of MZR, SFR, sSFR, and its colour and morphology suggest that this source is indeed a transitional dwarf galaxy at the end of its star formation period, as discussed in previous works \citep[e.g.][]{Peeples2008ApJ...685..904P}. More information is needed to conclude on this issue, however.

                        \begin{acknowledgements}
                                This work was supported by the project Evolution of Galaxies, of reference AYA2014-58861-C3-1-P, AYA2017-88007-C3-1-P and AYA2017-88007-C3-2-P, within the "Programa estatal de fomento de la investigación científica y técnica de excelencia del Plan Estatal de Investigación Científica y Técnica y de Innovación (2013-2016)" of the "Agencia Estatal de Investigación del Ministerio de Ciencia, Innovación y Universidades", and co-financed by the FEDER "Fondo Europeo de Desarrollo Regional". APG and MC are also supported by the Spanish State Research Agency grant MDM-2017-0737 (Unidad de Excelencia María de Maeztu  CAB). JAD is grateful for the support from the UNAM-DGAPA-PASPA 2019 program, and the kind hospitality of the IAC.
                                
                                Based on observations made with the Gran Telescopio Canarias (GTC), installed in the Spanish Observatorio del Roque de los Muchachos of the Instituto de Astrof\'isica de Canarias, on the island of La Palma.
                                
                                MALL is a DARK-Carlsberg Foundation Fellow (Semper Ardens project CF15-0384).
                                
                                This research uses data from the VIMOS VLT Deep Survey, obtained from the VVDS database operated by Cesam, Laboratoire d'Astrophysique de Marseille, France.
                                
                                Based on observations obtained with MegaPrime/MegaCam, a joint project of CFHT and CEA/IRFU, at the Canada-France-Hawaii Telescope (CFHT) which is operated by the National Research Council (NRC) of Canada, the Institut National des Science de l'Univers of the Centre National de la Recherche Scientifique (CNRS) of France, and the University of Hawaii. This work is based in part on data products produced at Terapix available at the Canadian Astronomy Data Centre as part of the Canada-France-Hawaii Telescope Legacy Survey, a collaborative project of NRC and CNRS.
                                
                                This work had made extensive use of TOPCAT and STILTS software \citep{Taylor2005}.

                        \end{acknowledgements}
                        
                        %
                        %
                        \bibliographystyle{aa} 

                        
                        %
                        %
                        %
                        
                        \begin{appendix} 
                                \section{Inverse deconvolution and uncertainty estimates}
                                \label{App_decInv}

                                \begin{figure*}[t!]
                                        \begin{center}
                                                \includegraphics[width=0.24\textwidth]{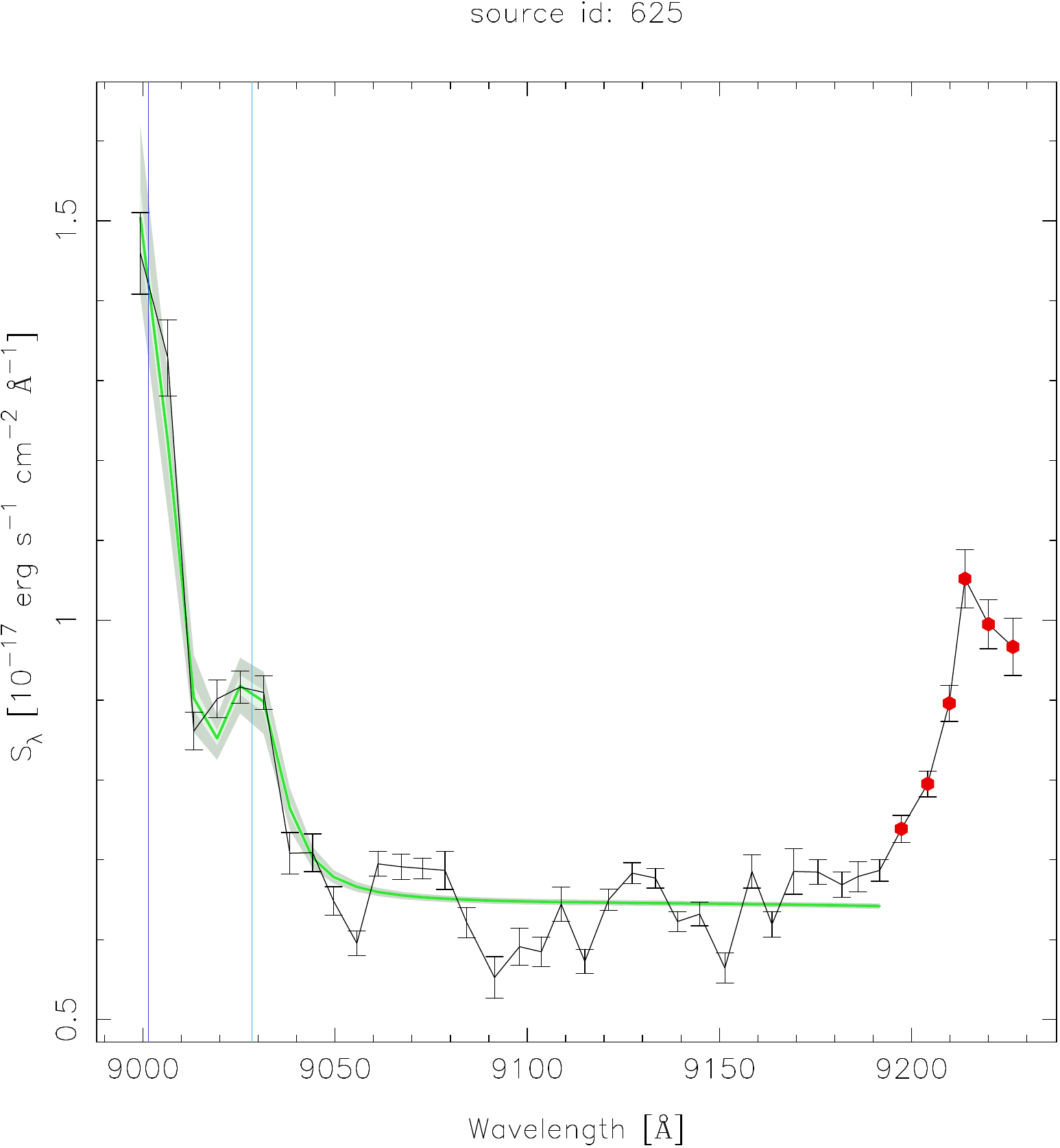}
                                                \includegraphics[width=0.24\textwidth]{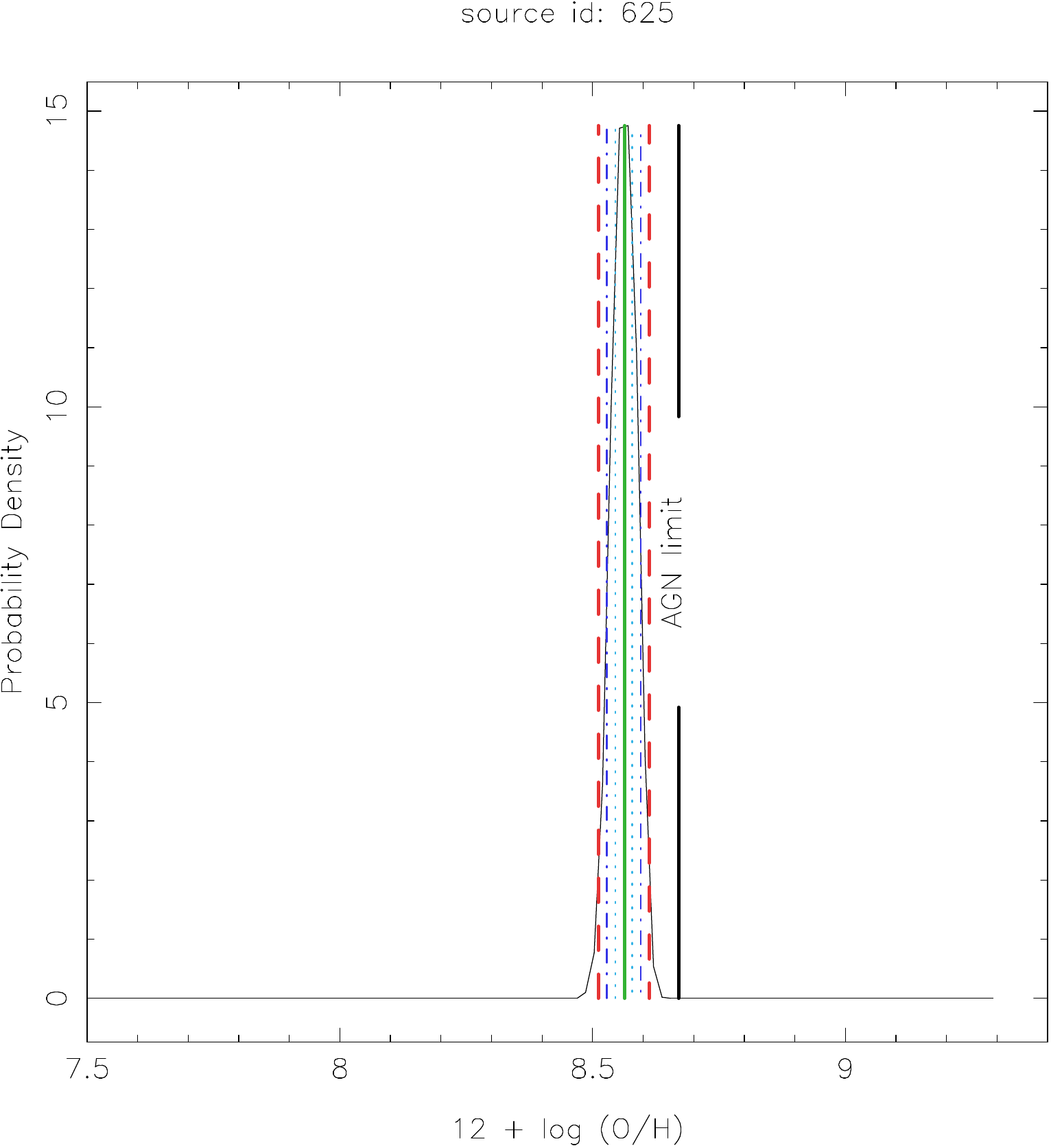}
                                                \includegraphics[width=0.24\textwidth]{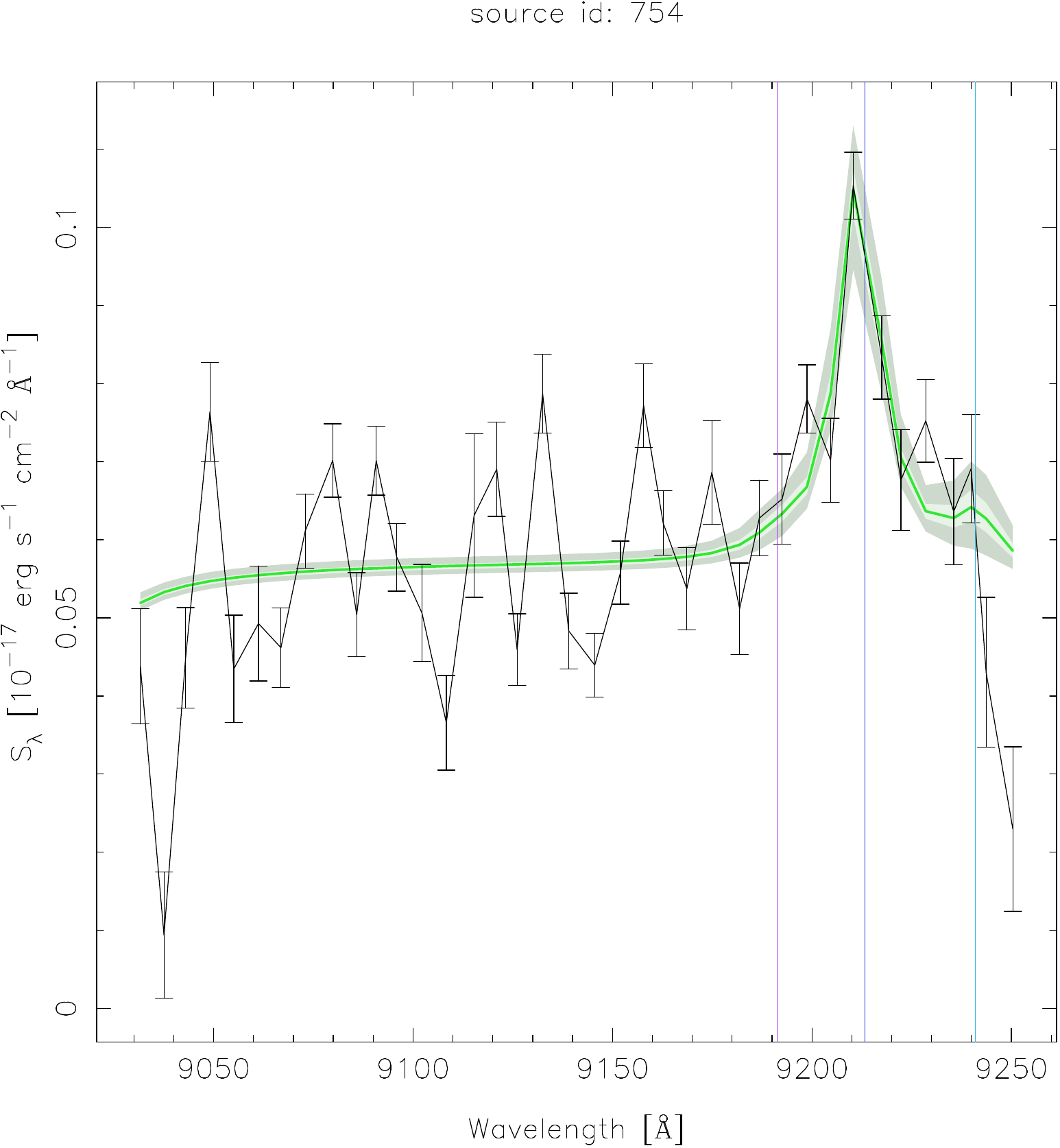}
                                                \includegraphics[width=0.24\textwidth]{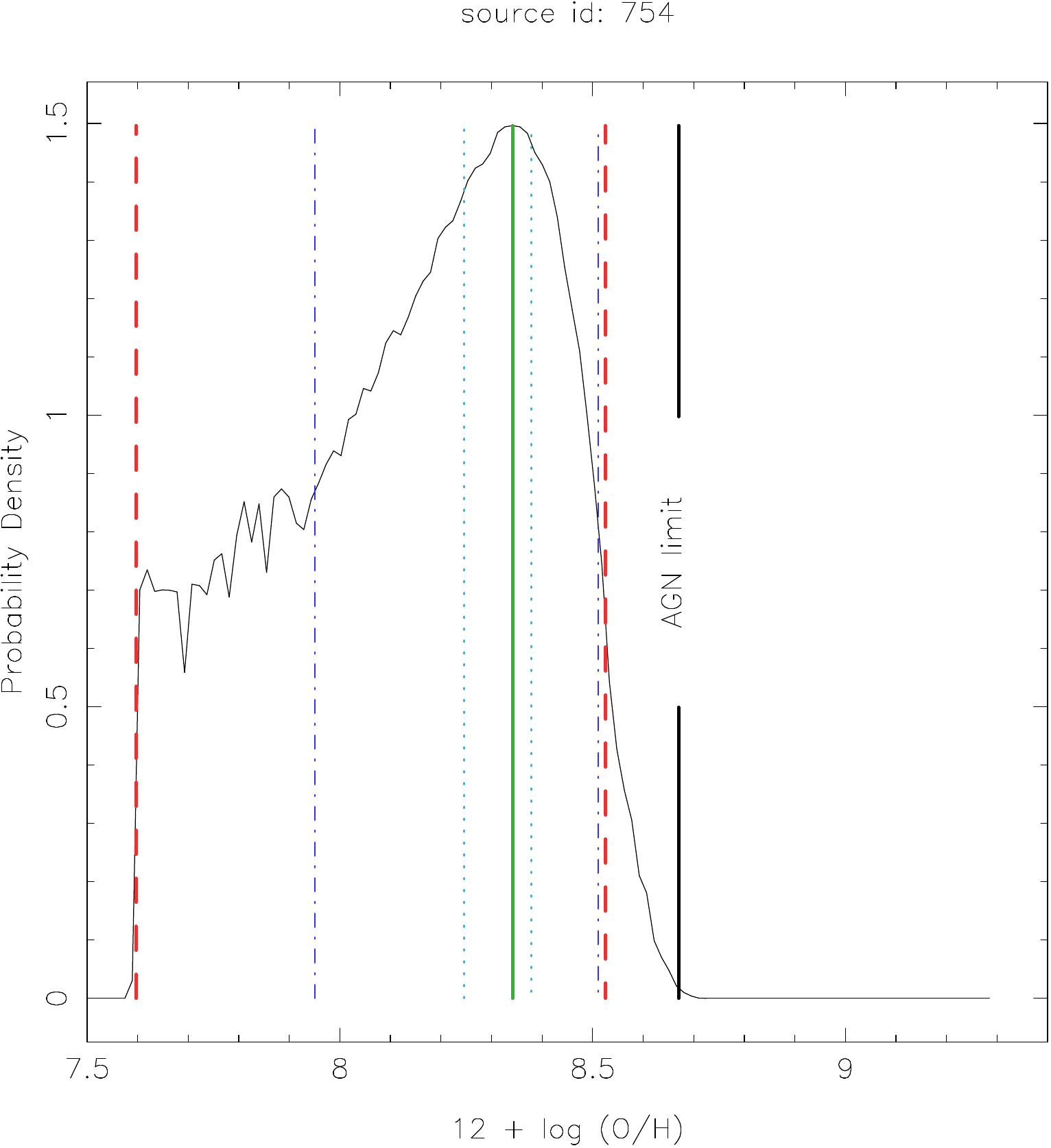} \\
                                                \includegraphics[width=0.24\textwidth]{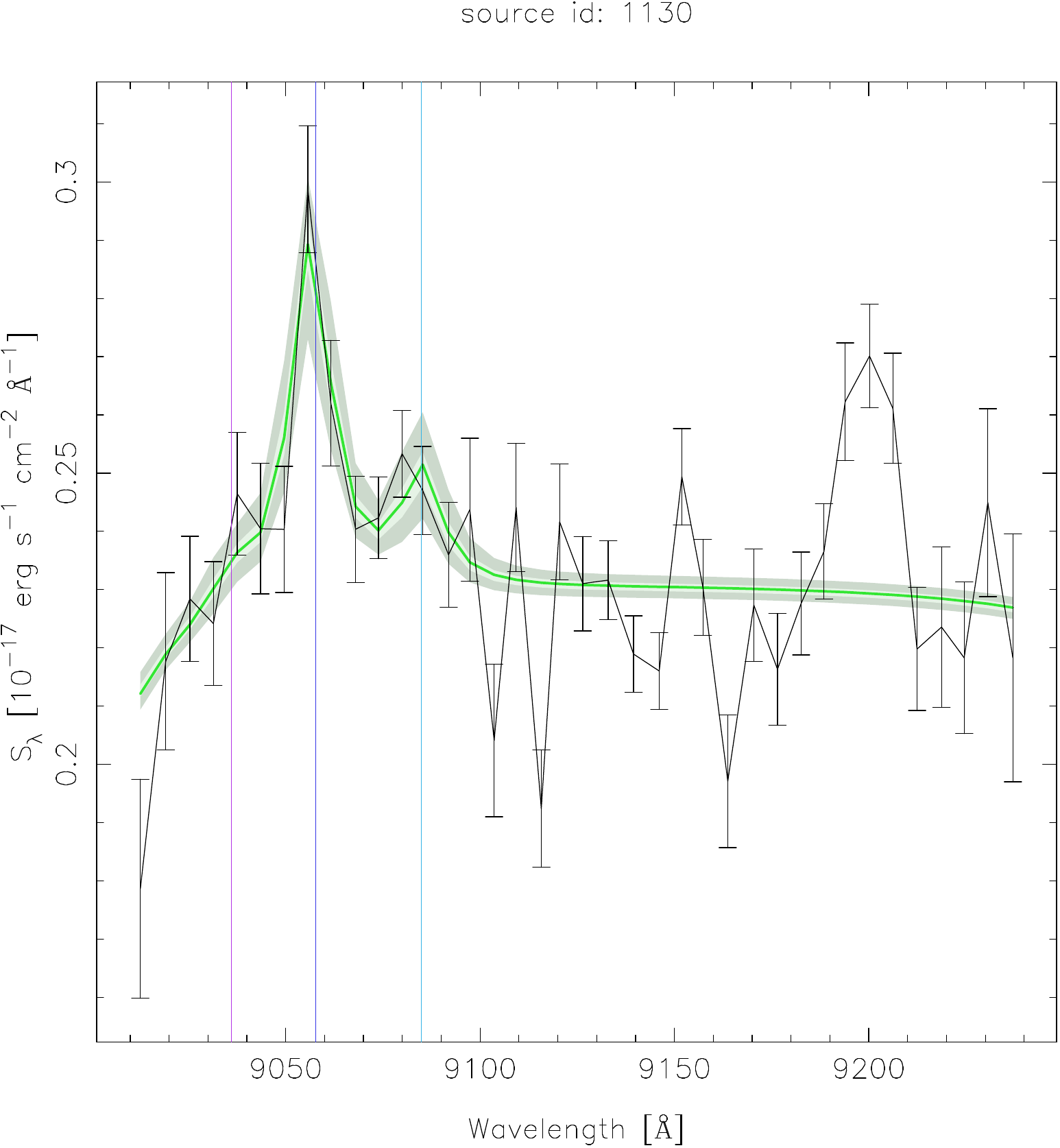}
                                                \includegraphics[width=0.24\textwidth]{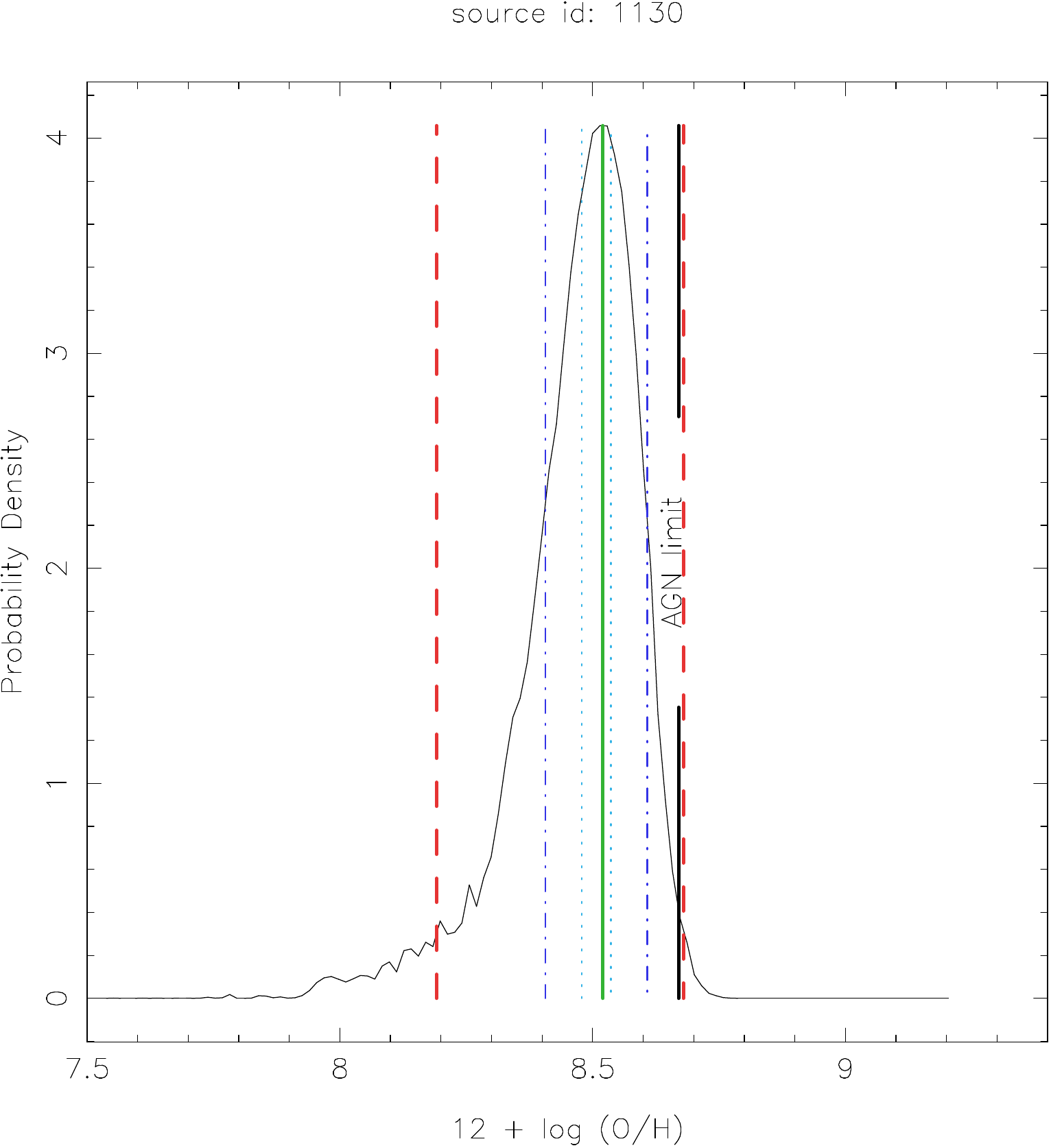}
                                                \includegraphics[width=0.24\textwidth]{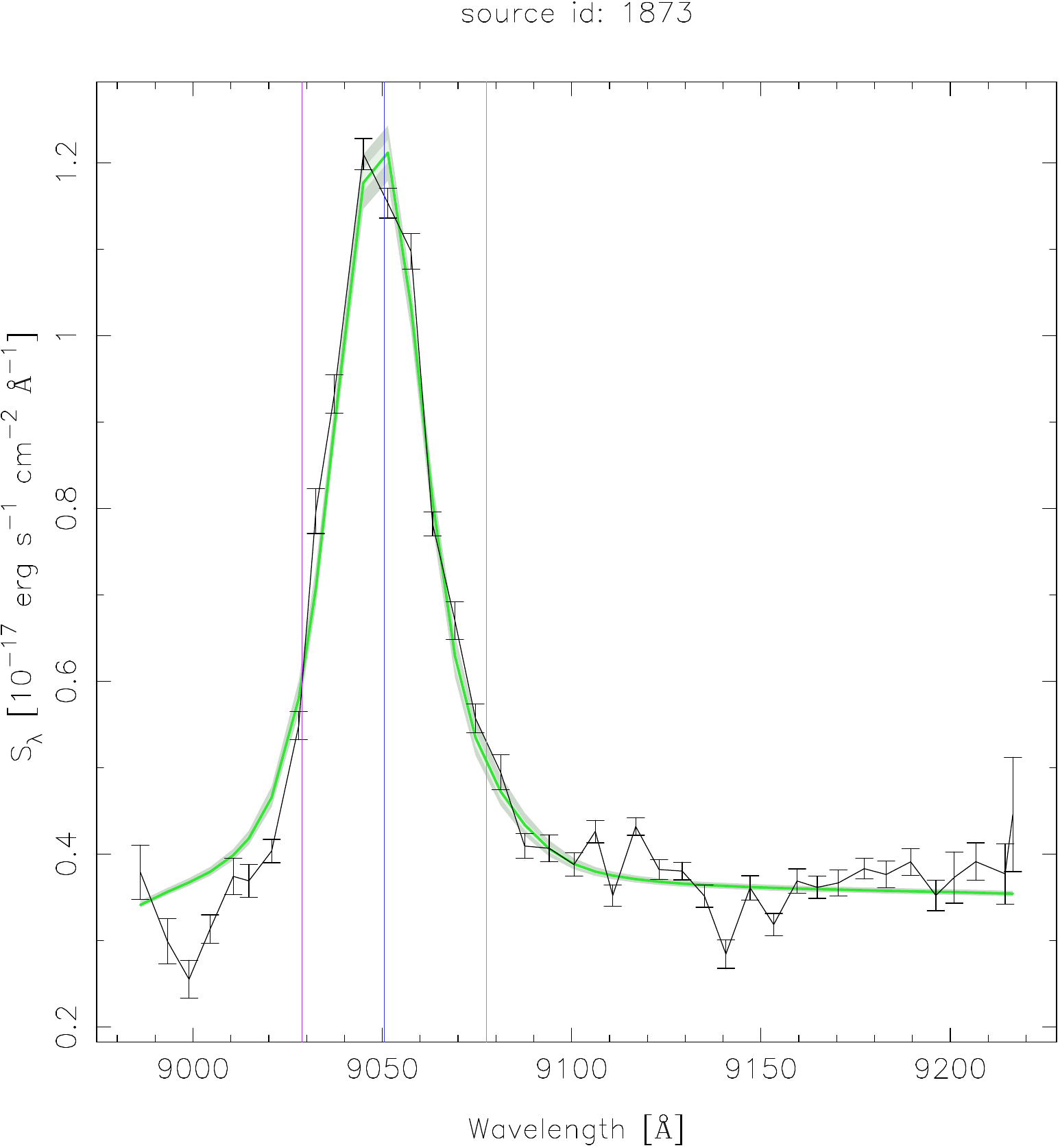}
                                                \includegraphics[width=0.24\textwidth]{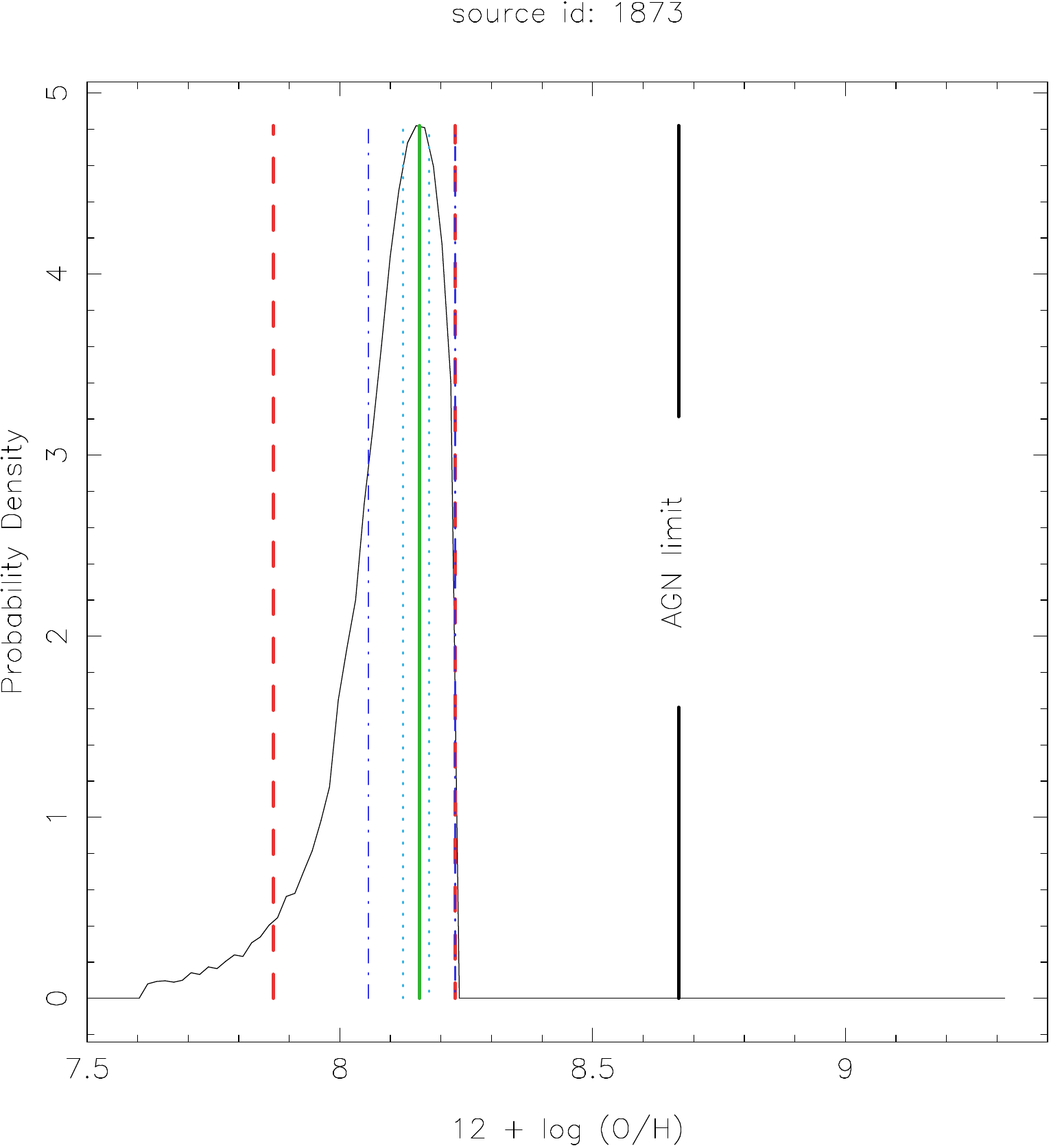} \\
                                                \includegraphics[width=0.24\textwidth]{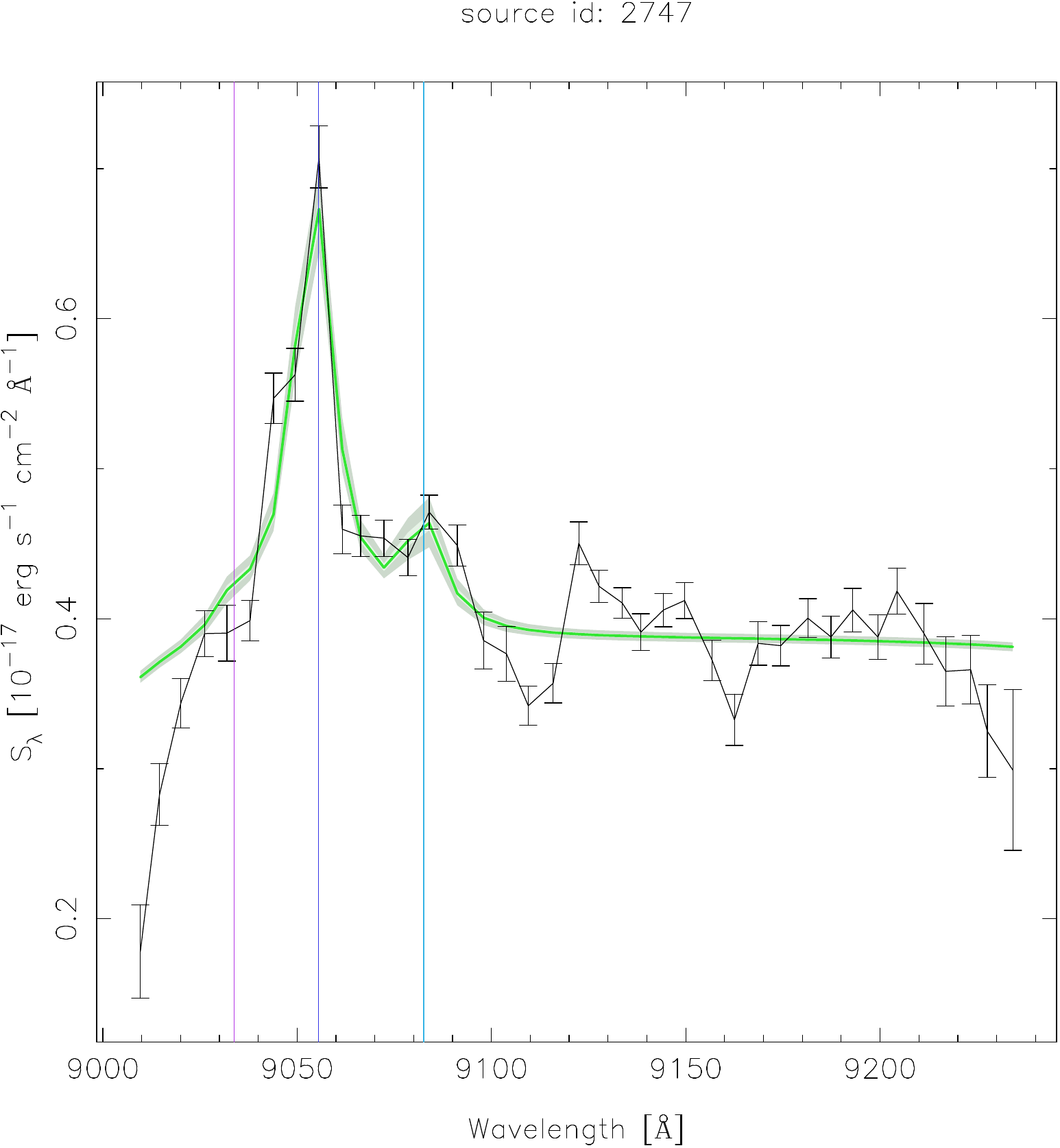}
                                                \includegraphics[width=0.24\textwidth]{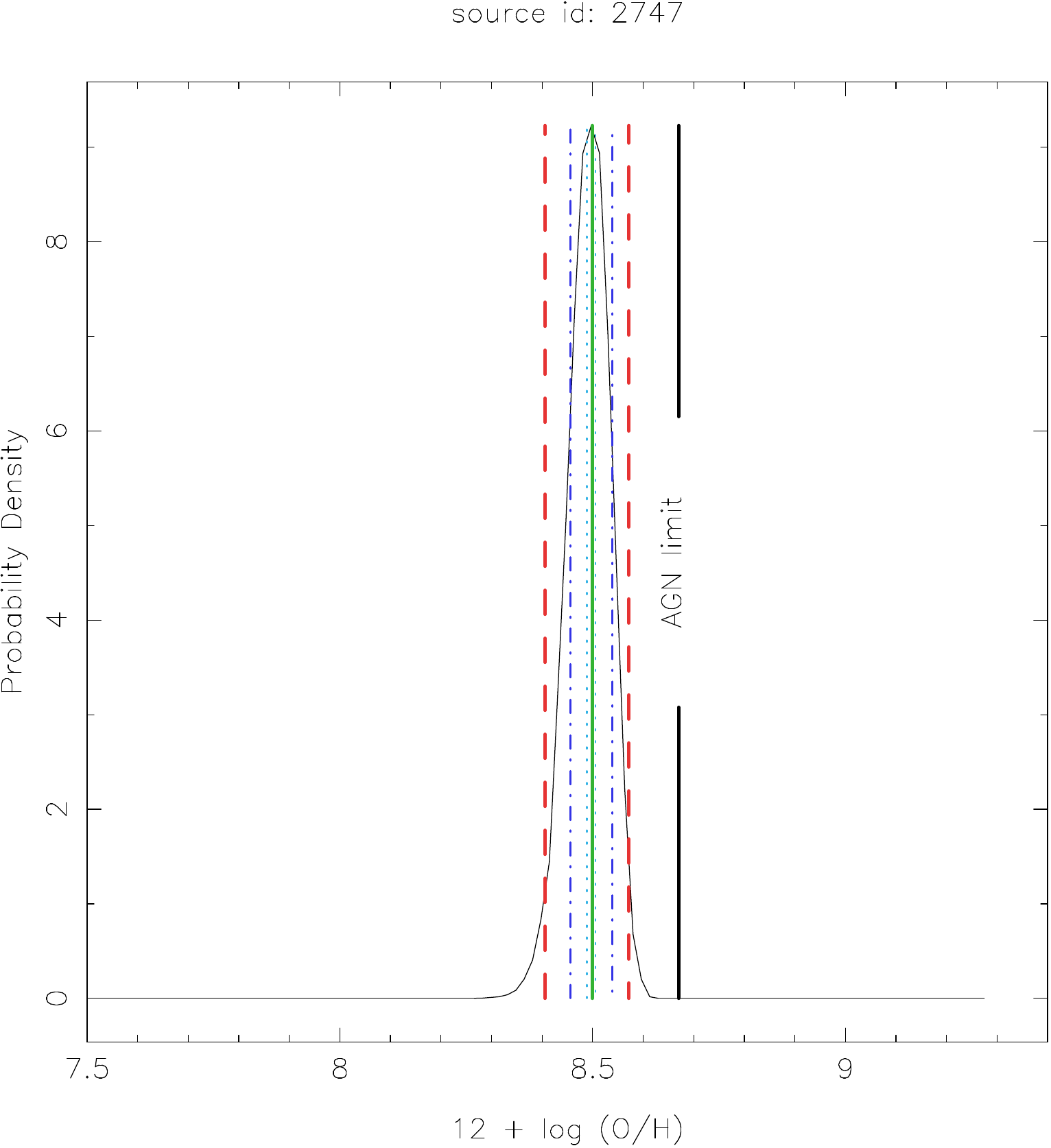}
                                                \includegraphics[width=0.24\textwidth]{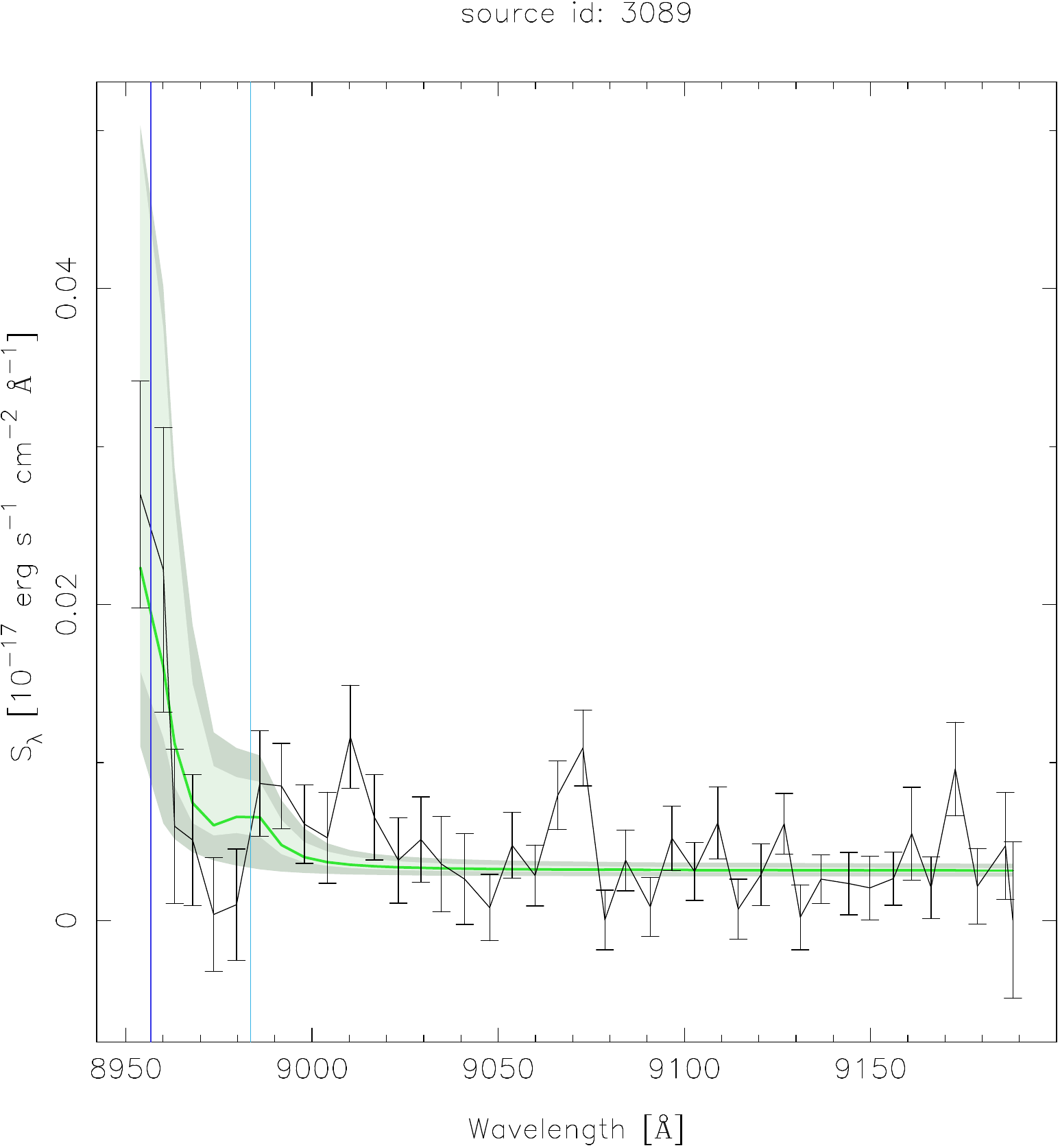}
                                                \includegraphics[width=0.24\textwidth]{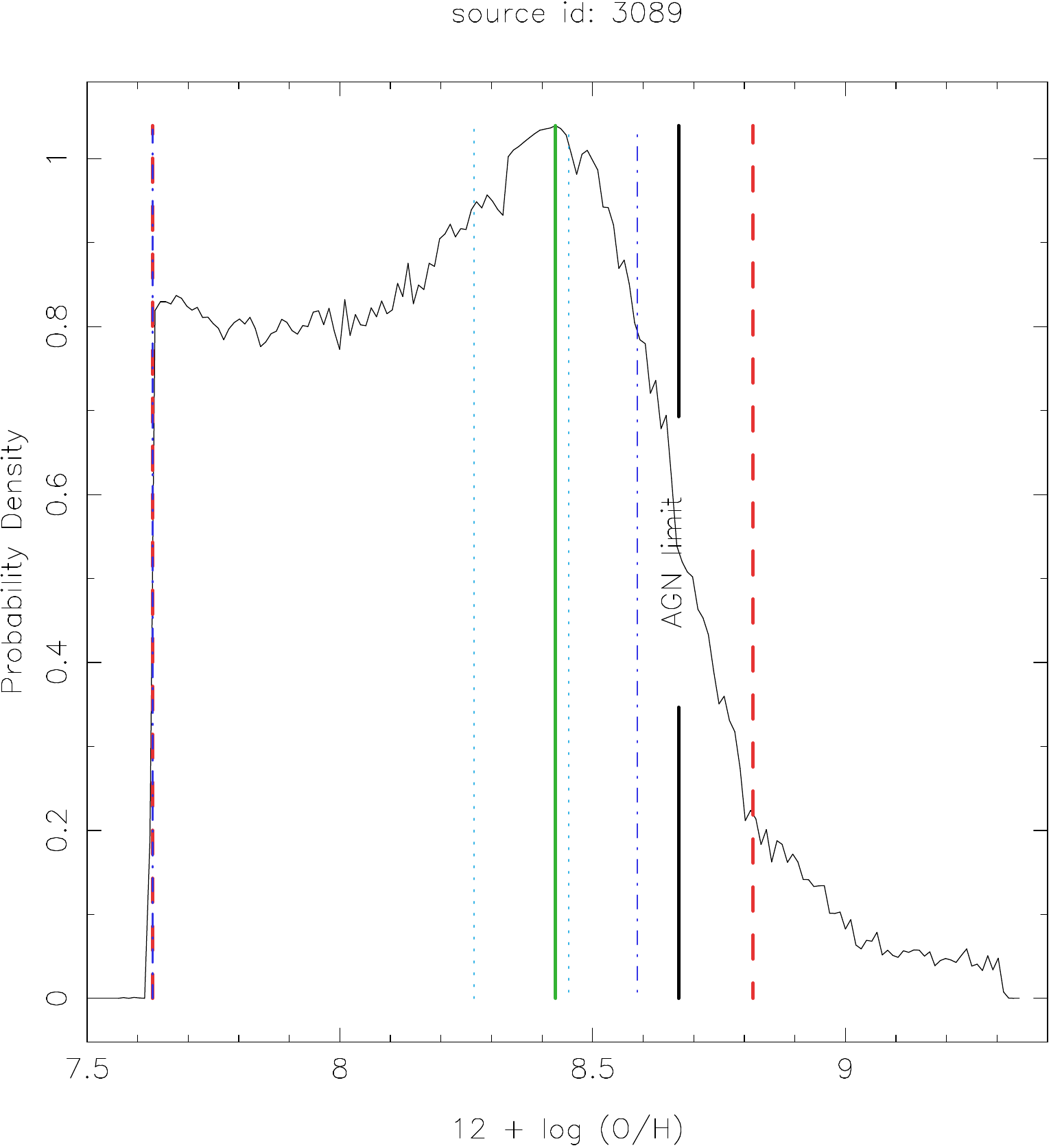} \\
                                                \includegraphics[width=0.24\textwidth]{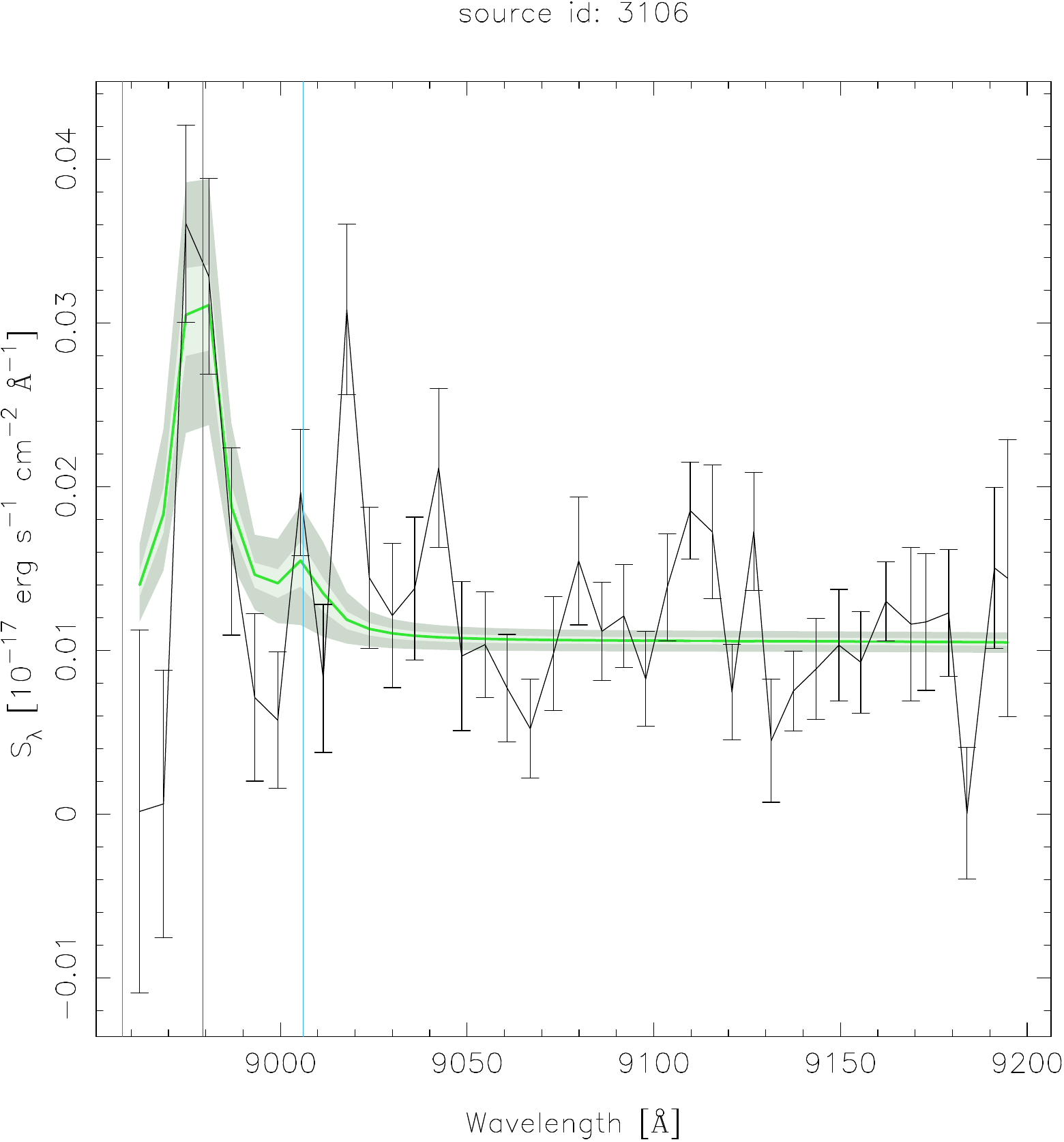}
                                                \includegraphics[width=0.24\textwidth]{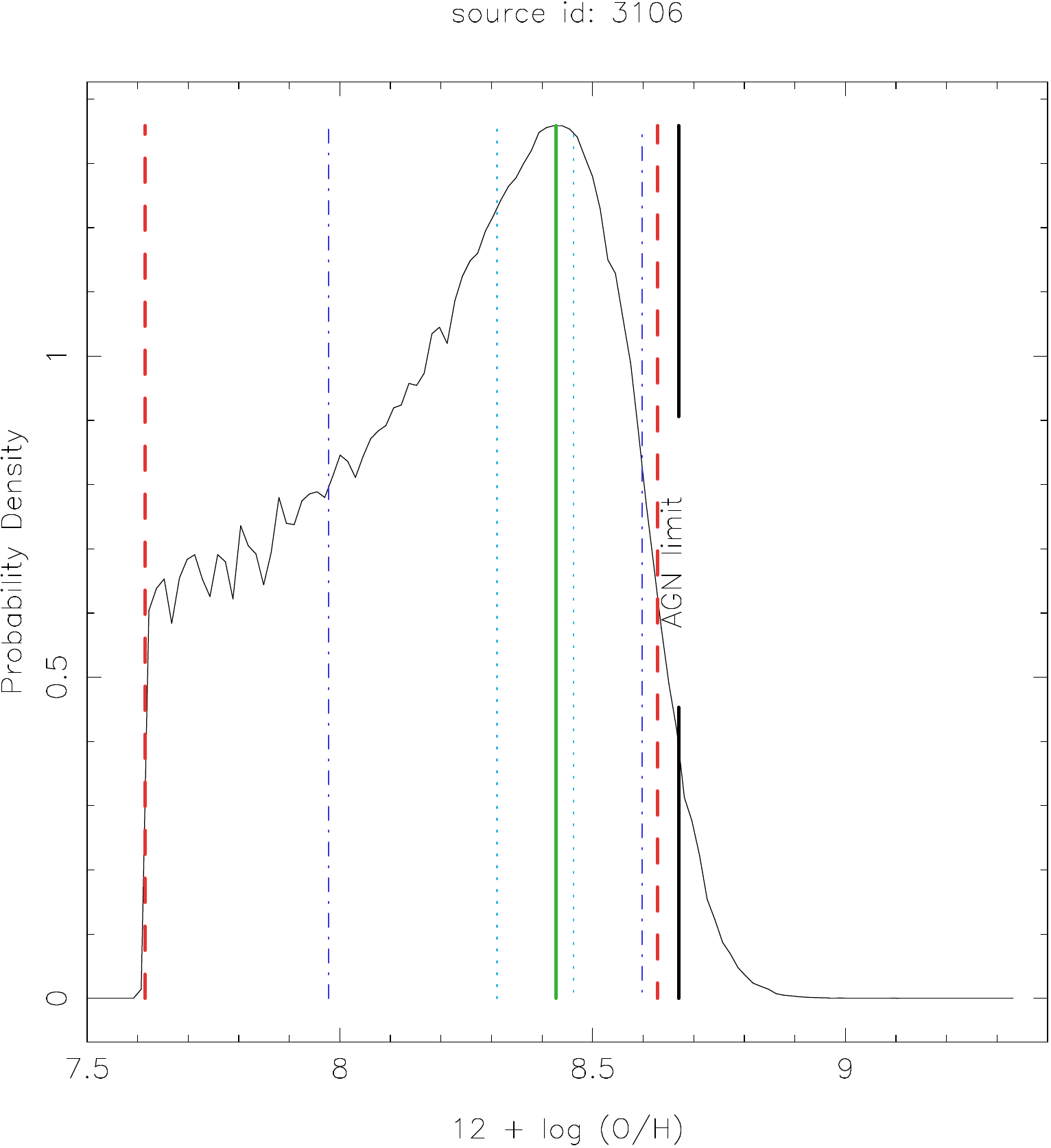}
                                                \includegraphics[width=0.24\textwidth]{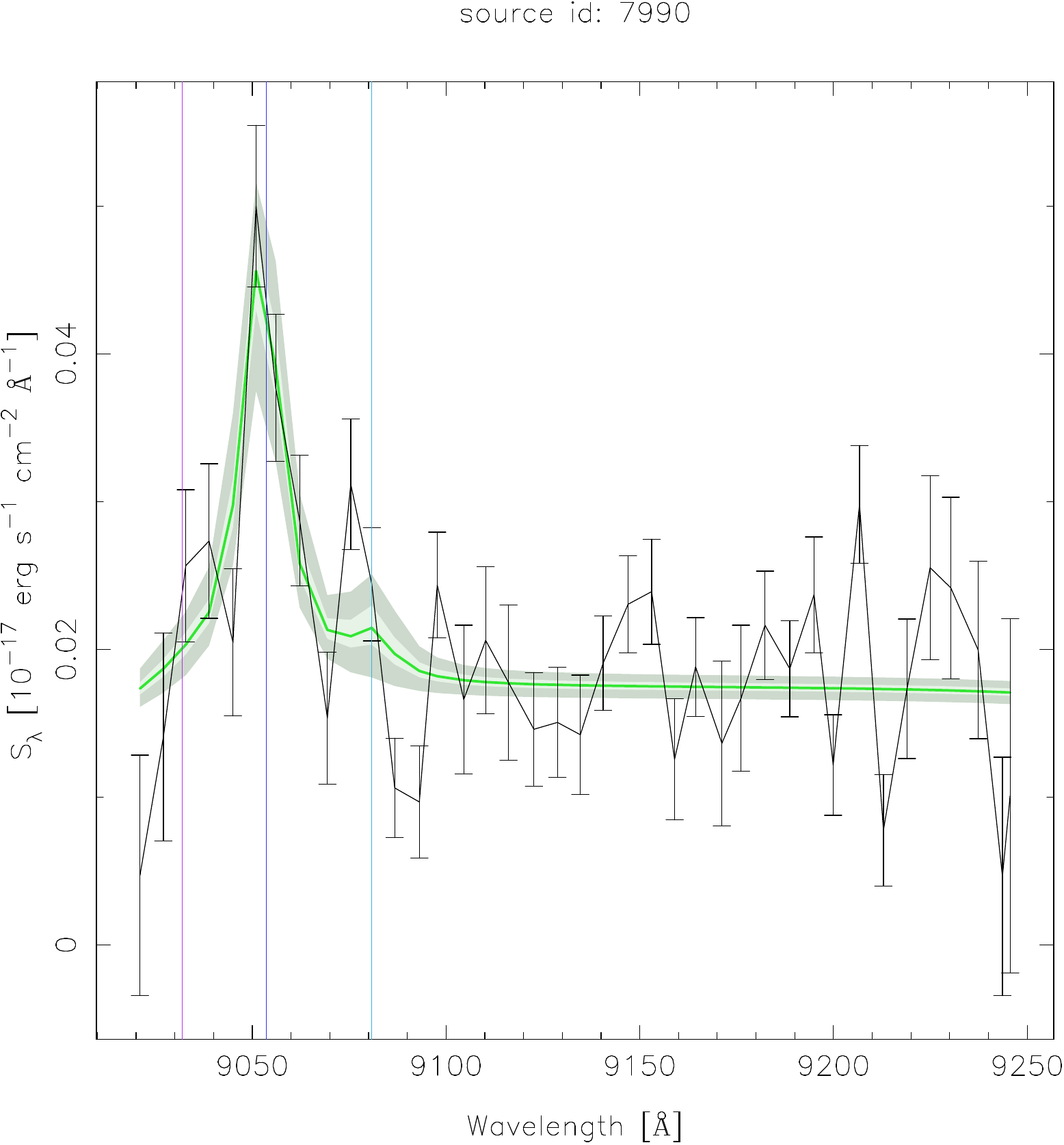}
                                                \includegraphics[width=0.24\textwidth]{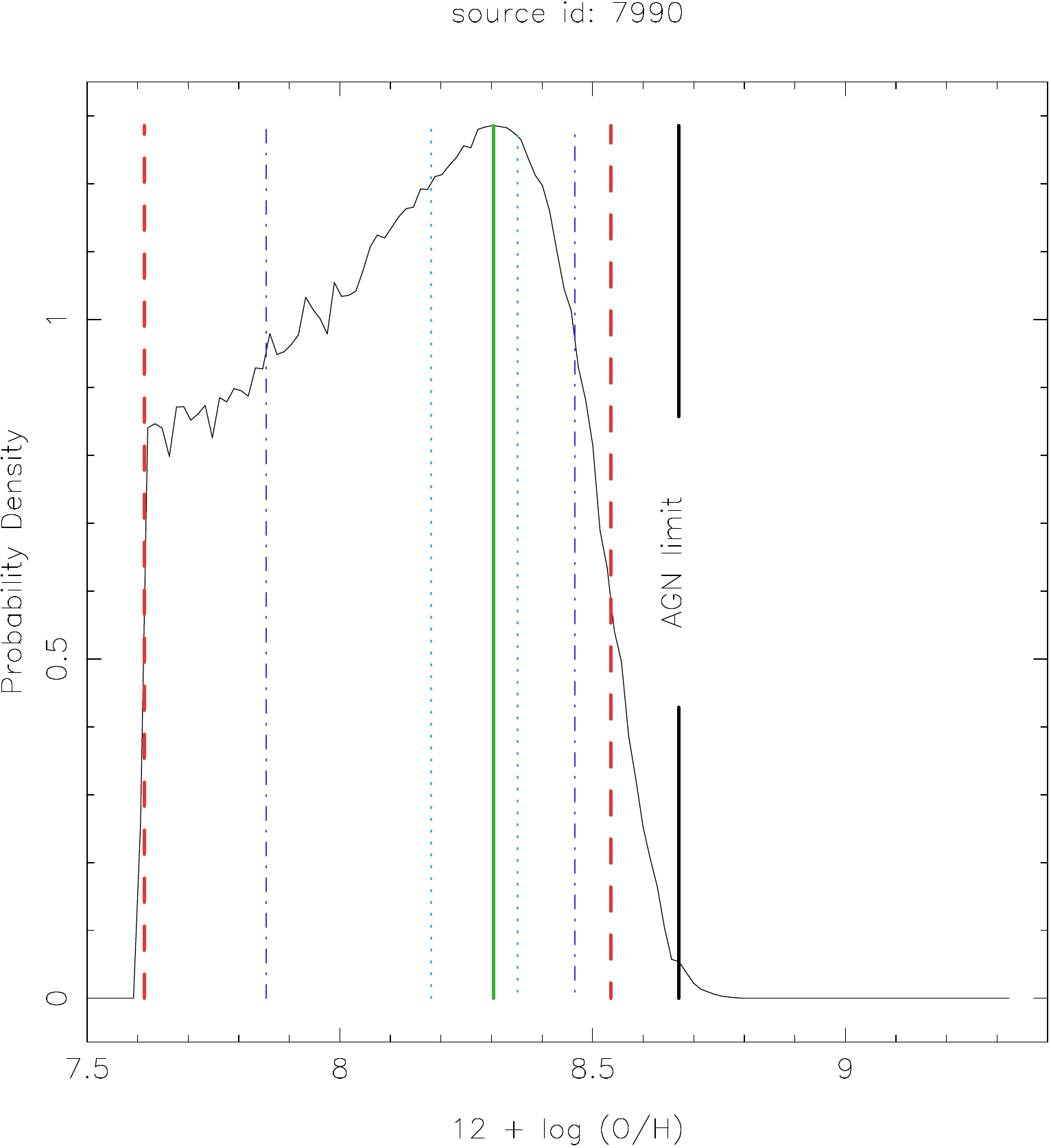}
                                                \caption{\label{fig:Appeps1} 
                                                        Observed PS, the best-fit PS, and the envelope of the PS simulations where all parameters are in the 25\% confidence interval (light green) and 68\% confidence interval (light grey) for the first eight SFG for which an \HAN\ line is detected. Violet, blue, and cyan vertical lines in the PS figures show the position of the \Nii $\lambda$6548, \HA,\ and \Nii $\lambda$6583 line, respectively, for the best $z$ obtained in by the fit. The red points that appear in some observed PS were removed during the fitting process because they correspond to the \Sii\ doublet or may be artefacts. To the right of each PS is the PDF of $12 + \log (\mathrm{O/H})$ that was obtained from the Monte Carlo simulations where the mode (our reference value), the 25\%, 68\%, and 90\% confidence intervals around the mode are shown as vertical lines; the black vertical line in these plots shows the empirical division between SFR and AGN at $12 + \log (\mathrm{O/H}) = 8.67$ (N2$=-0.4$). }
                                        \end{center}
                                \end{figure*}
                                
                                \begin{figure*}[t!]
                                        \begin{center}
                                                \includegraphics[width=0.24\textwidth]{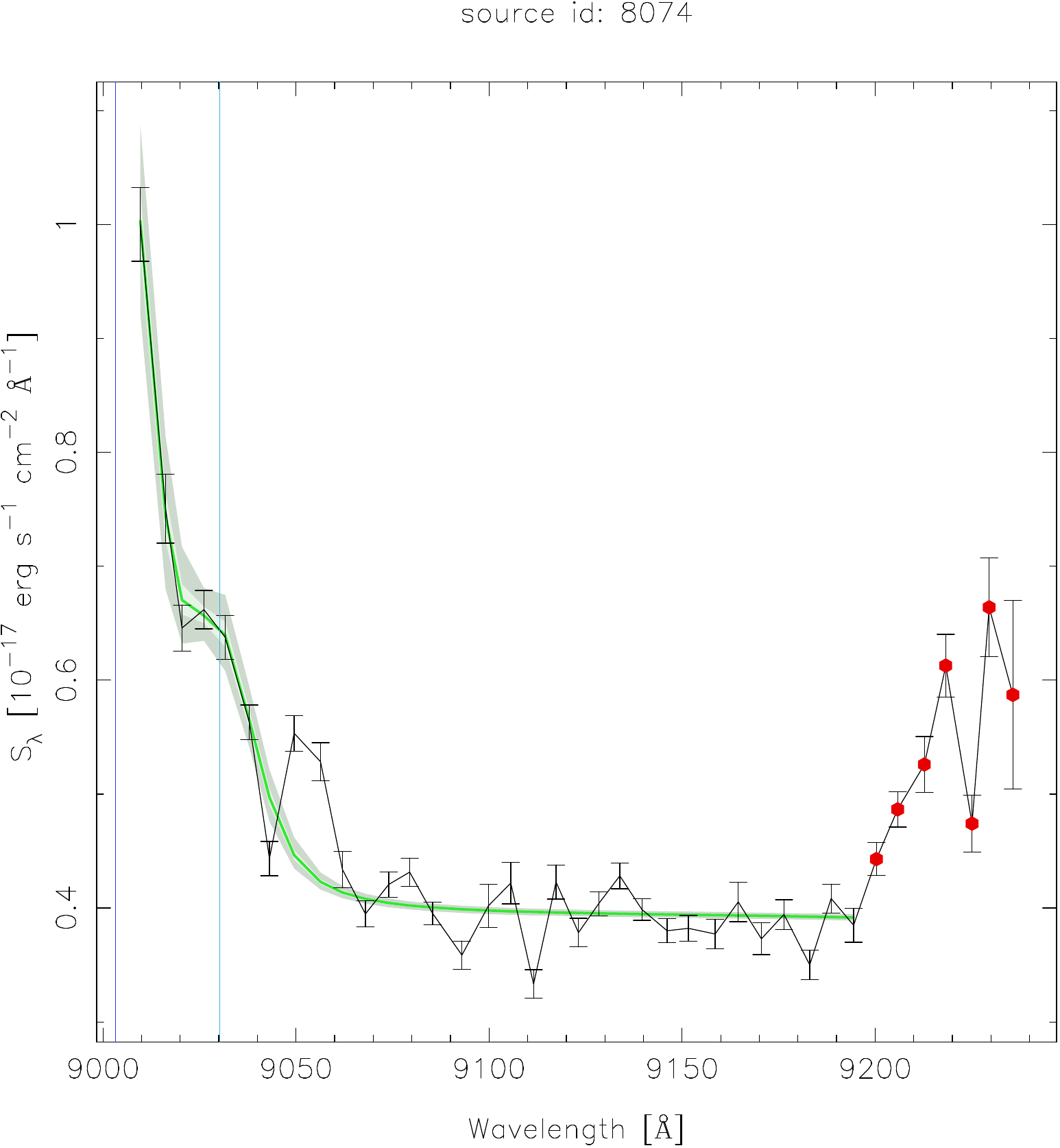}
                                                \includegraphics[width=0.24\textwidth]{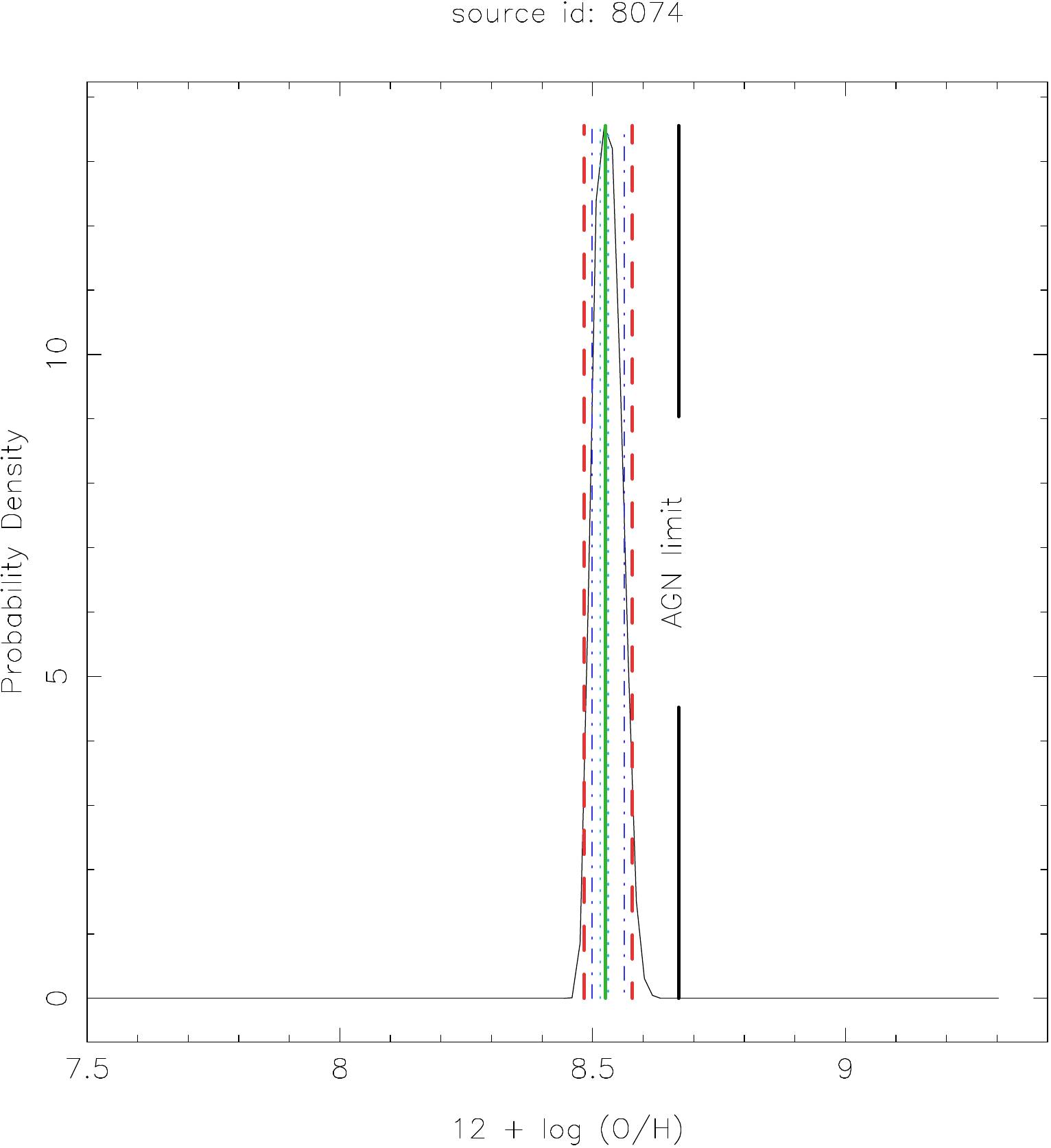} 
                                                \includegraphics[width=0.24\textwidth]{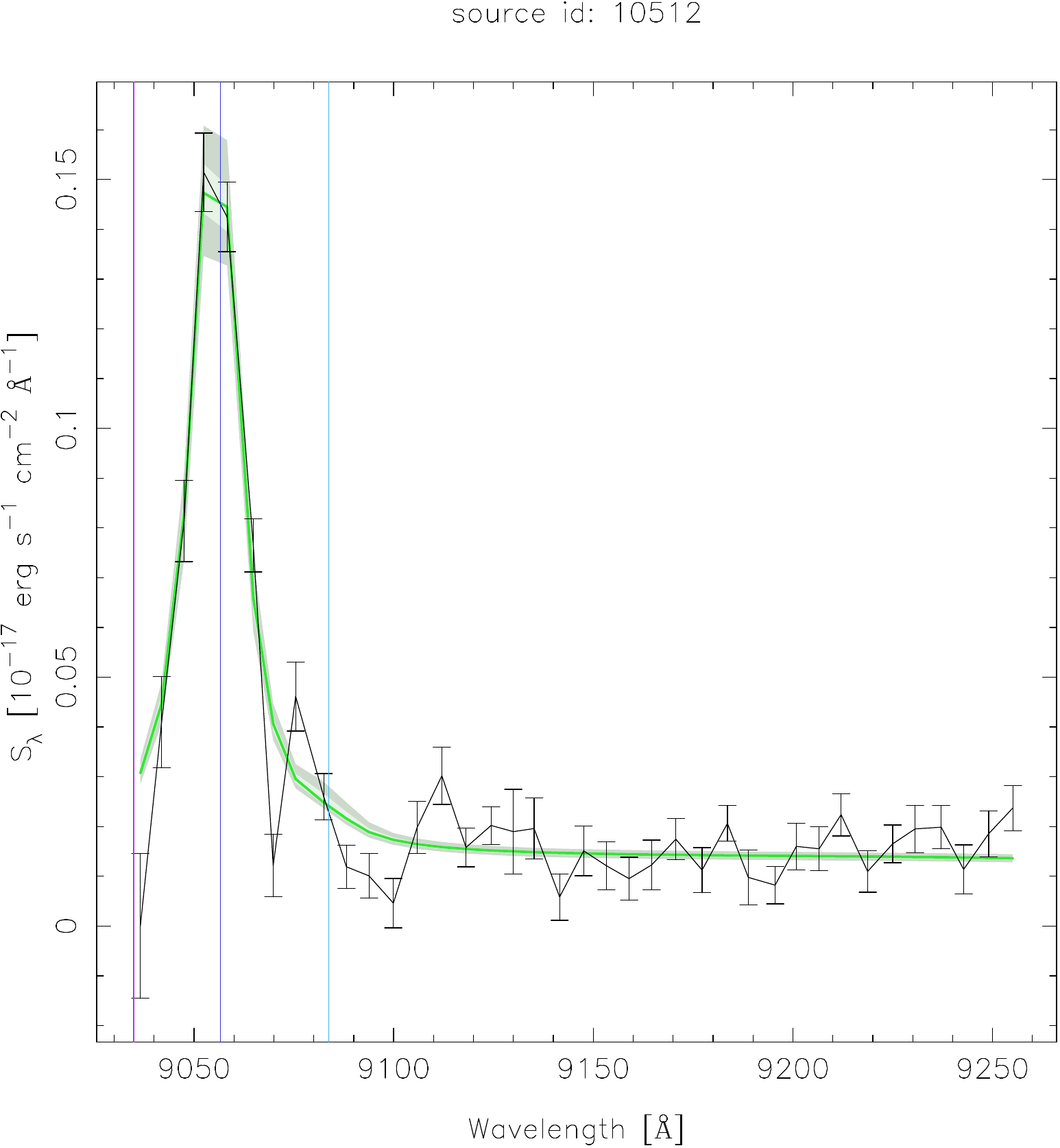}
                                                \includegraphics[width=0.24\textwidth]{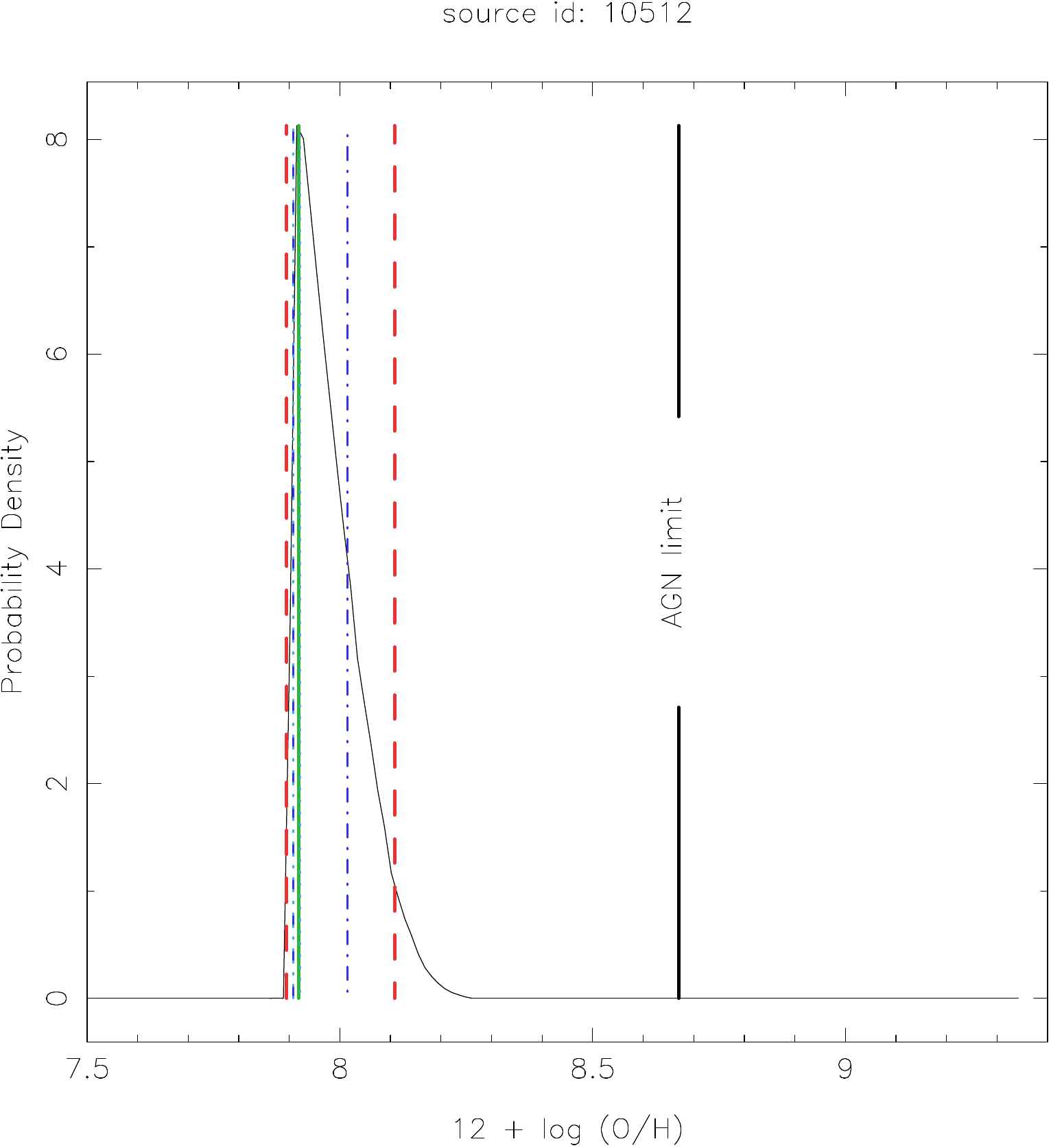}
                                                \caption{\label{fig:Appeps2} As Fig. \ref{fig:Appeps1} for the remaining two sources with a detected \Nii\ line.}
                                        \end{center}
                                \end{figure*}
                                
                                \begin{figure*}[t!]
                                        \begin{center}
                                                \includegraphics[width=0.24\textwidth]{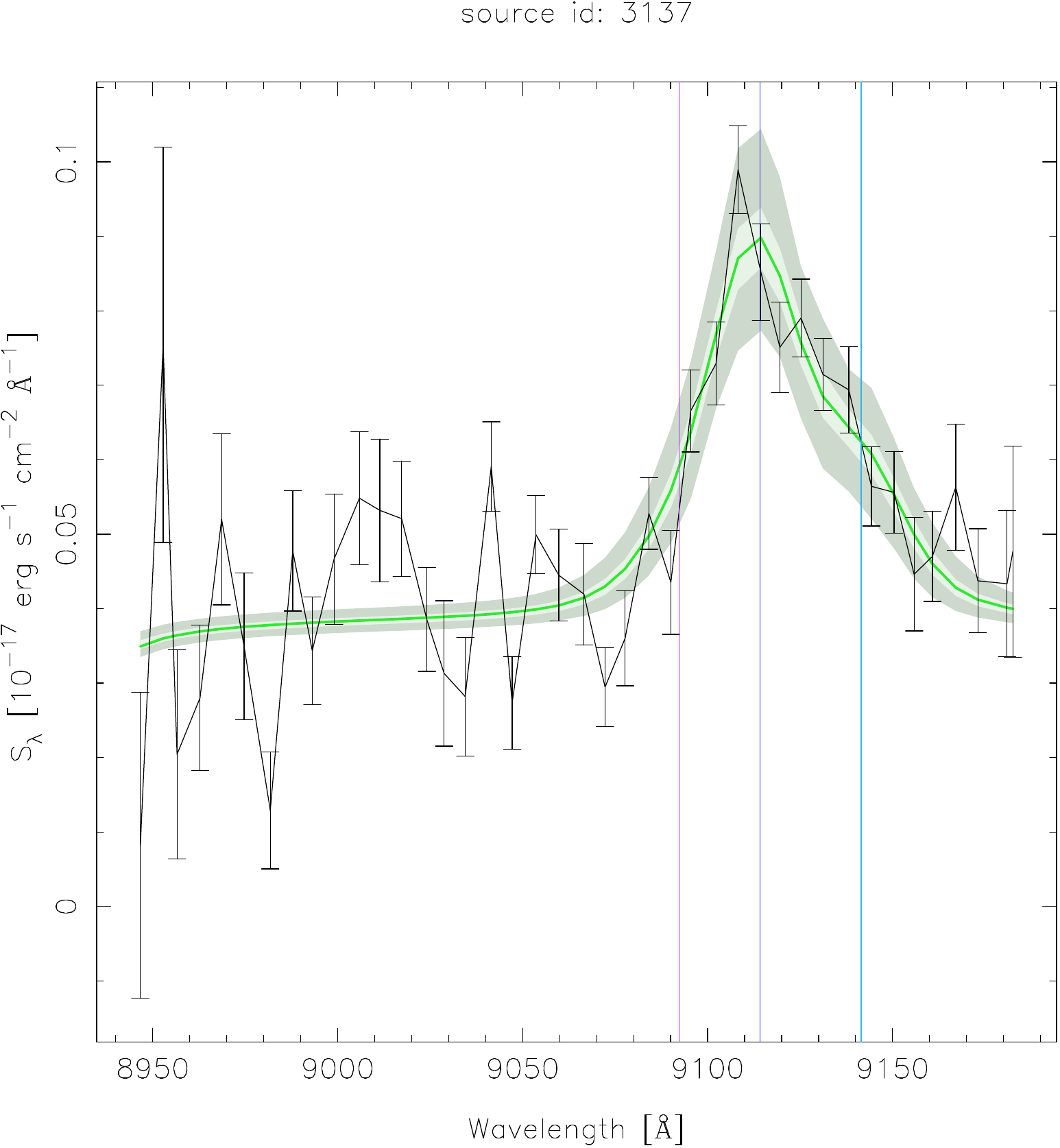}
                                                \includegraphics[width=0.24\textwidth]{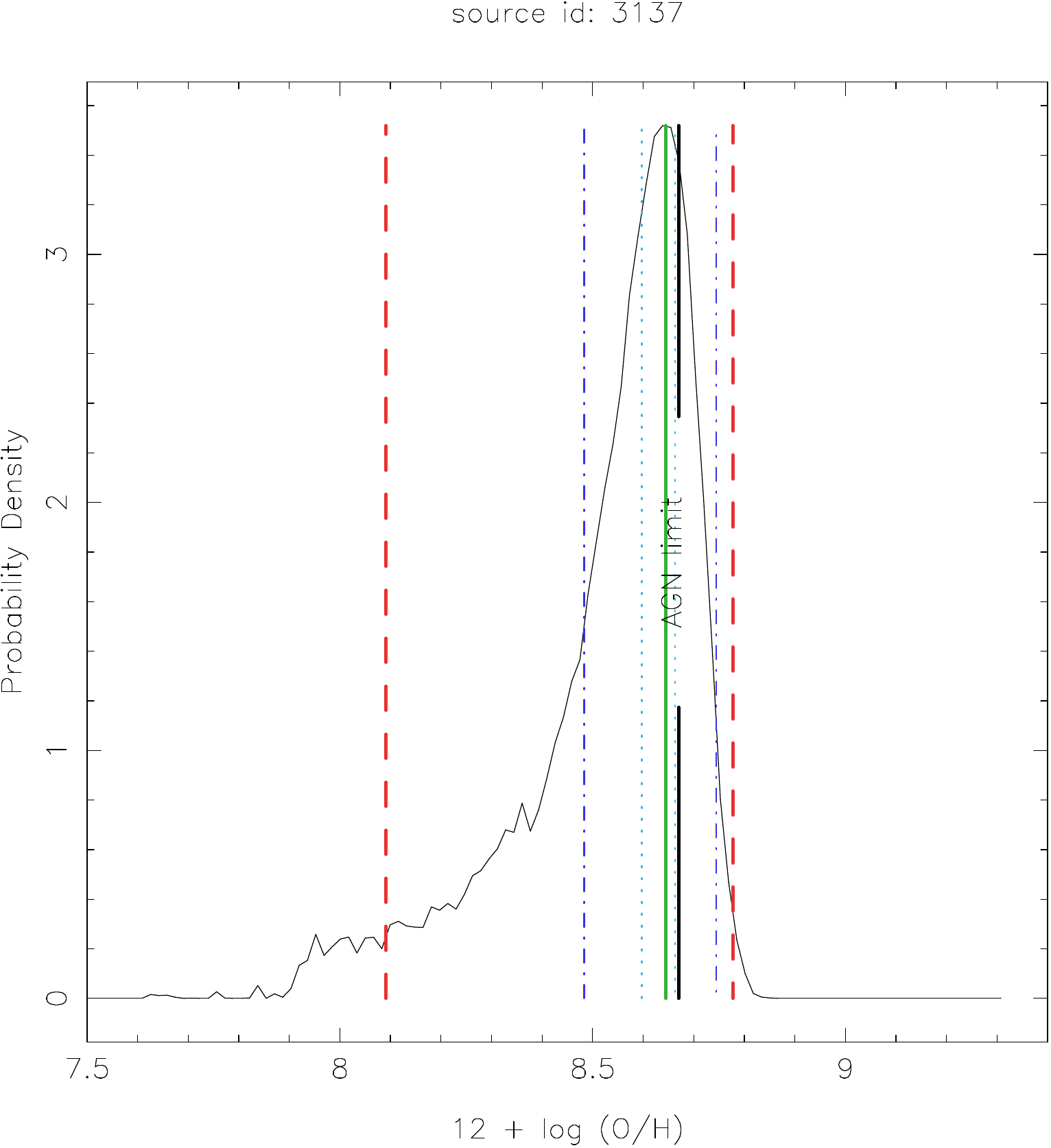} \\
                                                \caption{\label{fig:AppepsTrans} As Fig. \ref{fig:Appeps1}, but for the possible transitional dwarf galaxy (source {\tt id: 3137}).}
                                        \end{center}
                                \end{figure*}

                                \begin{figure*}[t!]
                                        \begin{center}
                                                \includegraphics[width=0.24\textwidth]{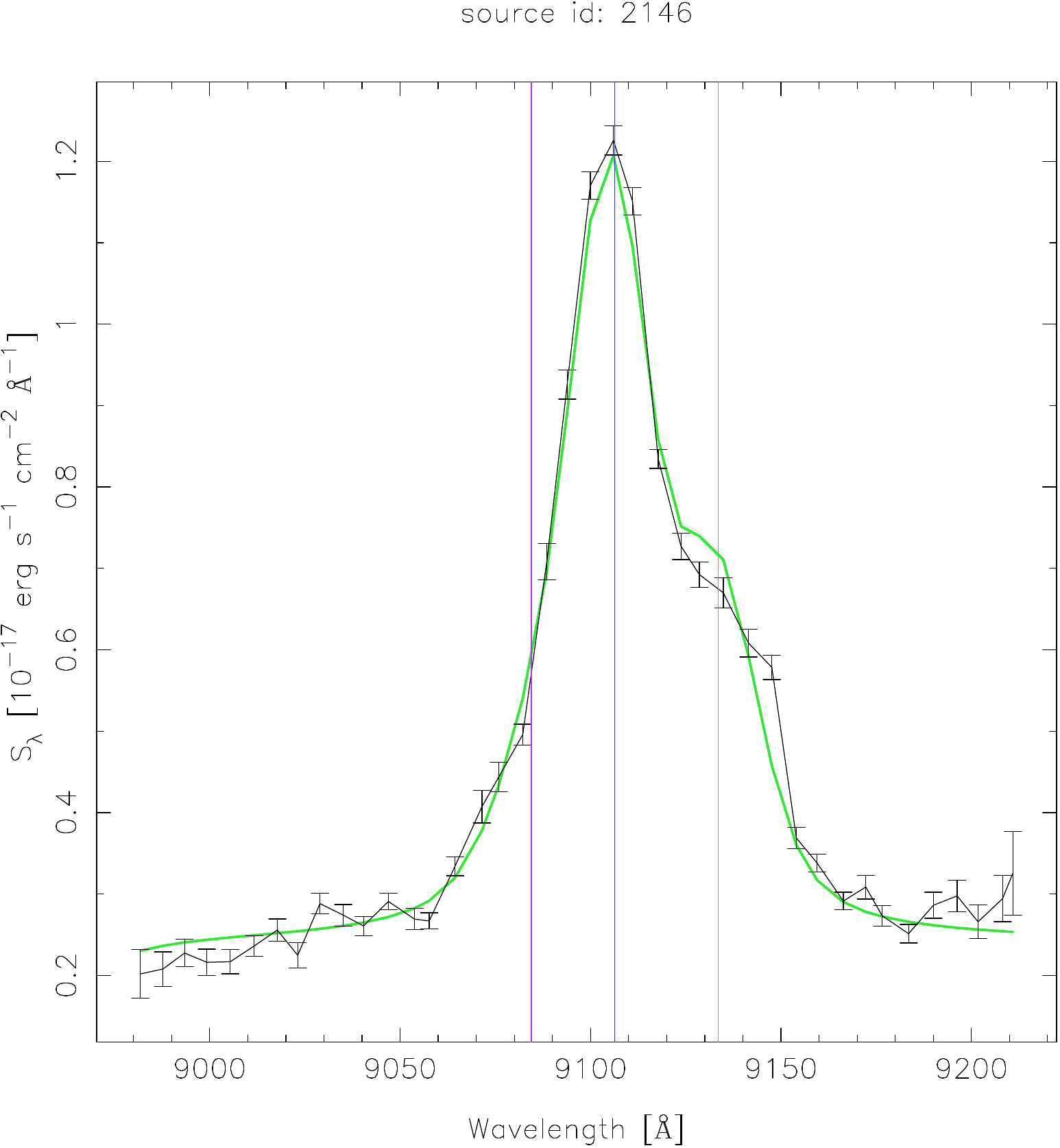}
                                                \includegraphics[width=0.24\textwidth]{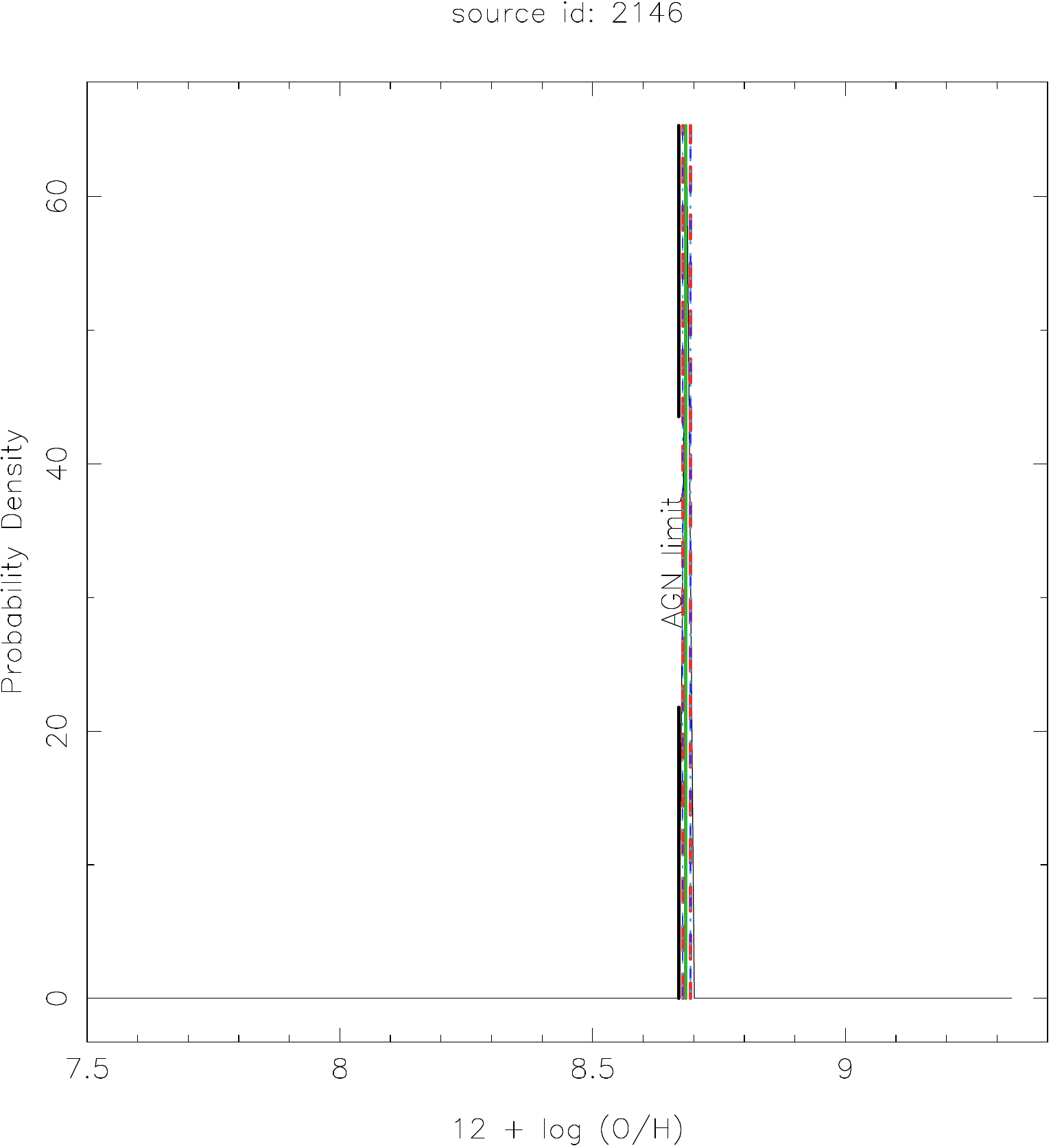}
                                                \includegraphics[width=0.24\textwidth]{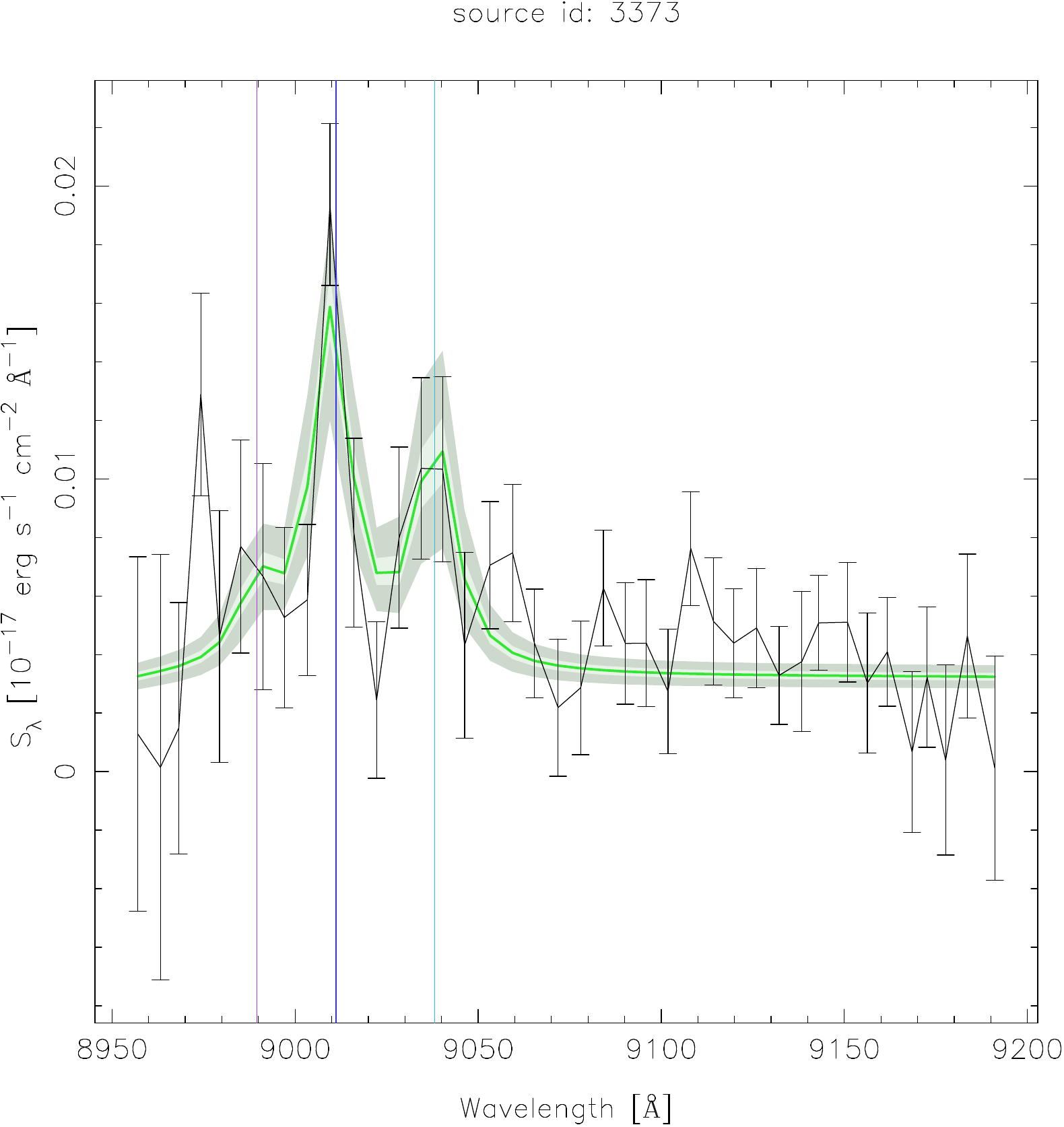}
                                                \includegraphics[width=0.24\textwidth]{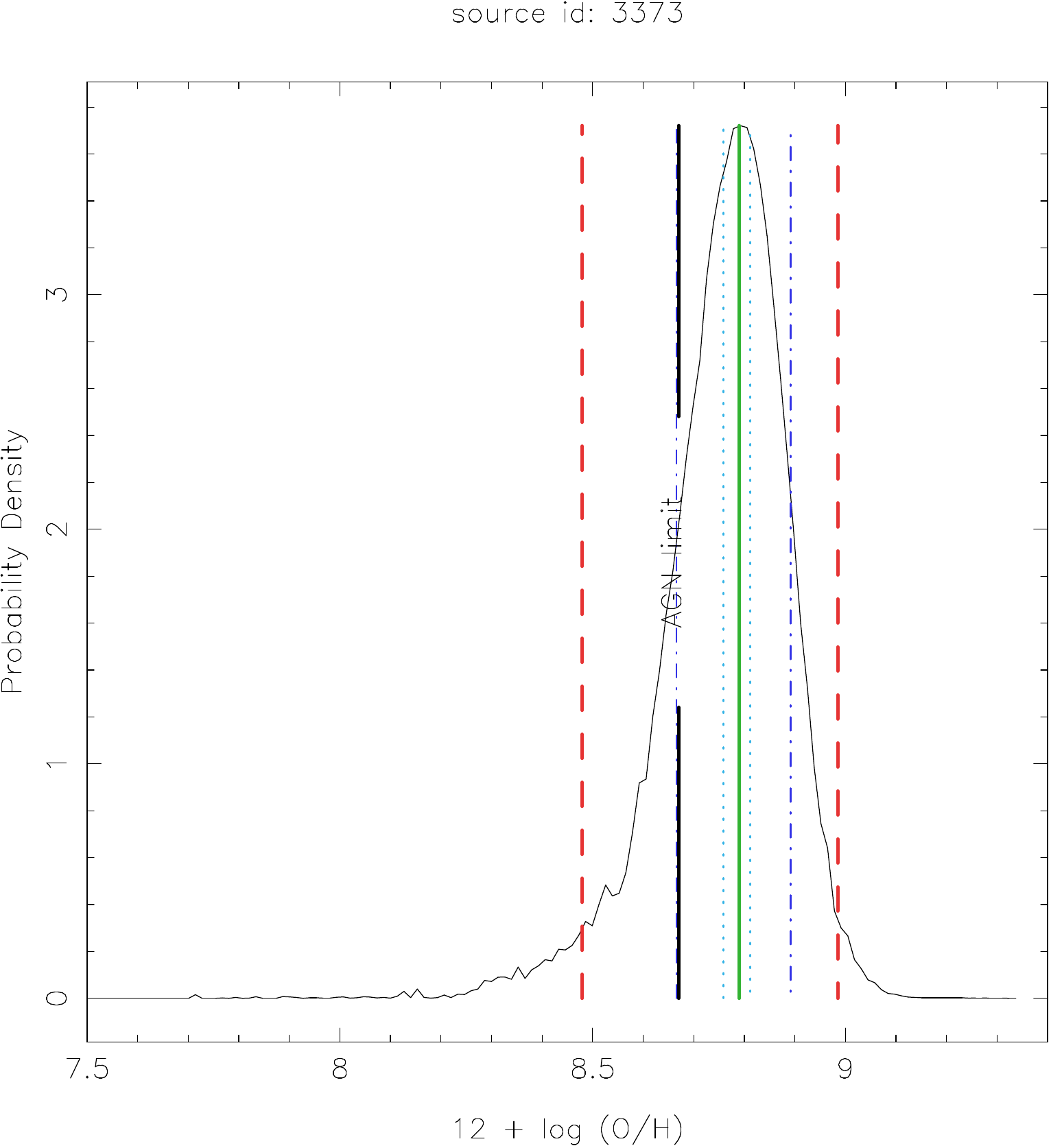}\\
                                                \includegraphics[width=0.24\textwidth]{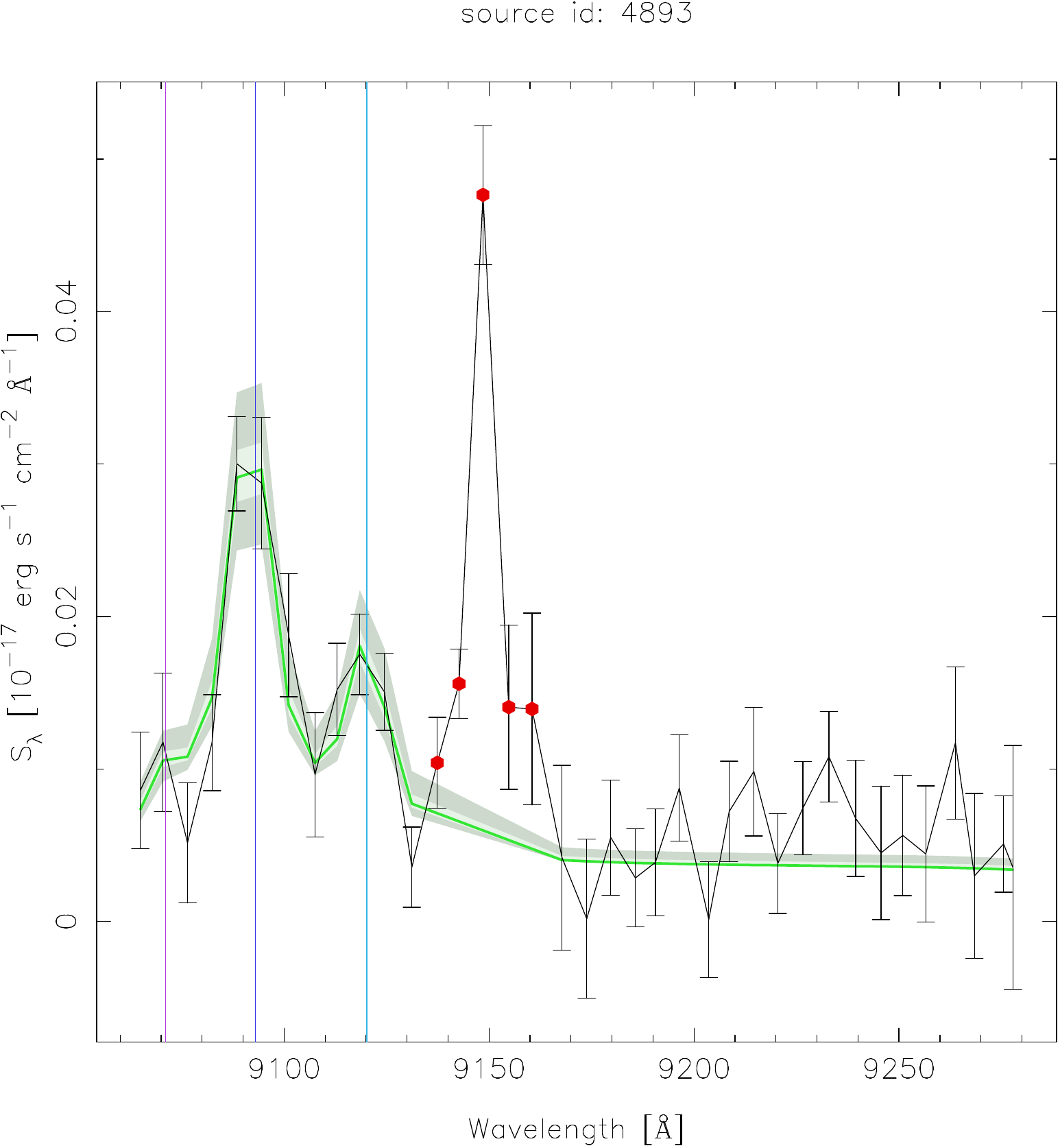}
                                                \includegraphics[width=0.24\textwidth]{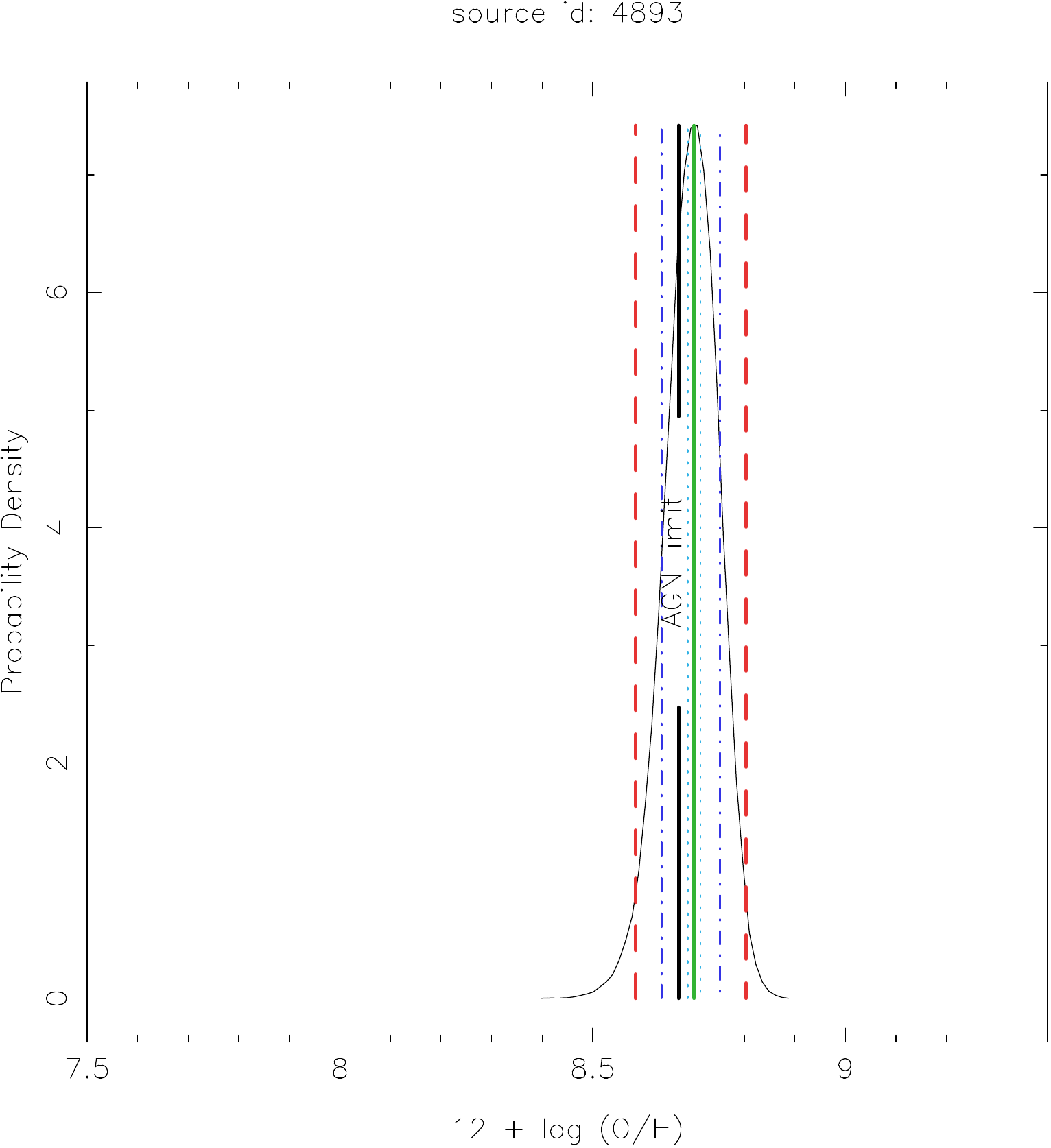}
                                                \caption{\label{fig:AppepsAGN} As Fig. \ref{fig:Appeps1}, but for the three AGN candidates.}
                                        \end{center}
                                \end{figure*}

                                \begin{figure*}[t!]
                                        \begin{center}
                                                \includegraphics[width=0.24\textwidth]{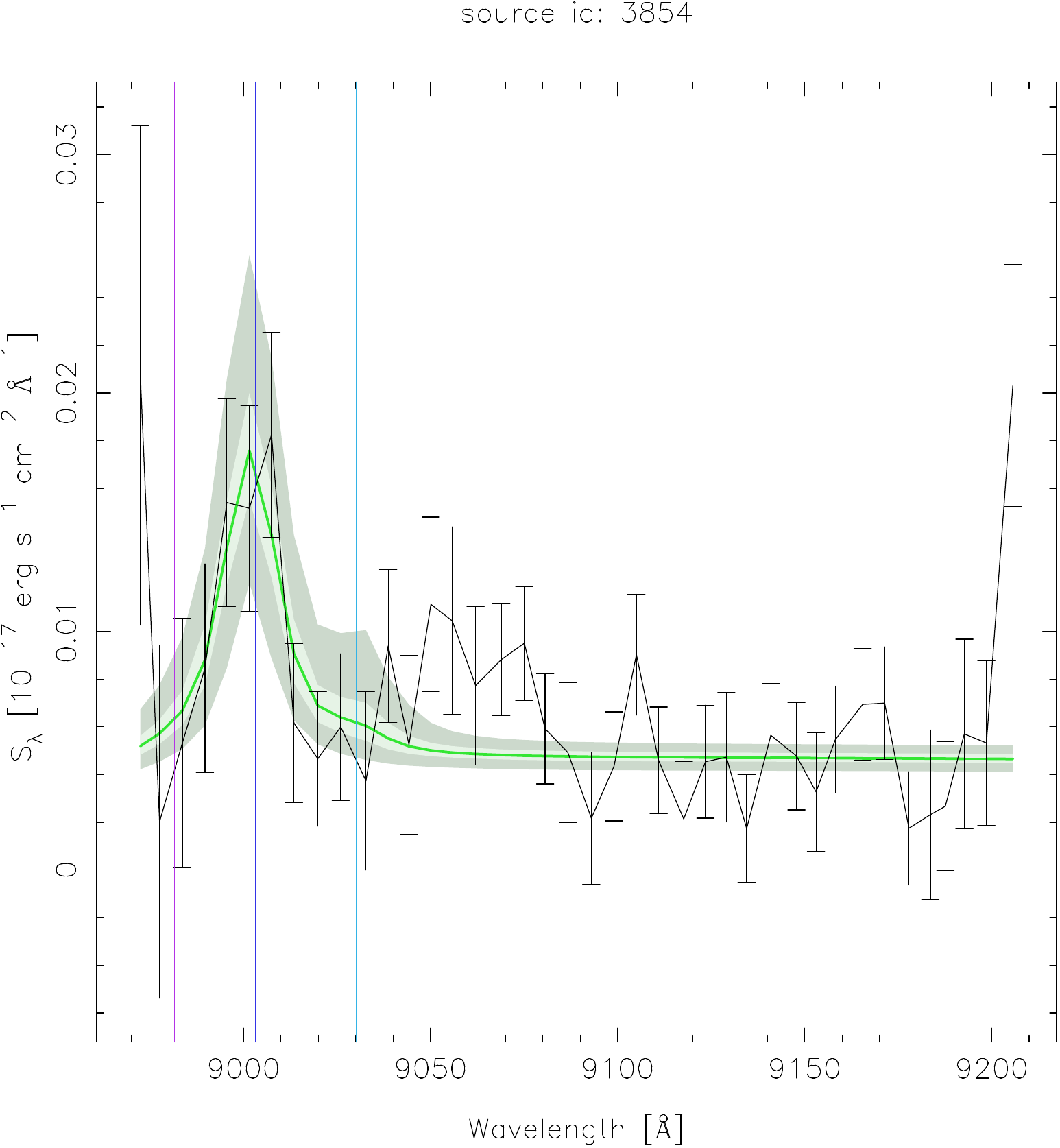}
                                                \includegraphics[width=0.24\textwidth]{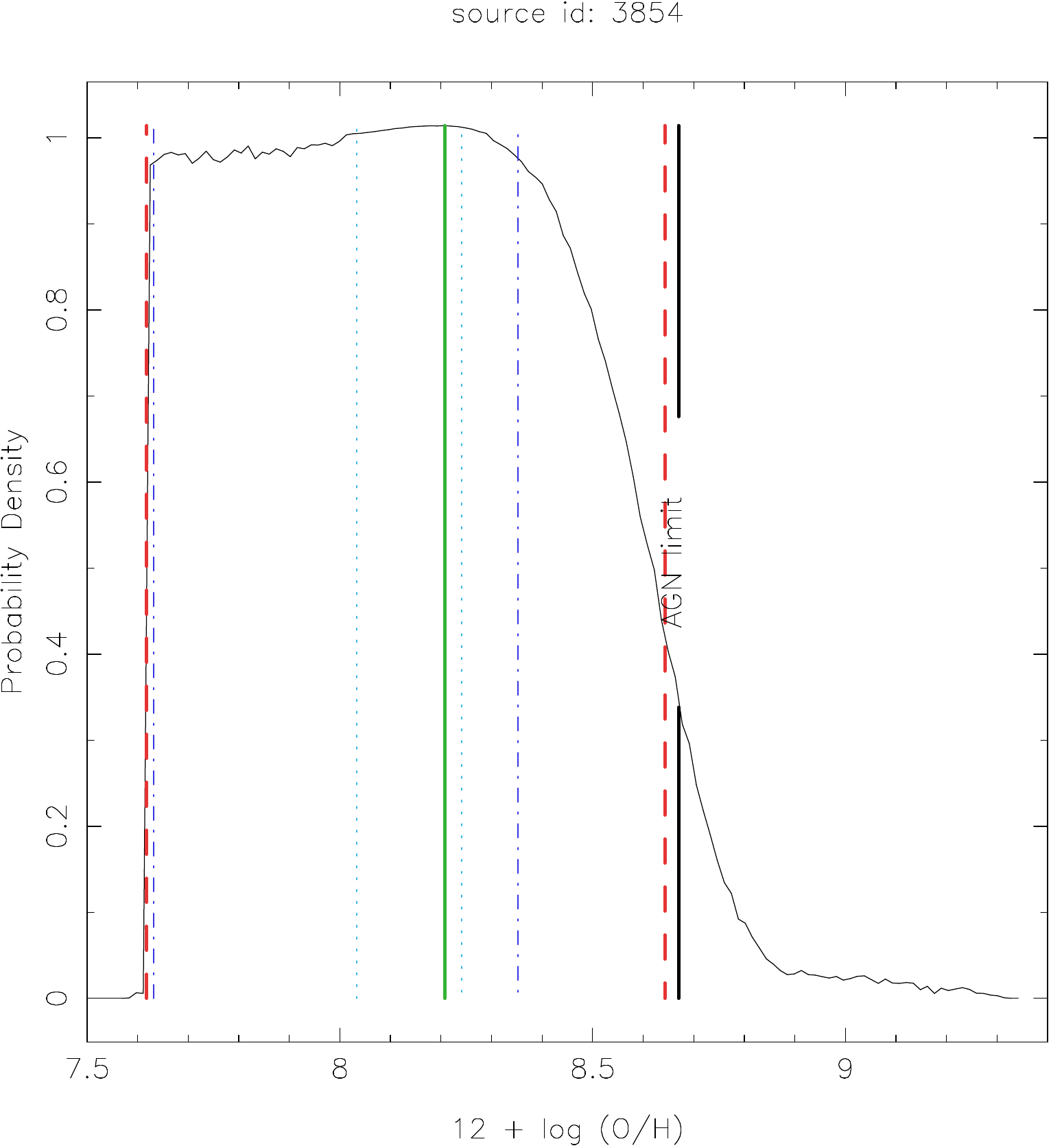}
                                                \includegraphics[width=0.24\textwidth]{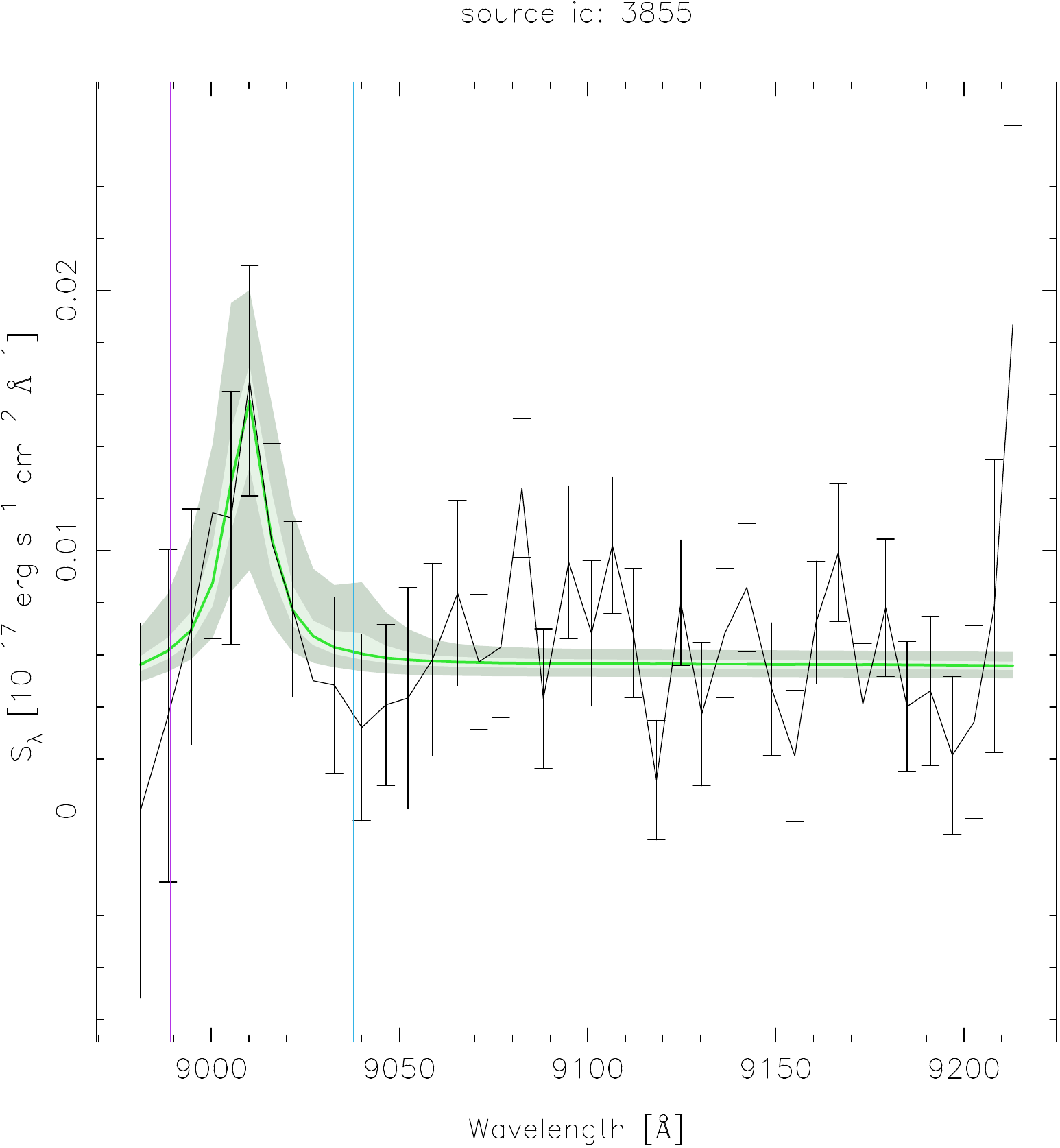}
                                                \includegraphics[width=0.24\textwidth]{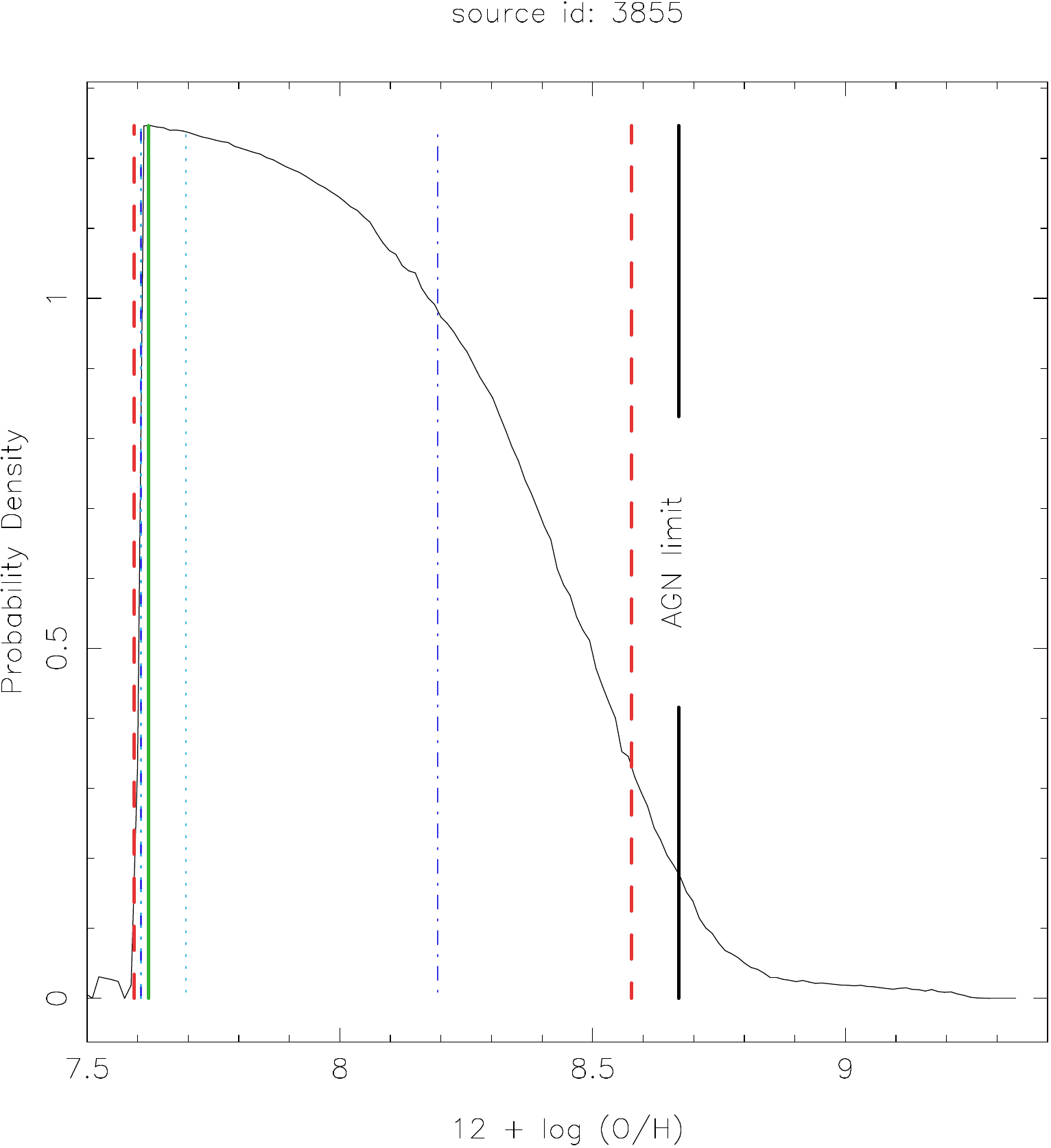} \\
                                                \includegraphics[width=0.24\textwidth]{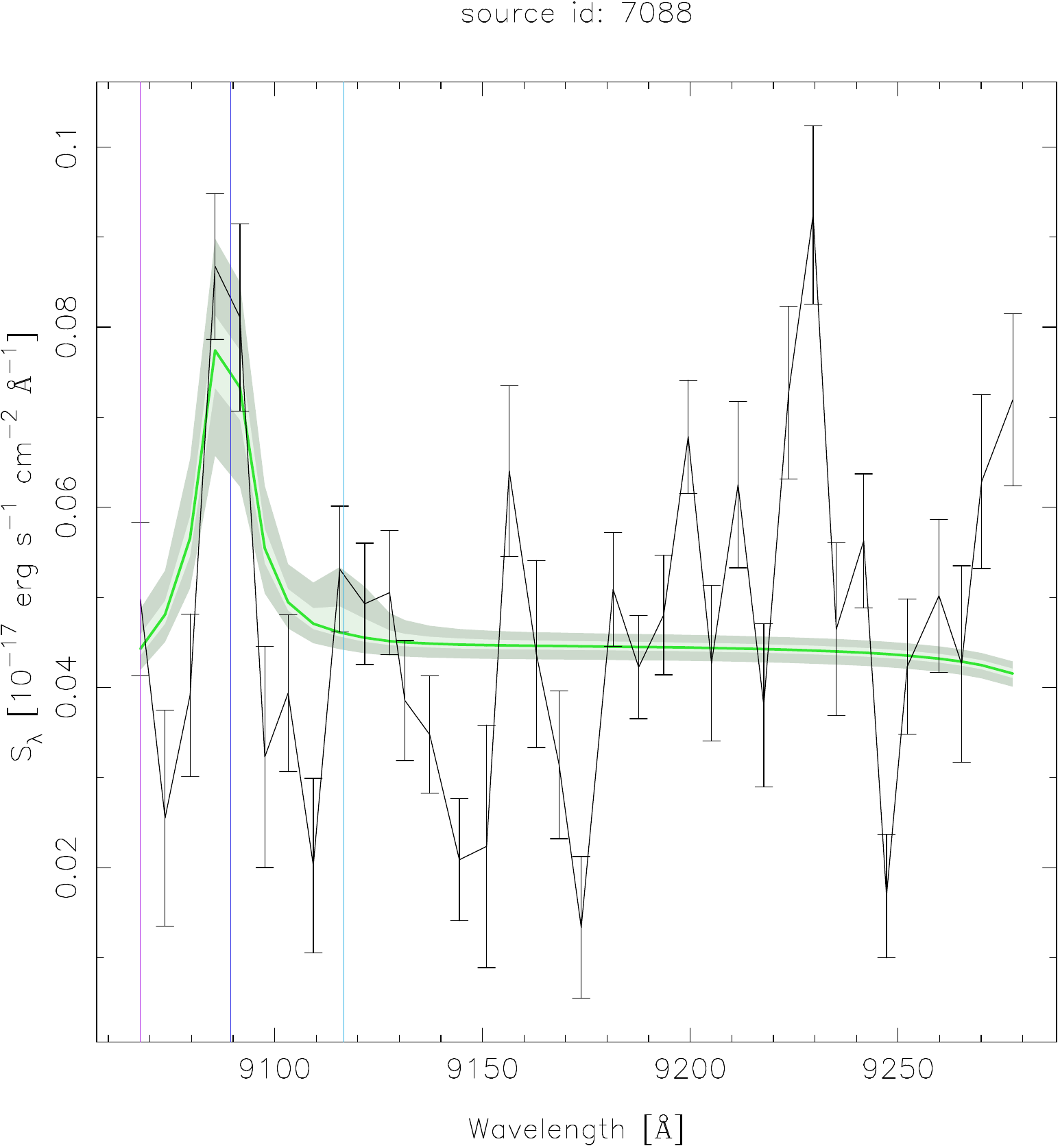}
                                                \includegraphics[width=0.24\textwidth]{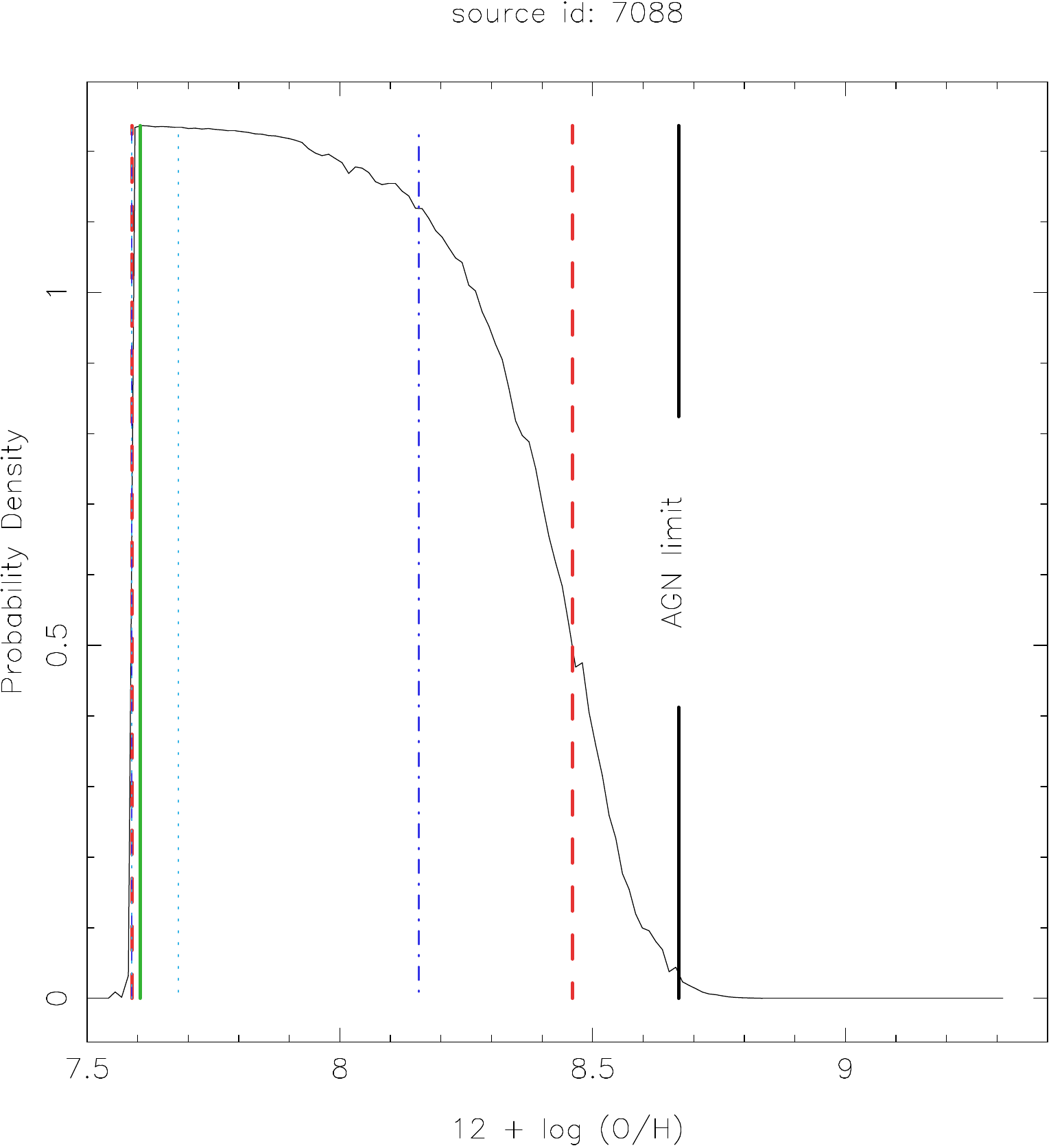}
                                                \includegraphics[width=0.24\textwidth]{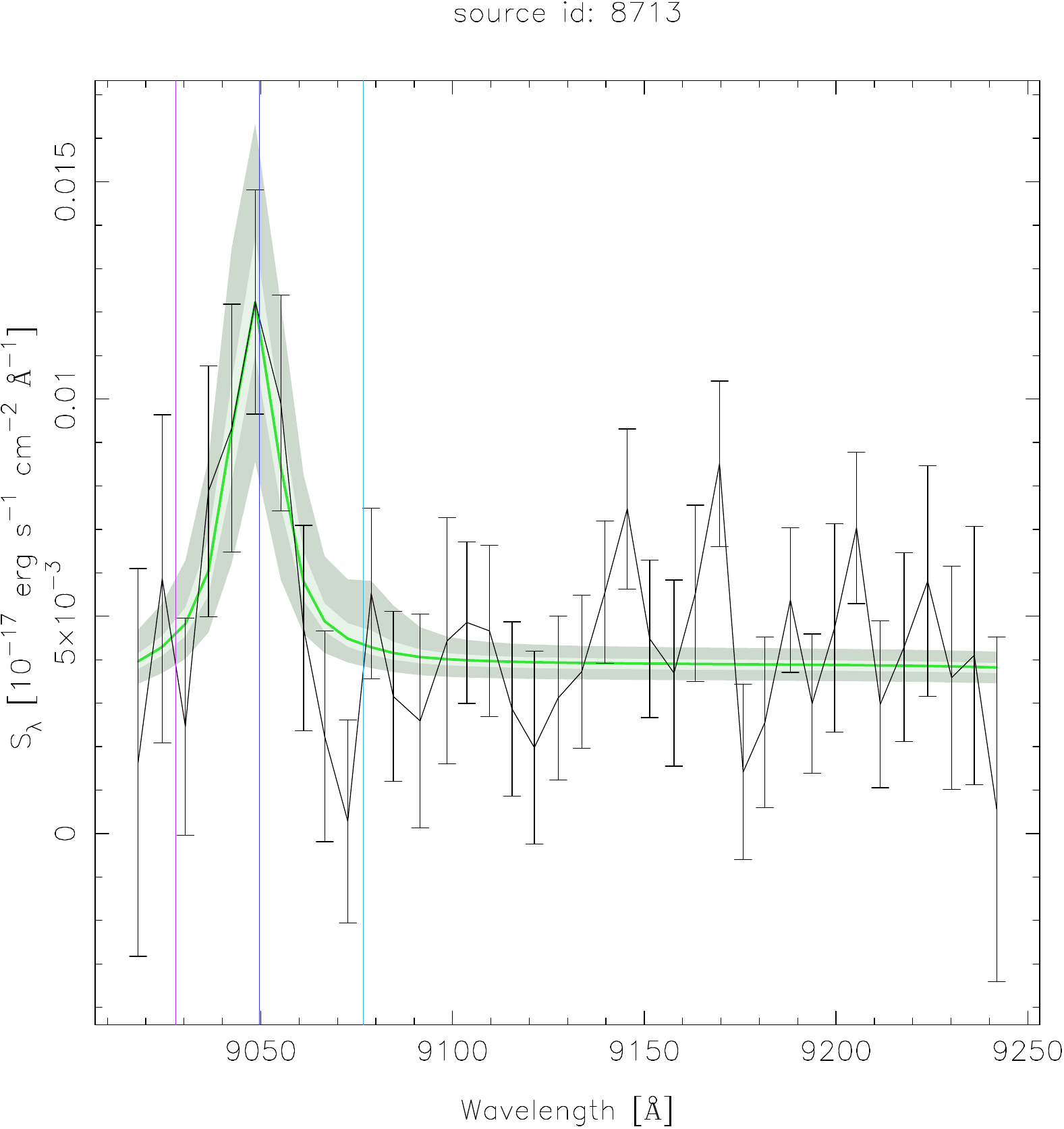}
                                                \includegraphics[width=0.24\textwidth]{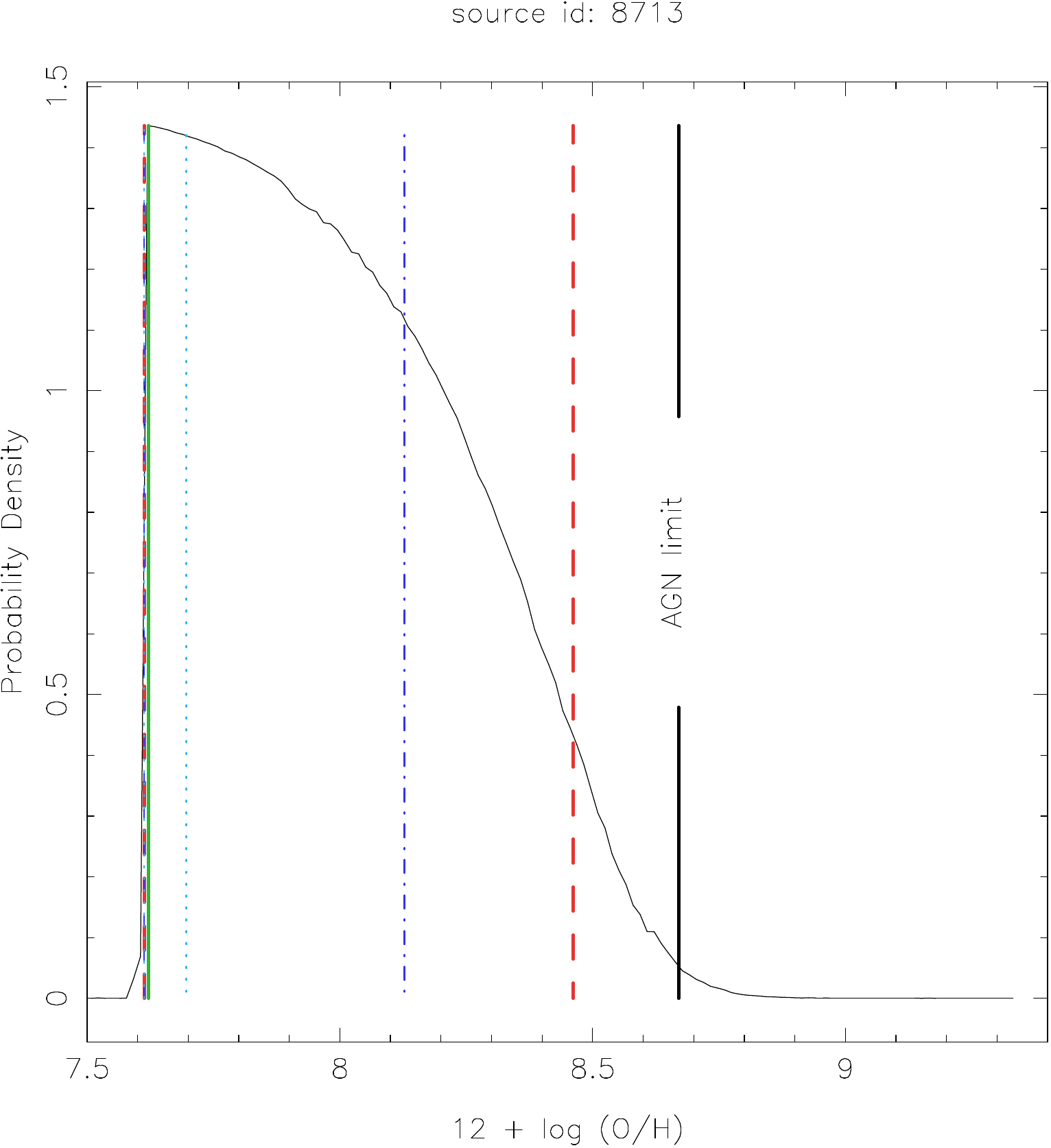} \\
                                                \includegraphics[width=0.24\textwidth]{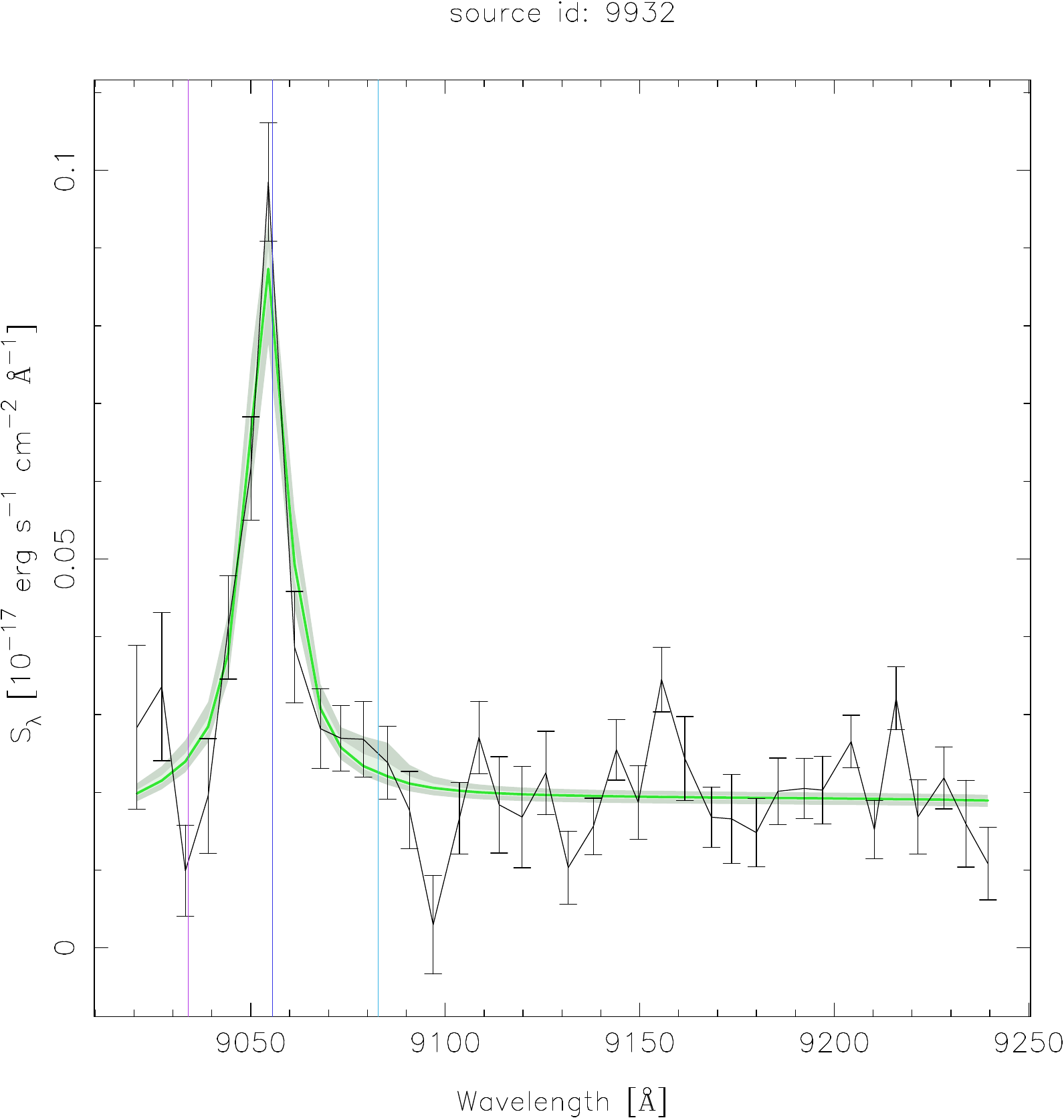}
                                                \includegraphics[width=0.24\textwidth]{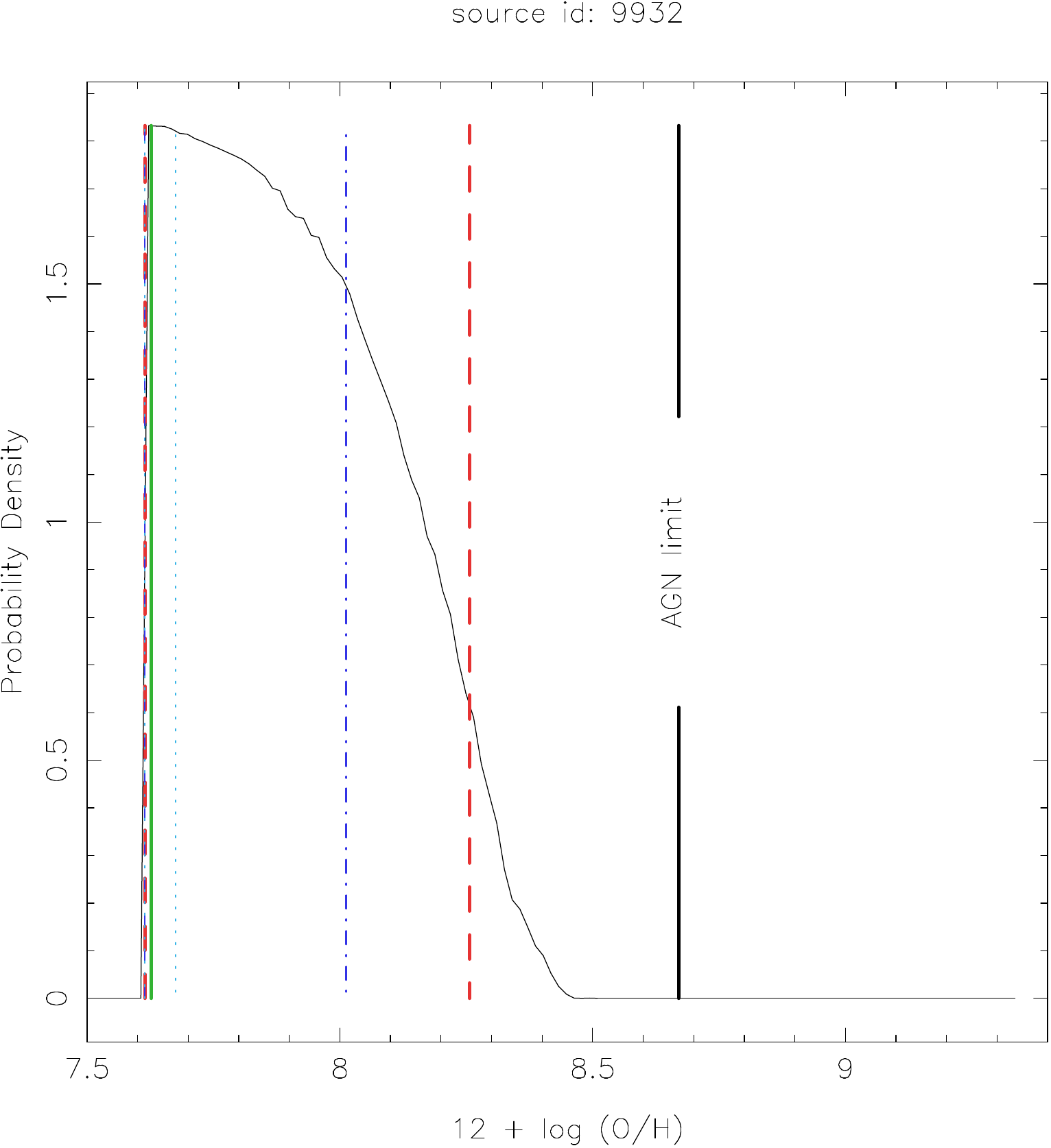}
                                                \caption{\label{fig:Appeps3} As Fig. \ref{fig:Appeps1}, but for the five sources whose \Nii\ line flux is lower than the OTELO line flux limit (low-flux sources).}
                                        \end{center}
                                \end{figure*}

                                Throughout this paper we made use of the inverse deconvolution of the pseudo-spectra (PS). In this appendix we describe how this deconvolution was performed.
                                
                                For a given PS we obtain (a) the maximum value of the PS $f_\mathrm{PS}^\mathrm{max}$ and it observed error $\sigma_\mathrm{PS}^\mathrm{max}$, and (b) an estimate of the continuum level defined as the median flux of the PS $f_\mathrm{c}$ and the deviation $\sigma_\mathrm{med,c}$ around this value (mean of the square root of the differences around $f_\mathrm{c}$). We also have a redshift guess value {\zguess} that is obtained from visual inspection of the PS in our GUI web-based tool, and it is assumed to be accurate up to the third digit, that is, with an uncertainty $\Delta z\,=\,0.001$. Finally, we define the wavelength region of the PS where the lines can be detected in such a way that it contains at least 2.5 slices (15 \AA\ being 6 \AA\ the average width of the RFT) below the redshifted \Nii$\lambda$6548 line and 2.5 slices above the redshifted \Nii$\lambda$6583 line:
                                
                                \begin{eqnarray}
                                \lambda_\mathrm{min} & =&  6548 \cdot ( 1 + z_{\rm GUESS} - \Delta z) - 15 \nonumber \\
                                \lambda_\mathrm{max} & =&  6583 \cdot ( 1 + z_{\rm GUESS} + \Delta z) + 15 \nonumber \\
                                \end{eqnarray}

                                We assumed a model spectrum as a rest-frame spectrum defined by Gaussian profiles of the \Nii$\lambda$6548, \HA,\ and \Nii$\lambda$6583 lines defined by their amplitude and a common line width for three lines, and a constant continuum level. The model to be compared with the observational data is therefore defined by following parameters: $f_\mathrm{mod}(z,f_c,\sigma_\mathrm{line},f_{\mathrm{N~II}~\lambda 6548}, f_\mathrm{H\alpha}, f_{\mathrm{N~II}~\lambda 6583})$. We note that there is an intrinsic assumption that this type of model is correct and that there is no additional component in the PS. 
                                In two cases (source {\tt id: 625} in Fig. \ref{fig:Appeps1}, and {\tt id: 8074} in Fig. \ref{fig:Appeps2}) the PS show \HAN\ and \Sii\ lines. For these we masked the points that correspond to \Sii\ emission lines. In other cases (source {\tt id: 4893} in Fig. \ref{fig:AppepsAGN}) the PS showed an additional component (probably an artefact) that was also removed.
                                
                                We performed $10^6$ Monte Carlo simulations over the model parameter space using values of {\zguess}, $f_\mathrm{c}$, $\sigma_\mathrm{med,c}$, $f_\mathrm{PS}^\mathrm{max}$ , and $\sigma_\mathrm{PS}^\mathrm{max}$ to constrain the parameter space (see below). Each simulation was convolved with the RFT filter system, and a synthetic PS $f_{\mathrm{PS mod},i}$ for each of the $N$ slices of the PS was obtained. This synthetic PS was compared with the observed PS, and the following likelihood function was obtained for each model and was compared with each point $i$ in the PS:
                                
                                \begin{equation}
                                {\cal L}(f_{\mathrm{mod}}| f_{\mathrm{PS}}) \propto \Pi_{i=1}^{N} w_i \exp(-\frac{\chi^2_i}{2})
                                ,\end{equation}
                                
                                \noindent where $\chi^2$ is given in the usual way,
                                
                                \begin{equation}
                                \chi^2_i =  \left(\frac{f_{\mathrm{mod},i} - f_{\mathrm{obs},i}}{\sigma_{\mathrm{obs},i}}\right)^2 
                                ,\end{equation}
                                
                                \noindent and $w_i$ is a normalized weight function defined as 
                                
                                \begin{equation}
                                w_i = \begin{cases}
                                \mathrm{MAX}(1,\frac{f_{\mathrm{obs},i}}{f_\mathrm{c, mod}}) & \mathrm{if} \, \lambda_i \in [\lambda_\mathrm{min},\lambda_\mathrm{max}] \\
                                1 & \mathrm{otherwise}
                                \end{cases}
                                .\end{equation}
                                
                                The use of the weight function aims to give a greater weight in the fit to the points where the line is presented and are above the continuum level, and in particular, to fit the core of the \HA\ line. In practical cases, the weight function only has an effect in the objects with faint lines.
                                
                                In order to sample the likelihood function, the first 30\% (300\,000) of the simulations made use of the following priors:
                                
                                \begin{enumerate}
                                        \item The redshift \z\ follows a flat distribution in the range {\zguess}$\pm\,0.001.$ 
                                        \item  The dispersion of the line, $\sigma_\mathrm{line}$, varies following a flat distribution from 20 to 500 km/s, which is translated into a range of 0.43 to 11 \AA\ for the \HA\ wavelength. This value is used for the three lines.
                                        \item The continuum value follows a flat distribution in the interval $[\mathrm{MAX}(0,f_\mathrm{c} - \sigma_\mathrm{med,c}),f_\mathrm{c} + \sigma_\mathrm{med,c}].$ 
                                        \item The rest-frame $f_\mathrm{H\alpha}$ line flux varies following a flat distribution in the interval $[f_\mathrm{c}, f_\mathrm{{max}}],$ where $f_\mathrm{{max}}$ is given by
                                        \begin{align} 
                                        f_\mathrm{max} = &(f_\mathrm{med,c} + f^{\mathrm{PS}}_{\mathrm{max}} + \,\,\,\,\,\,3\,\sigma(f_{\mathrm{PS}}^{\mathrm{max}})) \cdot \nonumber \\ 
                                        &\sigma_\mathrm{line\, max} \cdot (1 + z_\mathrm{max})\cdot\sqrt{2\pi}. \label{eq.RFHA} 
                                        \end{align}
                                        \item  The rest-frame $f_{\mathrm{N~II}~\lambda 6583}$ line flux varies following a flat distribution between $0.006 \cdot f_\mathrm{H\alpha}$ and $f_\mathrm{max}$.
                                        \item $\text{The }f_{\mathrm{N~II}~\lambda 6548}$ line flux is fixed to $1/3 \cdot f_{\mathrm{N~II}~\lambda 6583.}$
                                        
                                \end{enumerate}
                                
                                When these 300\,000 simulations were obtained, we computed a preliminary likelihood function.
                                This likelihood was sampled in a regular interval with 200 bins over the range in which each parameter varied.
                                
                                This likelihood function was marginalized for each parameter (i.e. we made a projection of the PDF on each particular parameter and the corresponding PDF was normalized). We obtained the marginalized likelihood using a regular interval of 200 bins over the range in which each parameter varied, and we obtained the range that contains 99\% of the simulations around the value with the maximum likelihood. The next 20\% (200\,000) of the simulations were restricted to the parameters that sample this 99\% confidence interval. This process was repeated five more times (using 10\% (100\,000) of the simulations at each step) with the limits obtained by the 85\%, 75\%, 68\%, 40\%, and 10\% confidence intervals around the most probable value.
                                
                                After this process we had a final PDF shape for each parameter, as well as all the required additional quantities such as the EWs, the N2 parameter, or the metallicities. We note that the asymmetry of the PDF implies that the mean and the standard deviation (squared of the variance) are not informative to describe the distribution. For this reason we used the mode of the resulting PDF as the reference value, and we obtained the region around the mode that included 10\%, 25\%, 50\%, 68\%, and 90\% of the area. In the main text the error bars show the 25\% and 68\% confidence interval, which also gives an idea of the asymmetry of the associated distribution.
                                
                                We show in Figs. \ref{fig:Appeps1} and \ref{fig:Appeps2} the observed PS, the best-fit PS, and the envelope of the PS simulations where all parameters are in the 25\% confidence interval (light green) and 68\% confidence interval (light grey) for the SFG for which the  \HAN\ line was detected. We also show the PDF of $12 + \log (\mathrm{O/H})$ that we obtained from the Monte Carlo simulations for each source where the mode (our reference value), the 25\%, 68\%, and 90\% confidence intervals around the mode are shown in vertical lines. We also show with a black vertical line the empirical division between SFG and AGN at $12 + \log (\mathrm{O/H}) = 8.67$ (N2$\,=\,-0.4$). Figures \ref{fig:AppepsTrans} and \ref{fig:Appeps3} show similar information for the possible transitional dwarf galaxy {\tt id: 3137}, and for sources whose \Nii\ line flux is lower than the OTELO line flux limit (low-flux sources), respectively; we note that for the latter, the PDF of $12 + \log (\mathrm{O/H})$ has a step-like shape that reflects the large uncertainty in the \Nii\ line flux. Finally, Figure \ref{fig:AppepsAGN} show the same information for AGN sources; we kept the PDF of $12 + \log (\mathrm{O/H})$  for comparison with the previous plots.
                                
                                We show in Table \ref{tab:Results} the modal values and 68\% confidence intervals around the mode for
                                        $f_\mathrm{H\alpha}$ and  EW(H$\alpha$) (after correction for stellar absorption), $f_{\mathrm{N~II}~\lambda 6583}$, and  $12 + \log (\mathrm{O/H})$. See Section \ref{sec:metallicity} for details about the uncertainties of $z_{\rm OTELO}$.
                                
                                \section{Uncertainty in stellar masses and SFR}
                                \label{App_uncert_mass}
                                
                                As explained in Sect. \ref{sec:methods}, stellar masses were computed using the prescription of LS18, which makes use of colour \gmenosi\ and the $i$ absolute magnitude. The associated uncertainty was computed making use of the $R$ subroutine {\tt propagate,} which makes an estimate of the error by a first- and second-order propagation of any input recipe, as well as by Monte Carlo simulations. The uncertainty quoted in Table \ref{tab:Results} corresponds to the variance obtained from 10\,000 the Monte Carlo simulations using that sub-routine.
                                
                                Finally, the uncertainty in the SFR quoted in Table \ref{tab:Results} requires some cautions. First of all, the SFR requires the use of an extinction value. We used the value that was obtained with the best-fit solution with the \texttt{LePhare} software and used empirical galaxy templates. This choice implies that the extinction value is a lower limit because the intrinsic extinction of the templates are unknown. In addition,  \texttt{LePhare} does not provide an evaluation of the uncertainty in $E(B-V)$, therefore this uncertainty was not included in the SFR. Finally, we recall that our $f_\mathrm{H\alpha}$ uncertainty does not correspond to a variance (where standard propagation theory would apply), but to a confidence interval. Taking these considerations into account, we computed the uncertainty in the SFR by applying the same formulae as we used to obtain the nominal SFR (the extinction correction used by \citealt{Ly2012ApJ...747L..16L} and the SFR estimate using the \citealt{Kennicutt2012ARA&A..50..531K} relation) to the $f_\mathrm{H\alpha}$ values that define the 68\% confidence interval. We stress that the uncertainty values would be underestimated because they do not include all the possible sources of uncertainty (interstellar extinction in particular).

                                \onecolumn
                                
                                
                                \onecolumn
                                \begin{landscape}
                                        \begin{longtable}{lllllllllll}
                                                \caption{\label{tab:Results} Catalogue of the OTELO \HA\ SFG sample. AGN candidates are separated in the lower part of the table. Values (i.e. metallicity and SFR) for these sources need to be taken with caution.}\\
                                                \hline\hline
                                                idobj & EBV  &  z$_{\rm OTELO}^{(a)}$ & Mass & $12 + \log (\mathrm{O/H})^{(b,c)}$ & $f_\mathrm{H\alpha}$$^{(b,d)}$ &  $f_{\mathrm{N~II}~\lambda 6583}$$^{(b)}$ & EW(H$\alpha$)$^{(b,d)}$ & SFR$_{\rm H\alpha}$$^{(e)}$ & n & r$_{\rm e}$    \\
                                                &  [mag] & & [log(M$_{*}$/M$_{\odot}$)] &  & [$10^{-17}$ erg s$^{-1}$ cm$^{-2}$]& [$10^{-17}$ erg s$^{-1}$ cm$^{-2}$]& [\AA] &  [\msun yr$^{-1}$] & & [kpc] \\
                                                \hline
                                                \endfirsthead
                                                \multicolumn{4}{c}%
                                                {\tablename\ \thetable\ -- \textit{Continued from previous page}} \\
                                                \hline
                                                idobj & EBV  &  z$_{\rm OTELO}^{(a)}$ & Mass &$12 + \log (\mathrm{O/H})^{(b,c)}$ & $f_\mathrm{H\alpha}$$^{(b,d)}$ &  $f_{\mathrm{N~II}~\lambda 6583}$$^{(b)}$ & EW(H$\alpha$)$^{(b,d)}$ & SFR$_{\rm H\alpha}$$^{(e)}$ & n & r$_{\rm e}$    \\
                                                &  [mag] & & [log(M$_{*}$/M$_{\odot}$)] &  & [$10^{-17}$ erg s$^{-1}$ cm$^{-2}$]& [$10^{-17}$ erg s$^{-1}$ cm$^{-2}$]& [\AA] &  [\msun yr$^{-1}$] & & [kpc]    \\
                                                \hline
                                                \endhead
                                                \hline \multicolumn{4}{r}{\textit{Continued on next page}} \\
                                                \endfoot
                                                \hline
                                                \endlastfoot
                                                \multicolumn{11}{l}{Sources with \Nii$\lambda$6583 above OTELO line flux limit.}\\
                                                \multicolumn{11}{l}{}\\
                                                625$^{(1)}$  &0.1&0.371& $9.80\pm 0.10$ & $8.56^{+0.03}_{-0.04}$ & $23.5^{+1.3}_{-1.1}$    & $5.21^{+0.41}_{-0.53}$    & $35.6^{+2.1}_{-1.6}$    & $1.12^{+0.06}_{-0.05}$ & ----- & ----- \\ 
                                                754          &0.3&0.404& $9.33\pm 0.12$ & $8.34^{+0.17}_{-0.39}$ & $1.14^{+0.10}_{-0.09}$  & $0.119^{+0.092}_{-0.080}$ & $19.6 \pm 1.7$          & $0.21 \pm 0.02$        & $2.61 \pm 0.09$ &1.29\\ 
                                                1130         &0.1&0.380& $9.38\pm 0.10$ & $8.52^{+0.09}_{-0.11}$ & $1.80^{+0.18}_{-0.16}$  & $0.39^{+0.14}_{-0.13}$   & $7.65 \pm 0.76$         & $0.09 \pm 0.01$        & $1.92 \pm 0.02$ &3.841\\ 
                                                1873         &0.1&0.379& $9.39\pm 0.10$ & $8.16^{+0.07}_{-0.10}$ & $32.07^{+0.51}_{-0.54}$ & $1.60^{+0.48}_{-0.38}$   & $88.2^{+1.8}_{-1.6}$    & $1.60 \pm 0.02$        & $2.15 \pm 0.02$ &6.199\\ 
                                                2747         &0.1&0.380& $9.33\pm 0.10$ & $8.50^{+0.04}_{-0.04}$ & $6.95^{+0.30}_{-0.33}$  & $1.38^{+0.25}_{-0.21}$    & $17.57^{+0.78}_{-0.93}$ & $0.35 \pm 0.02$       & $1.78$         &0.832\\ 
                                                3089         &0.0&0.365& $6.89\pm 0.14$ & $8.43^{+0.16}_{-0.80}$ & $0.41^{+0.53}_{-0.09}$  & $0.060^{+0.046}_{-0.060}$ & $126^{+168}_{-33}$      & $0.011^{+0.014}_{-0.002}$ & $1.95\pm 1.00$ &0.475\\ 
                                                3106         &0.0&0.368& $8.03\pm 0.17$ & $8.43^{+0.17}_{-0.45}$ & $0.52\pm 0.10$          & $0.077^{+0.053}_{-0.060}$ & $48.5^{+9.9}_{-11}$     & $0.014 \pm 0.003$      & $7.75 \pm 0.10$ &2.931\\ 
                                                3137$^{(2)}$ &0.1&0.389& $8.27\pm 0.10$ & $8.64^{+0.10}_{-0.16}$ & $1.88^{+0.26}_{-0.18}$  & $0.67^{+0.17}_{-0.22}$   & $47.8^{+7.4}_{-5.4}$  & $0.10 \pm 0.01$        & $1.25 \pm 0.01$ &0.702\\ 
                                                7990         &0.0&0.379& $7.97\pm 0.11$ & $8.30^{+0.16}_{-0.45}$ & $0.63\pm 0.08$          & $0.056^{+0.054}_{-0.046}$ & $35.1^{+4.9}_{-5.4}$   & $0.018 \pm 0.002$      & $1.47 \pm 0.07$ &1.854\\ 
                                                8074         &0.1&0.372& $9.44\pm 0.10$ & $8.53^{+0.04}_{-0.03}$ & $27.0^{+1.1}_{-1.6}$   & $5.58 \pm 0.41$           & $67.2^{+2.7}_{-4.4}$   & $1.29^{+0.05}_{-0.08}$ & $0.65 \pm 0.01$ &3.528\\ 
                                                10512        &0.0&0.380& $8.05\pm 0.11$ & $7.92^{+0.10}_{-0.01}$ & $3.29^{+0.12}_{-0.13}$ & $0.06^{+0.06}_{-0.00}$    & $231^{+19}_{-18}$       & $0.094^{+0.003}_{-0.004}$ & $0.73 \pm 0.07$ &0.84\\ 
                                                \hline 
                                                \multicolumn{11}{l}{Sources whose \Nii$\lambda$6583 is lower than the OTELO line flux limit.}\\
                                                \multicolumn{11}{l}{}\\
                                                3854         &0.3&0.372& $7.71\pm 0.13$ & $8.21^{+0.14}_{-0.58}$ & $0.332^{+0.085}_{-0.082}$ & $0.020^{+0.063}_{-0.020}$ & $69^{+24}_{-20}$      & $0.05 \pm 0.01 $   & $ 1.24 \pm 0.26$ & 1.41\\ 
                                                3855         &0.0&0.373& $7.35\pm 0.12$ & $7.62^{+0.57}_{-0.02}$ & $0.22\pm 0.07$            & $0.001^{+0.041}_{-0.001}$ & $37^{+16}_{-13}$      & $0.006 \pm 0.002$  & $ 1.37 \pm 0.25$ & 1.017\\ 
                                                7088         &0.1&0.385& $8.76\pm 0.11$ & $7.61^{+0.55}_{-0.02}$ & $0.89^{+0.13}_{-0.16}$    & $0.01^{+0.11}_{-0.01}$    & $19.5^{+3.6}_{-3.8}$  & $0.046^{+0.007}_{-0.008}$ & $2.47 \pm 0.06$ & 2.142\\ 
                                                8713         &0.0&0.379& $7.47\pm 0.13$ & $7.62^{+0.51}_{-0.01}$ & $0.2\pm 0.05$             & $0.001^{+0.022}_{-0.001}$ & $50^{+16}_{-13}$      & $0.006\pm 0.001$          & $ 0.95 \pm 0.15$ & 1.835\\ 
                                                9932         &0.0&0.380& $8.22\pm 0.11$ & $7.63^{+0.39}_{-0.01}$ & $1.40^{+0.11}_{-0.12}$    & $0.008^{+0.073}_{-0.008}$ & $71.0^{+6.8}_{-7.7}$  & $0.04 \pm 0.003$          & $0.86 \pm 0.05$ & 2.878\\ 
                                                \hline 
                                                \hline
                                                \multicolumn{11}{l}{AGN candidates.}\\
                                                \multicolumn{11}{l}{}\\
                                                2146         &0.3&0.387& $9.54\pm 0.10$ & $8.68 \pm 0.01$        & $30.88 \pm 0.20$        & $12.92^{+0.20}_{-0.00}$ & $123.22^{+0.36}_{-0.53}$  & $5.06 \pm 0.03$     & $ 2.18 \pm 0.03$ & 1.069\\ 
                                                3373         &0.0&0.373& $7.12\pm 0.13$ & $8.79^{+0.10}_{-0.12}$ & $0.25^{+0.05}_{-0.04}$  & $0.157 \pm 0.05$        & $74^{+22}_{-17}$         & $0.007 \pm 0.001$   & -----  & -----  \\ 
                                                4893         &0.1&0.385& $7.41\pm 0.21$ & $8.70^{+0.05}_{-0.07}$ & $0.61 \pm 0.06$         & $0.20 \pm 0.05$         & $165^{+16}_{-32}$        & $0.032 \pm 0.003$ & ----- & ----- \\ 
                                                \hline  
                                        \end{longtable}
                                        \raggedright\small{
                                                $^{(1)}$ Object selected via DEEP2 redshift, see Section \ref{sec:sample_selection} for details.\\
                                                $^{(2)}$ Possible transitional dwarf galaxy. \\
                                                $^{(a)}$ Uncertainty on \zotelo\ is $\pm\,0.001\,(1+z_{\rm OTELO})$.\\
                                                $^{(b)}$ Uncertainty defined as the 68\% confidence interval around the mode.\\
                                                $^{(c)}$ Uncertainty in metallicity does not include calibration uncertainty.\\
                                                $^{(d)}$ Corrected for stellar absoption.\\
                                                $^{(e)}$ Lower limit; see Appendix \ref{App_uncert_mass} for the uncertainty evaluation.
                                        }
                                        
                                \end{landscape}


                        \end{appendix}
                        
\end{document}